 \newcommand{\hii}{\relax \ifmmode {\mbox H\,{

shape ii}}\else H\,{\scshape ii}\fi}
\newcommand{\mi}{\relax \ifmmode {\mu{\mbox m}}\else $\mu$m\fi}
\newcommand{\ha}{\relax \ifmmode {\mbox H}\alpha\else H$\alpha$\fi}
\newcommand{\hb}{\relax \ifmmode {\mbox H}\beta\else H$\beta$\fi}
\newcommand{\degree}{\hbox{$^\circ$}}
\newcommand{\sii}{\relax \ifmmode {\mbox S\,{\scshape ii}}\else S\,{\scshape ii}\fi}
\newcommand{\siii}{\relax \ifmmode {\mbox S\,{\scshape iii}}\else S\,{\scshape iii}\fi}
\newcommand{\nii}{\relax \ifmmode {\mbox N\,{\scshape ii}}\else N\,{\scshape ii}\fi}
\newcommand{\oi}{\relax \ifmmode {\mbox O\,{\scshape i}}\else O\,{\scshape i}\fi}
\newcommand{\oii}{\relax \ifmmode {\mbox O\,{\scshape ii}}\else O\,{\scshape ii}\fi}
\newcommand{\oiii}{\relax \ifmmode {\mbox O\,{\scshape iii}}\else O\,{\scshape iii}\fi}

\newcommand{\rdostres}{\relax \ifmmode {\,\mbox{R}}_{\rm 23}\else \,\mbox{R}$_{\rm 23}$\fi} 

\newcommand{\gsim}{\hbox{\rlap{\lower.55ex\hbox{$\sim$}} \kern-.3em
\raise.4ex \hbox{$>$}}}
\newcommand{\lsim}{\hbox{\rlap{\lower.55ex\hbox{$\sim$}} \kern-.3em
\raise.4ex \hbox{$<$}}}

\documentclass{aa}  

\usepackage{graphicx}
\usepackage{txfonts}
\begin{document}

   \title{The dependence of oxygen and nitrogen abundances on stellar mass from the CALIFA survey
   \thanks{Based on observations collected at the Centro Astron\'omico
      Hispano Alem\'an (CAHA) at Calar Alto, operated jointly by the Max-Planck
      Institut f\"ur Astronomie and the Instituto de Astrof\'{\i}sica de Andaluc\'{\i}a (CSIC).}
        }

\titlerunning{O/H and N/O gradients in CALIFA galaxies}


   \author{
        E. P\'erez-Montero  \thanks{epm@iaa.es} \inst{\ref{iaa}}
          \and
          R. Garc\'\i a-Benito \inst{\ref{iaa}}
          \and
          J.M. V\'\i lchez \inst{\ref{iaa}}
          \and
          S. F. S\'anchez \inst{\ref{iaa},\ref{unam}}
          \and
          C. Kehrig \inst{\ref{iaa}}
          \and
          B. Husemann \inst{\ref{leib}}
          \and
            S. Duarte Puertas \inst{\ref{iaa}}
            \and
            J. Iglesias-P\'aramo \inst{\ref{iaa}}
            \and
             L. Galbany \inst{\ref{mia-chile},\ref{u-chile}}
             \and
              M. Moll\'a \inst{\ref{ciemat}}
              \and
             C. J. Walcher \inst{\ref{leib}}
             \and
              Y. Ascas\'\i bar \inst{\ref{uam}}
              \and
              R. M. Gonz\'alez Delgado \inst{\ref{iaa}}
              \and
              R. A. Marino \inst{\ref{zurich},\ref{ucm}}
              \and
               J. Masegosa \inst{\ref{iaa}}
                \and
                 E. P\'erez \inst{\ref{iaa}}
                 \and
                 F. F. Rosales-Ortega \inst{\ref{inaoe}}
                 \and
                 P. S\'anchez-Bl\'azquez \inst{\ref{uam},\ref{uc-chile}}
                 \and
                 J. Bland-Hawthorn \inst{\ref{sydney}}
                 \and
                  D. Bomans \inst{\ref{bochum}}
                  \and
                  \'A. R. L\'opez-S\'anchez \inst{\ref{aat},\ref{mua}}
                   \and
                   B. Ziegler \inst{\ref{viena}}
                   \and
           {The CALIFA collaboration} 
}

 \offprints{E. P\'erez-Montero \email{epm@iaa.es}}

   \institute{
\label{iaa} Instituto de Astrof\'\i sica de Andaluc\'\i a - CSIC. Apdo. de correos 3004, E-18080, Granada, Spain
 \and
 \label{unam}Instituto de Astronom\'\i a,Universidad Nacional Auton\'oma de Mexico, A.P. 70-264, 04510, M\'exico,D.F.
  \and
  \label{leib} Leibniz-Institut f\"ur Astrophysik Potsdam (AIP), An der Sternwarte
16, D-14482 Potsdam, Germany.
 \and
 \label{mia-chile} Millennium Institute of Astrophysics. Chile
 \and
  \label{u-chile} Departamento de Astronom\'\i a, Universidad de Chile, Camino El Observatorio 1515, Las Condes, Santiago, Chile
        \and
        \label{ciemat} Departamento de Investigaci\'on B\'asica, CIEMAT, Avda. Complutense 40 E-28040 Madrid, Spain.
       \and     
       \label{uam}Departamento de F\'isica Te\'orica, Universidad Aut\'onoma de Madrid, 28049 Madrid, Spain.
       \and
        \label{zurich} Department of Physics, Institute for Astronomy, ETH Z\"urich, CH-8093, Z\"urich, Switzerland
        \and
        \label{ucm} Departamento de Astrof\'{i}sica y CC$.$ de la Atm\'{o}sfera, Facultad de CC$.$ F\'{i}sicas,  Universidad Complutense de Madrid, Avda.\,Complutense s/n, 28040 Madrid, Spain.
               \and
       \label{inaoe} Instituto Nacional de Astrof{\'i}sica, {\'O}ptica y Electr{\'o}nica, Luis E. Erro 1, 72840 Tonantzintla, Puebla, Mexico
        \and
        \label{uc-chile} Instituto de Astrof\'\i sica, Facultad de F\'\i sica, Pontificia Universidad Cat\'olica de Chile, 4860 Av. Vicu\~na Mackenna, Santiago, Chile. 
        \and
        \label{sydney}  Sydney Institute for Astronomy, School of Physics A28,
University of Sydney, NSW 2006, Australia.
        \and
        \label{bochum} Astronomical Institute of the Ruhr-University Bochum Universitaetsstr. 150, 44801 Bochum, Germany.
        \and
        \label{aat} Australian Astronomical Observatory, PO Box 915, North Ryde, NSW 1670, Australia
        \and
        \label{mua} Department of Physics and Astronomy, Macquarie University, NSW 2109, Australia
        \and
        \label{viena} University of Vienna, T\"urkenschanzstrasse 17, 1180 Vienna, Austria.
}              

   \date{Received: ---, Accepted: ---}

 
  \abstract
   {The study of the integrated properties of star-forming galaxies is central
to understand their formation and evolution. Some of these properties are extensive 
and therefore their analysis require totally covering and spatially resolved 
observations. Among these properties, metallicity can be defined in spiral discs by means of 
integral field spectroscopy (IFS) of individual \hii\ regions. The simultaneous analysis of the abundances
of primary elements, as oxygen, and secondary, as nitrogen, also provides clues about the star
formation history and the processes that shape the build-up of spiral discs.}
   {Our main aim is to analyse simultaneously O/H and N/O abundance ratios in \hii\ 
regions in different radial positions of the discs in a large sample of spiral galaxies to obtain the
slopes and the characteristic abundance ratios that can be related to their integrated properties.
}   
   {We analysed the optical spectra of individual selected \hii\ regions extracted from a sample of 
350 spiral galaxies of the CALIFA survey. We calculated total O/H abundances and, for the first time, 
N/O ratios using the semi-empirical routine {\sc Hii-Chi-mistry}, which, according to
\cite{hcm}, is consistent with the direct method and reduces the uncertainty
in the O/H derivation using [\nii] lines owing to the dispersion in the O/H-N/O relation.
Then we performed linear fittings to the abundances as a function 
of the de-projected galactocentric distances.}
   {
The analysis of the radial distribution both for O/H and N/O in the non-interacting
galaxies reveals that both average slopes are negative, but  a non-negligible fraction of
objects have a flat or even a positive gradient (at least 10\% for O/H and 4\% for N/O).
The slopes normalised to the effective radius appear to have a slight dependence on the total stellar mass 
and the morphological type, as late low-mass
objects tend to have flatter slopes. No clear relation is found, however, to explain the presence of inverted
gradients in this sample, and there is no dependence between the average slopes and the
presence of a bar.
The relation between the resulting O/H and N/O linear fittings at the effective radius
is much tighter (correlation coefficient $\rho_s$ = 0.80)
than between O/H and N/O slopes ($\rho_s$ = 0.39) or for O/H and N/O
in the individual \hii\ regions ($\rho_s$ = 0.37).
These O/H and N/O values at the effective radius also
correlate very tightly (less than 0.03 dex of dispersion) with total luminosity and 
stellar mass. The relation with other integrated properties, such as star formation rate, colour, or
morphology, can be understood only in light of the found relation with mass.
}
   {}


   \maketitle
%

\section{Introduction}

The study of galaxy formation and evolution through different cosmological epochs
is currently a hot topic of research. The amount of available data coming from large ground- and
space-based telescopes at increasingly higher redshift ($z$) and better spatial resolution is growing
substantially thanks to many deep sky surveys, such as zCOSMOS \citep{zcosmos}, 
MASSIV \citep{massiv}, MANGA \citep{manga}, or SAMI \citep{sami}.
Among the different integrated properties of galaxies to investigate, metal content is of special
interest to understand the evolutionary status and pathway followed towards the
current state of galaxies.
Metallicity ($Z$) is a crucial parameter since its value is the fruit of the conversion
of unprocessed gas onto stars, which transform light into heavier elements and
spew these elements out into the interstellar medium (ISM) or even into the intergalactic medium (IGM) to take potentially part in a new loop of star formation and chemical enrichment. Then, $Z$ traces 
the age, star formation history, and efficiency of galaxies, and it is therefore linked 
to their other integrated properties (e.g. \citealt{rh94}).
A thorough dissection of the observed integrated properties in the galaxies
of the local Universe gives us the opportunity to look deeper into the fundamental relations 
observed in a Universe at larger $z$ under a new light. 

In this way, the integrated values of extensive properties and the
spatial distribution of intensive properties, which are not
always homogeneous, can be investigated in local galaxies.  
In the case of $Z$, it is known to correlate with
the total luminosity (e.g. \citealt{lequeux79, skillman89,lamareille04}) and 
the total stellar mass of galaxies (e.g. \citealt{t04}).
These correlations were found to exist even in spiral galaxies, where spectroscopy
of individual \hii\ regions helps to trace a negative gradient of $Z$ (e.g. \citealt{gs87, vce92,
zaritsky94, marino12}). 
Despite the slopes, these gradients are found to be steeper for fainter
galaxies \citep{garnett97}.
A common mass-independent slope can be obtained when
the slope is  divided by the effective or the isophotal radius (e.g. \citealt{diaz89, S14_grad}) and
one can recover for each galaxy a characteristic $Z$ value that also correlates with
their integrated properties (e.g. \citealt{pcv04, mk06}). This normalisation has been also 
used by several authors to propose a common slope for all non-interacting disc galaxies
(e.g. \citealt{bresolin09, rosales11, bresolin12, S14_grad, ho15, marino15}).
This could be consistent with a common inside-out scenario of growth
of the disc owing to the fall of gas from the halo in all galaxies (e.g. \citealt{pichon11}).

However there are some indications that the steepness of the gradient of $Z$ in galaxy discs,
and hence the derivation of a characteristic value, can be influenced by other factors.
Among these, the presence of a bar or the morphological
type has been proposed (e.g. \citealt{vce92, zaritsky94}), although also see \cite{S14_grad}.
There is also evidence for the presence of flat or even inverted gradients 
in interacting galaxies (e.g. \citealt{rich12,lopezs15}) or 
in some low-mass and/or gas-rich galaxies
(e.g. \citealt{werk10, werk11,moran12}). 
On the other hand, \cite{carton15} find steeper gradients in gas-rich galaxies.
According to the study of the radial distribution of
metals at higher $z$ the fraction of galaxies 
with an inverted or a flat gradient could be higher owing to a higher 
rate of interactions and mergers in a denser Universe \citep{cresci10, queyrel12,troncoso14}.

When $Z$ is obtained from the optical spectra of \hii\ regions it is
usually expressed in terms of the abundance of a primary element as oxygen (12+log(O/H),
hereafter O/H).
Nevertheless, exploring  
the spatial variation of the chemical abundance ratio of
a primary element or a secondary element, such as nitrogen
(log(N/O), hereafter N/O) at same time, can shed some  light on the
origin of the flattening or the inversion of chemical gradients in discs and the
global relations with the galactic integrated properties.
\cite{edmunds90} shows that this kind of ratio could be relatively unaffected by
hydrodynamical effects, such as infalls of metal-poor gas or 
outflows of enriched material.
This makes the relation between N/O and the integrated stellar mass 
(hereafter MNOR; \citealt{pmc09}) correlate even in those galaxies whose $Z$ is dissevered
from its star formation history \citep{kh05}, such as is the case in
some extreme emission line galaxies (e.g. \citealt{amorin10, pm11}).
In this way the MNOR, contrary to the mass-metallicity relation (MZR), 
does not have any dependence on the integrated star formation rate (SFR) \citep[SFR;][]{pm13}. 

Besides the study of the spatial variation of N/O can help
to derive more accurate abundance gradients 
in spatially resolved galaxies.
Since N spreads over a much wider range of variation than O (e.g. 
\citealt{thuan10}) its study can be achieved at a better precision from
the analysis of the emission lines from \hii\ regions.
In addition, possible deviations from the typical O/H-N/O relation
(i.e. growing N/O for growing O/H in the production regime of
secondary N) can also be useful to explore variations in the star 
formation efficiency \citep{molla06}, the stellar yields, or initial 
mass function (IMF) changes across discs (e.g. \citealt{mattsson09}),
as most  O is produced by short-living
massive stars and most of N is ejected to the ISM by long-living low- and
intermediate-mass stars in the metal-rich discs of galaxies.

Inasmuch as we aim to investigate the behaviour of the abundance
gradients in local galaxies, we need a sufficiently large sample of
objects covered in two dimensions with integral field spectroscopy (IFS)
in the optical range,
as is the case of CALIFA (Calar Alto Legacy Integral
Field Area survey; \citealt{S12}). 
The sample of galaxies studied in the CALIFA survey is well characterised and
all integrated properties are well studied and can be correlated with the
resulting characteristic chemical abundances.
In Section 2 of this paper we describe the CALIFA
survey and how individual \hii\ regions were extracted from the data cubes.

The rest of the paper is organised as follows:
In Section 3.1 we give the details of the stellar continuum subtraction,
emission-line measurement and, in section 3.2, the selection of star-forming \hii\ regions.
The determination of O/H and N/O abundances is
described in Section 3.3. 
This task was carried out using the model-based, semi-empirical 
code {\sc Hii-Chi- mistry}. As shown in \cite{hcm}, this method is 
totally consistent with the direct method when it is compared with  
\hii\ regions with available empirical electron temperatures.
This method allows the simultaneous study of O/H and N/O abundance ratios without
assuming that any previous relation between them is taken from a calibration 
sample, as occurs when a strong-line method is used based on [\nii]
emission lines. 
The global relation between O/H and N/O for
all \hii\ regions is presented in Section 4.
In Section 5 we describe the calculations of radial chemical variations fittings
and the statistical distributions of the resulting slopes
and the values at the effective radius. We also analyse the possible correlations
between the resulting slopes, the characteristic abundances, and
other integrated properties of the galaxies, such as luminosity, stellar
mass, SFR, colour, morphological type, and the presence of a bar
in Section 6.
 In Section 7, we also discuss
in detail the case of those objects with a flat or inverted $Z$ gradient.
Finally, in Section 8, we summarise our results and present our
conclusions.


\section{Sample of galaxies and extraction of \hii\ regions}

The galaxies observed by CALIFA were selected from the SDSS survey
encompassing a wide range of their integrated properties, including mass, luminosity, colours,
and morphologies \citep{walcher14}. This gives us insight into many of their
spatially resolved properties with a statistically significant point 
of view in the local Universe (0.005 $< z <$ 0.03). 

The details of the survey, sample, observational strategy, and reduction 
are explained in \cite{S12} and \cite{GB15}. All galaxies were observed using PMAS \citep{Roth} 
in the PPAK configuration \citep{kelz}, covering a hexagonal field of view (FoV) 
of 74$\arcsec$ $\times$ 64$\arcsec$ sufficient to map the full optical extent of 
the galaxies up to two to three disc effective radii. This is possible because of 
the diameter selection of the sample \citep{walcher14}. The observing strategy 
guarantees a complete coverage of the FoV with a final pixel size 1\arcsec, 
corresponding to $\approx$ 1 kpc at the average $z$t of the survey.

The CALIFA mother sample consists of 939 galaxies, but for this work we
used a sample of 350 objects observed with the gratings V500 and V1200.
The observed wavelength range and spectroscopic resolution (3745–7500 \AA, $\lambda/\Delta\lambda 
\approx$ 850, for the low-resolution setup and 3650-4800 \AA, $\lambda/\Delta\lambda 
\approx$ 1500, for the high resolution setup) are more than sufficient to explore the most 
prominent ionised gas emission lines, from [\oii]$\lambda$ 3727 to [\sii]$\lambda$6731, 
on the one hand, and to deblend and subtract the underlying stellar population, on the other 
(e.g. \citealt{S12, Kehrig12, CF13}). The dataset was reduced using version 1.5 of the CALIFA pipeline, 
whose modifications with respect to the pipeline presented in \cite{S12} are described in detail in \cite{GB15}. 
In summary, the data fulfill the predicted quality-control requirements with a spectrophotometric accuracy 
that is better than 15\% everywhere within the wavelength range, both absolute and relative with a depth 
that allows us to detect emission lines in individual \hii\ regions as weak as $\approx$  10$^{-17}$ erg $\cdot$ s$^{-1}$ 
$\cdot$ cm$^{-2}$, and with a signal-to-noise ratio of S/N $\approx$ 3–5,
even in the case of [\oiii] auroral lines presented in \cite{m13}. 

The final product of the data reduction is a regular-grid data cube with $x$ 
and $y$ coordinates that indicate the right ascension and declination of the target 
and $z$ a common step in wavelength. The CALIFA pipeline also provides the propagated 
error cube, a proper mask cube of bad pixels, and a prescription of how to handle the 
errors when performing spatial binning (due to covariance between adjacent pixels after 
image reconstruction). These data cubes, together with the ancillary data described 
in \cite{walcher14}, are the basic starting points of our analysis.

Emission-line regions in the CALIFA galaxies were segregated in the resulting data cubes and their
corresponding spectra were extracted using the semi-automatic routine {\sc hiiexplorer}
\footnote{Available at \url {http://www.caha.es/sanchez/HII\_explorer}}.
The details of this routine are described well in \cite{S12b} and \cite{Fabian12}, but basically  the routine is based on 
the assumption that emission-line regions can be found as clumpy and isolated H$\alpha$ peaks
of emission of several arcseconds, which correspond to diameters of a few hundreds of
parsecs. The total number of selected emission-line regions from the 350 analysed galaxies
following this procedure is 15\,757.

 \section{Analysis of the extracted regions}

\subsection{Emission-line measurement}

The stellar underlying population spectral emission was fitted and removed from the extracted 
individual spectra of the extracted emission-line regions to avoid absorption components of
the Balmer emission lines. This continuum was fitted using the programme {\sc fit3d} v. 2.0
\footnote{Available at \url {http://www.astroscu.unam.mx/~sfsanchez/FIT3D/index.html}}
\citep{S06,S11}. This package fits the continuum as a linear combination of several
single stellar populations (SSPs) of different ages and $Z$ taken from the MILES project
\citep{Vazdekis, Falcon}. A stellar extinction law by \cite{Cardelli} was used
in the fitting process, assuming a value of R$_v$ = 3.1,  a simple dust screen
distribution, and that all SSPs have the same dust attenuation. 

Once the stellar continuum is subtracted, {\sc fit3d} performed a Gaussian function fitting to 
measure the fluxes of the
most prominent emission lines of the extracted spectra. To reduce the number of free
parameters, this algorithm forces the systemic velocity and the velocity dispersion to be the
same for emission lines produced by the same ion (e.g. [\oiii] 4959, 5007 \AA\AA) and
to apply a multi-component analysis for those blended emission lines. 
For the aim of this work, we used [\oii] 3727 \AA, \hb, [\oiii] 5007 \AA,
\ha, [\nii] 6584 \AA, and [\sii] 6717, 6731 \AA\AA\ with a signal-to-noise ratio better
than 3 as obtained using the above described automatic fitting procedure.

\subsection{Selection of star-forming \hii\ regions}

\begin{figure}
   \centering
   \includegraphics[width=9.5cm]{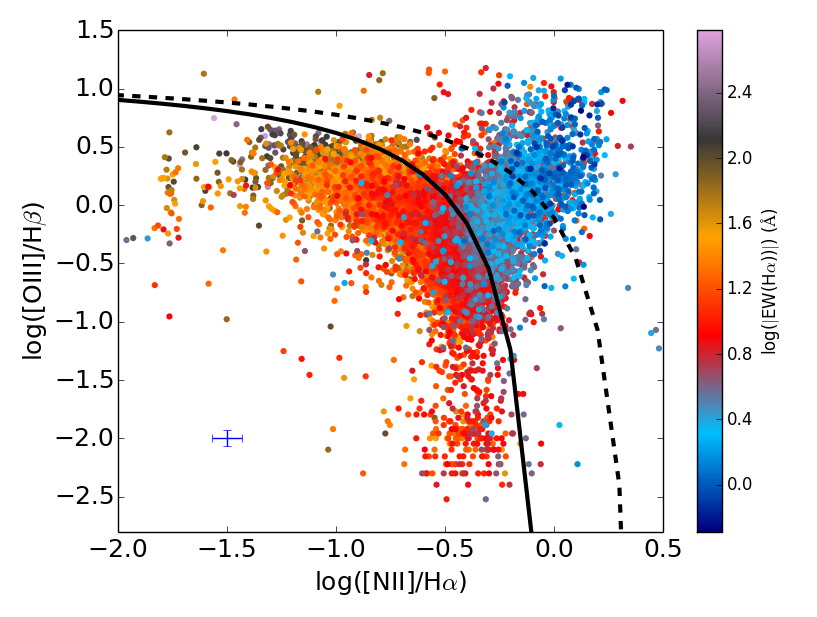}
      
   \caption{[\oiii]/\hb\ against [\nii]/\ha\ relation, one of the BPT diagrams, for the
15\,757 extracted emission-line regions in the analysed CALIFA galaxies. Colour scale shows
the H$\alpha$ equivalent width. The solid black line represents the empirical 
curve from \cite{kauffman03} to divide objects ionised by massive stars and
by non-thermal processes. The dashed line is the theoretical curve defined
by \cite{kewley01}. The lower left cross indicates the typical associated error in each axes.}
\label{o3-n2-wha}
\end{figure}

The determination of chemical abundances and most of the physical properties in ionised gaseous regions
are based on the assumption that the gas is ionised by a spectral energy distribution
dominated by very young massive stars during a process of ongoing star formation. 
Therefore, for the extracted objects in our sample of CALIFA galaxies, it is necessary 
to separate star-forming regions from other emission-line regions in which the 
main source of ionisation is different (i.e. AGNs, shocks, and post-AGB stars).

The most widely used method based on optical emission lines to detect star-forming regions is 
based on diagnostic diagrams of emission-line ratios sensitive to the excitation
of the gas (e.g. \citealt{BPT} (BPT), \citealt{VO87}). One of these diagrams is the extinction-independent relation
between [\oiii] 5007 \AA\/\hb\ and [\nii] 6584\AA/\ha,\ where different demarcation curves
are used to classify the ionising source. Figure \ref{o3-n2-wha} shows this diagram for the
15\,757 extracted emission-line regions in the 350 analysed CALIFA galaxies and 
the empirical curve defined by \cite{kauffman03} using SDSS galaxies and the
model-based curve defined by \cite{kewley01}. There is a general consensus that
below these two lines, emission-line objects can be considered star-forming regions, while those
above both curves are ionised by an AGN or shocks.
All regions in that part of the diagram were discarded for the subsequent analysis.

In addition, the regions between these two curves were proposed to be composite regions in which their ionisation 
has a mixed origin (i.e. star formation and nuclear activity) \citep{kewley06}, although we can also find 
pure star-forming regions with an overabundance of nitrogen, as in \cite{pmc09}, or regions that can be ionised by
evolved stars \citep{Kehrig12}.
Therefore, to select only star-forming \hii\ regions we also used the criterion described
by \cite{CF11} based on the use of EW(\ha), as a proxy of the weight of the 
old stellar population. This was measured on the spectra before the subtraction 
of the underlying stellar population over the continuum. As shown in Fig.\ref{o3-n2-wha} 
the distribution of EW(\ha) correlates with the position of the regions in the BPT diagram. 
In this way, as also carried out by \cite{S14_grad}, we discarded for our analysis 
those regions with EW(\ha) $>$ -6 \AA.
In addition, to ensure that only regions ionised by massive star formation
are used, we took the results from the SSP fitting to select regions 
with at least a mass fraction of 30\% of young stars (i.e. with an age younger than 500 Myr).
The number of selected star-forming regions in the 350 analysed CALIFA galaxies
using the above criteria is then 8\,196.

\subsection{Derivation of chemical abundances}

   \begin{figure}
   \centering
   \includegraphics[width=9.5cm]{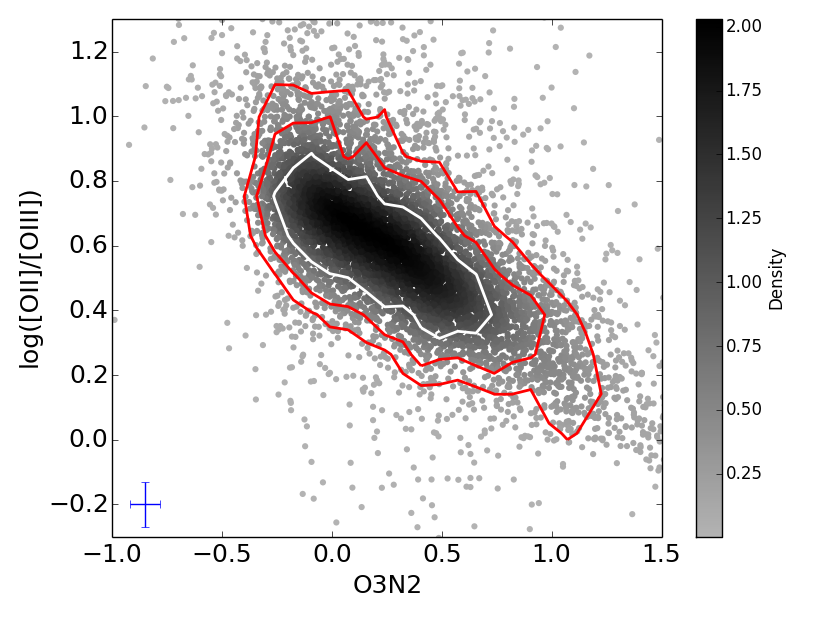}
      
   \caption{Relation between the index O3N2, which is a tracer of
$Z,$ and the emission line ratio [\oii]/[\oiii], which is a tracer of the excitation,
for those selected star-forming regions for which [\oii] lines were used
for the calculation of chemical abundances.  
Density of points over the mean value are represented according to the colour bar.
White solid line represents the 1$\sigma$ contour, while red lines represent 
2$\sigma$ and 3$\sigma$ contours. The left lower cross indicates the typical 
errors of the two ratios.}
\label{o3n2_o2o3}
\end{figure}


   \begin{figure}
   \centering
   \includegraphics[width=9.5cm]{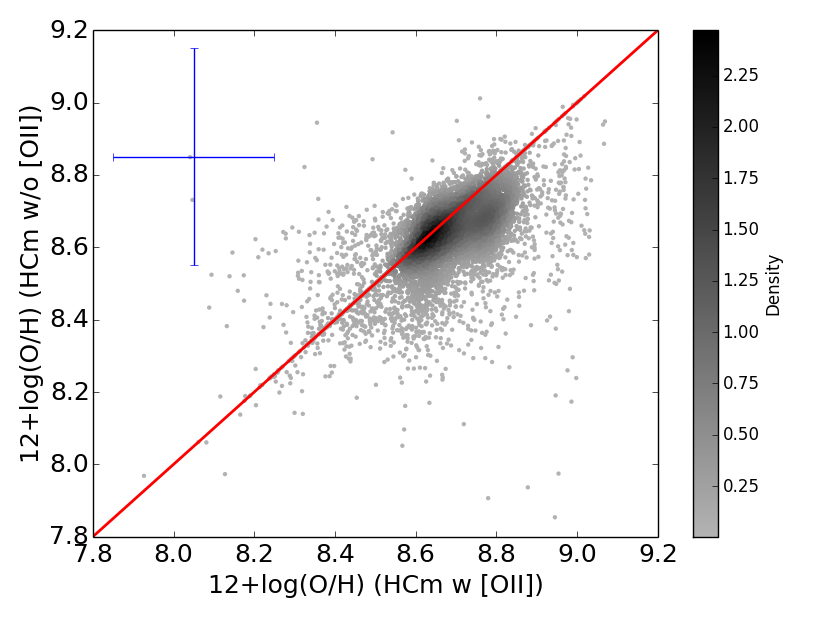}

   \caption{Comparison between the O/H abundances derived
from {\sc HCm} with and without taking [\oii] emission lines.
We only represent the \hii\ regions of the sample whose O/H uncertainty is
reduced when [\oii] is considered. 
Density of points over the mean value are represented according to the colour bar.
The red solid line represents the 1:1 relation.
The upper left cross represents the standard
deviation of both methods in relation to the direct method as carried out by \cite{hcm}.
}

        \label{OH_comp_hcm}
    \end{figure}
   \begin{figure*}[!t]
   \centering
   \includegraphics[width=9cm]{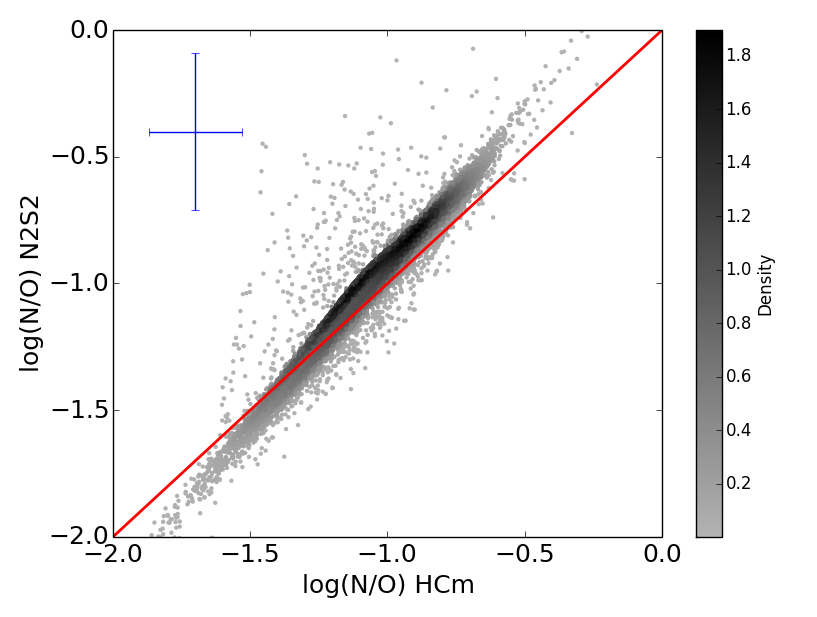}
   \includegraphics[width=9cm]{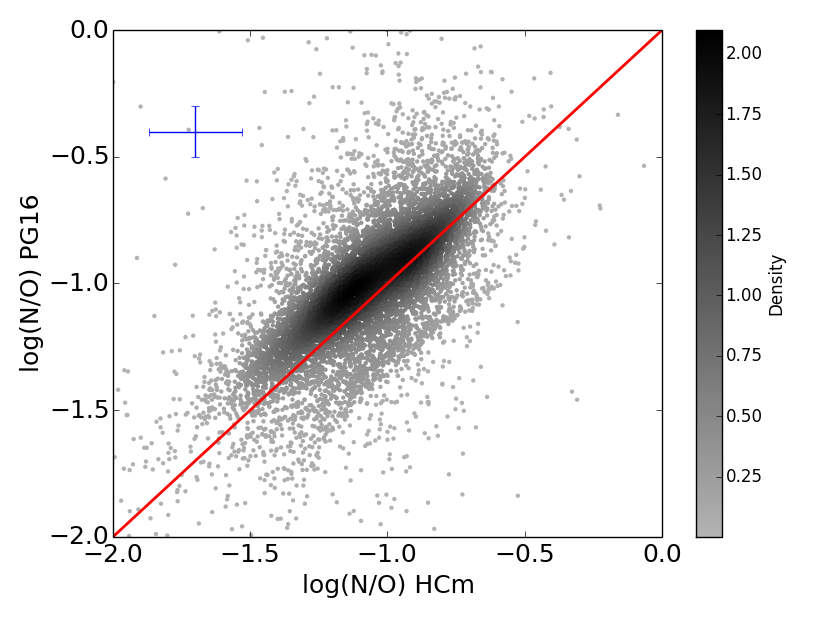}

   \caption{Comparison between N/O derived from {\sc HCm}
and from different strong-line methods. At left, 
for all selected \hii\ regions, with the N2S2 parameter 
calibrated by \cite{pmc09}. At right, the calibration by \cite{pg16}
based on [\nii] and [\oii] emission lines
for those regions whose [\oii] lines were used.
Density of points over the mean value are represented according to the colour bar.
The red solid line represents the 1:1 relation.
The upper left crosses represent the standard
deviation of the residuals to the direct method as carried out by \cite{pmc09} and \cite{hcm}.
}
         \label{NO_comp}
    \end{figure*}

   \begin{figure*}
   \centering
   \includegraphics[width=9cm]{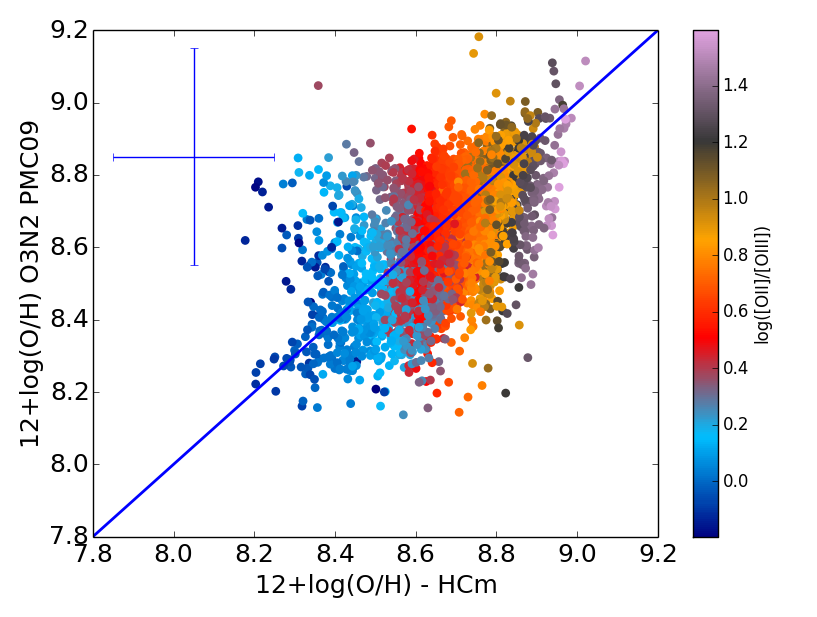}
   \includegraphics[width=9cm]{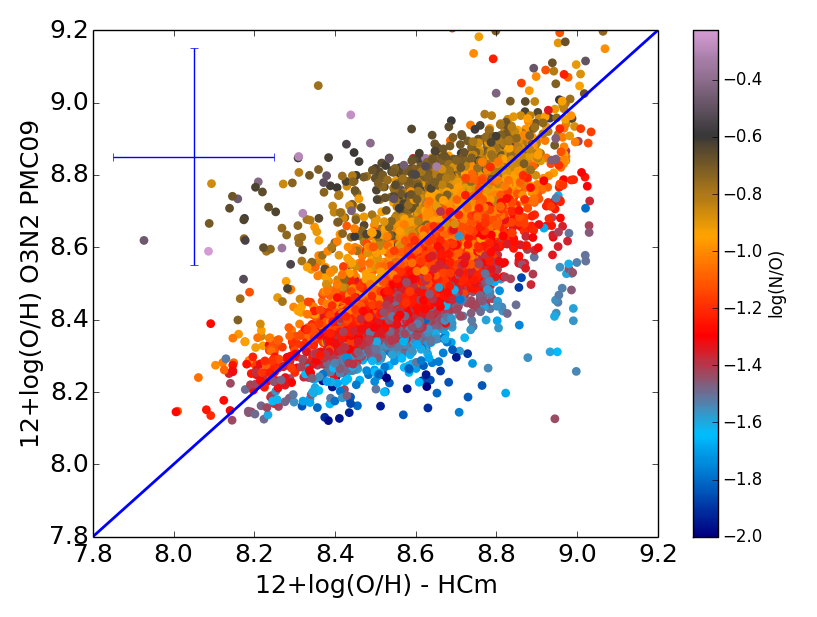}

   \caption{Comparison between O/H derived from {\sc HCm} and
from the empirical calibration of the O3N2 parameter given by \cite{pmc09}. 
The blue solid lines represent the 1:1 relation.
The left panel has a colour scale for the [\oii]/[\oiii] ratio and the right panel for N/O.
For the sake of consistency, the number of points in left panel is lower because [\oii] lines were not used for
the derivation of O/H.
The upper left crosses represent the standard deviations of the residuals to the direct 
method of both methods as compared to the direct method carried out by \cite{pmc09} and \cite{hcm}.
}
 
        \label{OH_comp_o3n2}
    \end{figure*}

Both elemental abundance ratios O/H and N/O were derived in the selected sample of \hii\ regions
using the programme {\sc Hii-Chi-mistry} v.2.0\footnote{Available at
\url{http://www.iaa.es/~epm/HII-CHI-mistry.html}} (hereafter {\sc HCm}) \citep{hcm}. This method
calculates these abundance ratios and the ionisation parameter (log$U$) as the
averages and standard deviations of the 1/$\chi^2$-weighted distributions of the
input conditions in a large grid of photoionisation models calculated with 
{\sc cloudy} v. 13.01 \citep{cloudy}.
The grid of models cover a wide range in the space of O/H ([6.9,9.1]
in steps of 0.1 dex), N/O ([-2.0,.0] in steps of 0.125 dex), and
log$U$ ([-4.0,-1.5] in steps of 0.25 dex). 
To avoid the clustering of the
resulting quantities around
the discrete values of the grid, however, we calculated linear interpolations
both in the O/H and the N/O axis   and, thereby, improved the resolution up to
0.02 dex and 0.025 dex, respectively.

The 1/$\chi^2$ weights are calculated as the quadratic
differences between certain observed emission-line ratios and the predictions 
made for the same ratios in each model of the grid. 
The observational input consists of the reddening-corrected
relative-to-\hb\ fluxes of [\oii] 3727 \AA, [\oiii] 4363, 5007 \AA\AA,
[\nii] 6584 \AA, and [\sii] 6717+6731 \AA\AA. 
The extinction correction of all lines was applied 
comparing the observed \ha/\hb\ flux ratio with the theoretical
value predicted by \cite{storey95} for standard conditions of density and
temperature (i.e. \ha/\hb\ = 2.86 for T$_e$ = 10$^4$ K and n$_e$ = 10$^2$ cm$^{-3}$) 
and assuming an extinction law by \cite{Cardelli} with $R_v$ = 3.1. We checked that 
the temperature and density variation across the analysed \hii\ regions introduces an 
additional uncertainty in the extinction correction that is much lower than the uncertainties 
associated with the emission-line measurements.

The analysis performed by \cite{hcm} shows that when the chemical abundances
derived using {\sc HCm} are compared with the abundances derived using the direct method 
in a sample of \hii\ regions and galaxies with a measurement of their electron
temperatures, the agreement is much better than the associated errors.
This agreement is also found even though the models use only stellar energy distributions of
massive clusters from {\sc popstar} \cite{popstar} with a single
age (1 Myr). However, the consistency between the abundances predicted by
the grid of models and the abundances calculated using the direct method can only be confirmed
in the sample of objects with electron temperature, so for our sample this can only be
fully proven when the corresponding auroral lines are measured with enough confidence.

Using {\sc HCm} to derive both O/H and N/O values in \hii\ regions has
three important advantages over other strong-line methods,
whether they have been empirically or theoretically calibrated. Firstly, it derives
N/O as a first step using appropriate emission-line ratios (i.e. N2O2, N2S2; \citealt{pmc09}), 
allowing the use of [\nii] emission lines to calculate O/H without
any prior assumption on the O/H-N/O relation. 
Secondly, the use of models allows us to obtain consistent chemical abundances
 in a $Z$ regime that cannot be properly calibrated using the 
direct method (i.e. oversolar abundances), which makes this method especially 
useful for the chemical study of \hii\ regions in spiral discs. Finally, 
{\sc HCm}  leads to values of the abundances that are consistent 
with the direct method regardless of the set of emission lines
taken as observed input, instead of using different strong-line
methods that are possibly calibrated in different ways for different available emission lines.  

Unfortunately,  the auroral [\oiii] 4363 \AA, which is needed to calculate electron temperatures,
was not measured with enough signal-to-noise in barely any of the selected \hii\ regions of our CALIFA
sample.
 This line could only be measured with confidence for 16 \hii\ regions  \citep{m13}.
For these regions, the agreement between the abundances derived from the direct method 
and the abundances derived from {\sc HCm} is better than the derived errors.
For the rest of the sample, in absence of the [\oiii] auroral line, {\sc HCm} 
uses a limited grid of empirically constrained models to provide chemical
abundances that are consistent with the direct method for the sample of \hii\ regions with electron
temperatures described in \cite{hcm}.
In this way, it is assumed to be an empirical law between O/H and log$U$ 
(i.e. higher log$U$ values for lower O/H and vice versa).
This O/H-log$U$ relation has been already observed in different 
samples of \hii\ regions (e.g. \citealt{ed85, de86})
and could be the effect of the use of a biased sample 
or an evolutionary sequence for the
empirical calibration of most strong-line methods.
This assumption is well justified 
in the case of our sample as can be seen in
Fig.\ref{o3n2_o2o3}, where it is represented the 
emission-line ratio [\oii]/[\oiii], which can be used as an 
indicator of the excitation of the ionised gas (e.g. \citealt{pmd05}), with the O3N2 
parameter (O3N2 = log([\oiii]/\hb $\cdot$ \ha/[\nii])), which
in metal-rich objects, correlates with $Z$ (e.g. \citealt{alloin}).
There is a clear trend towards lower values of [\oii]/[\oiii] (high log$U$) for
high O3N2 (i.e. low O/H) and vice versa.
Despite the assumed $Z$-log$U$ in the models,
{\sc HCm} allows a certain variation of log$U$ 
for each value of O/H, according to the observed relation
for \hii\ regions with electron temperatures in \cite{hcm} and
in agreement to some expected variation of log$U$ across galactic discs (see \citealt{ho15})
or the hardening of the ionising radiation \citep{pmv09}.

Using this constrained grid of models, the 
uncertainty in the final results is however higher than
with the presence of the [\oiii] auroral line.
 We considered the following
quadratic addition as the associated error in each \hii\ region:\\
\begin{equation} \label{Eq:eq_1}
\sigma^{2}_{HII} = \sigma^{2}_{HCm} + \sigma^{2}_{g} + \sigma^{2}_{res}
,\end{equation}
where $\sigma_{HCm}$ is the standard deviation of the resulting distribution
of abundances derived by {\sc HCm}, $\sigma_g$ is the improved 
interpolated resolution of the grid of models (i.e. 0.02 dex for O/H and 
0.025 dex for N/O), and $\sigma_{res}$ is the average residual with
the sample of objects with abundances derived using the direct
method reported by \cite{hcm} using a sample of \hii\ regions
with measured electron temperatures (i.e. 0.22 dex for
O/H and 0.16 dex for N/O) in absence of [\oiii] 4363\AA\ 
in the whole range.

Although all optical strong lines required by {\sc HCm} were measured 
in all objects, the [\oii] lines were not used for the calculation of the 
abundances in all of them (in a 29\% of the objects). 
When the standard deviation of the 1/$\chi^2$-weighted model-based
abundances were lower without [\oii], these were ruled out
to reduce the resulting uncertainties
(i.e. the [\oii] fluxes can be more uncertain because of a more critical
flux calibration in the blue part of the spectrum and to a larger 
wavelength baseline for reddening corrections).
A comparison between the O/H abundances derived by our method with and without [\oii]
for those \hii\ regions whose [\oii] were used is shown 
in Fig.\ref{OH_comp_hcm}. The mean residual is 0.04 dex, which is much lower
than the reported errors, so this does not affect the results of
this work very much.

A direct comparison between the chemical abundances obtained from {\sc HCm}
and the direct method cannot be established as the electron temperature cannot be
measured in almost any \hii\ regions of the CALIFA sample. However, we compare our
results with some other strong-line methods calibrated with the direct method sample.
The left panel of Fig. \ref{NO_comp} shows that the comparisons 
between the N/O  obtained from {\sc HCm} 
with the N/O from the calibration by \cite{pmc09} of the
N2S2 parameter (=log([\nii]/[\sii])) for all the
selected \hii\ regions. The right panel shows the comparison
with the N/O from the calibration by \cite{pg16} based on [\nii] and [\oii] lines
only for those objects whose [\oii] lines
were used in the {\sc HCm} calculations.
In both cases the average of the residuals is much lower than the
reported associated uncertainty (i.e. 0.05 dex in both cases),
but the standard deviation of these residual results is much lower for
N2S2 (0.10 dex) than for the method based on [\nii]/[\oii] calibrated
by \cite{pg16} (0.25 dex).



   \begin{figure}[t]
   \centering
   \includegraphics[width=9.5cm]{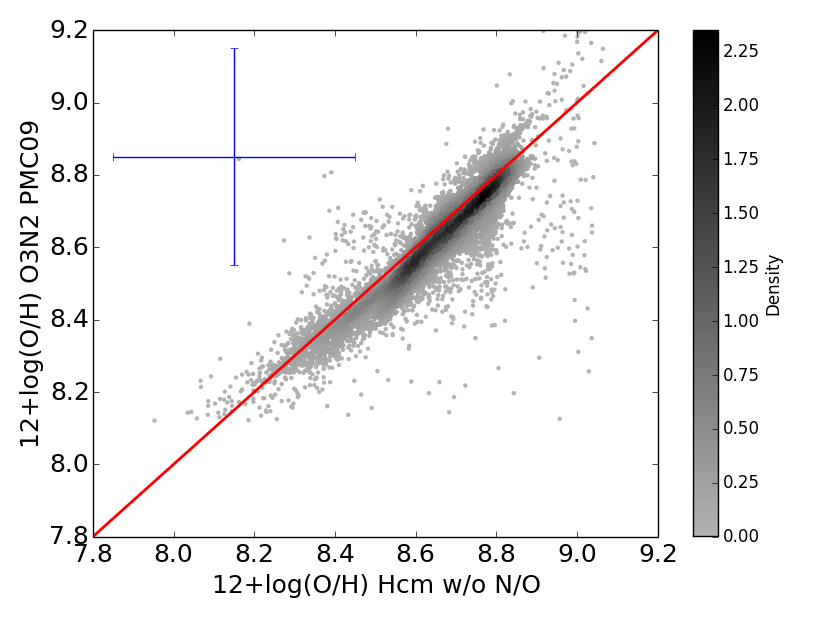}

   \caption{Comparison between the O/H abundances derived
from {\sc HCm} with the empirical calibration of the O3N2 parameter given by \cite{pmc09}.
In this case {\sc HCm} O/H values were calculated without a previous determination of N/O, 
but assuming the expected O/H-N/O relation.
Density of points over the mean value are represented according to the colour bar.
The red solid line represents the 1:1 relation.
The upper left cross represents the standard deviations of the residuals of both methods as
compared to the direct method carried out by \cite{pmc09} and \cite{hcm}.
}
\label{OH_comp_sNO}
    \end{figure}

Regarding O/H, in order to examine the impact of considering a
free O/H-N/O relation in the models as compared with other methods empirically
calibrated, in Fig. \ref{OH_comp_o3n2}
we show the comparison for the selected CALIFA star-forming regions
between the O/H derived by {\sc HCm}
and the empirical calibrations of the O3N2 parameter given by \cite{pmc09} that is
valid for O/H $\gtrsim$ 8.1. There is not a large systematic offset between the two methods 
(i.e. the average of the residuals is 0.04 dex) and the dispersion,
which is calculated as the standard deviation of the residuals, is 0.11 dex. 
To explore the origin of this dispersion, 
we include a colour scale 
representing other quantities in both panels of the figure.
In the left panel, only for those \hii\ regions whose [\oii] lines
were used for the abundance calculation, the emission line ratio [\oii]/[\oiii], 
which is a tracer of the gas excitation.
As in the case of Fig. \ref{o3n2_o2o3}, there is a clear trend towards  higher excitation for lower $Z$. 
However, there is not any clear indication that the dispersion can
be due to the excitation.
On the other hand, in the right panel, the colour scale indicates the N/O
ratio. As can be seen the agreement is better for lower $Z$ when 
N/O is lower, and for higher O/H when N/O is higher. Other combinations of
O/H and N/O in the same plane leads to larger deviations of the 1:1 relation
producing the observed dispersion.

In Fig. \ref{OH_comp_sNO} we show a similar comparison with O3N2 as calibrated
by \cite{pmc09} but, in this case, the O/H abundances derived using {\sc HCm}
are derived used a grid of models assuming the expected relation between O/H and N/O (i.e. in
the regime of secondary production of N, N/O grows with O/H; see right panel of Fig.3 in \citealt{hcm}).
In this case, the dispersion is very small; i.e. the standard deviation of the
residuals is reduced to 0.05 dex. Since most of the objects with
electron temperature used to calibrate strong-line methods follow the above
typical O/H-N/O relation, the agreement is good when it is assumed.
However, as in the case of our sample of CALIFA star-forming regions, the
dispersion is enhanced as a consequence of a uncertain N/O value, if it is not previously
determined.

   \begin{figure}
   \centering
   \includegraphics[width=9.5cm]{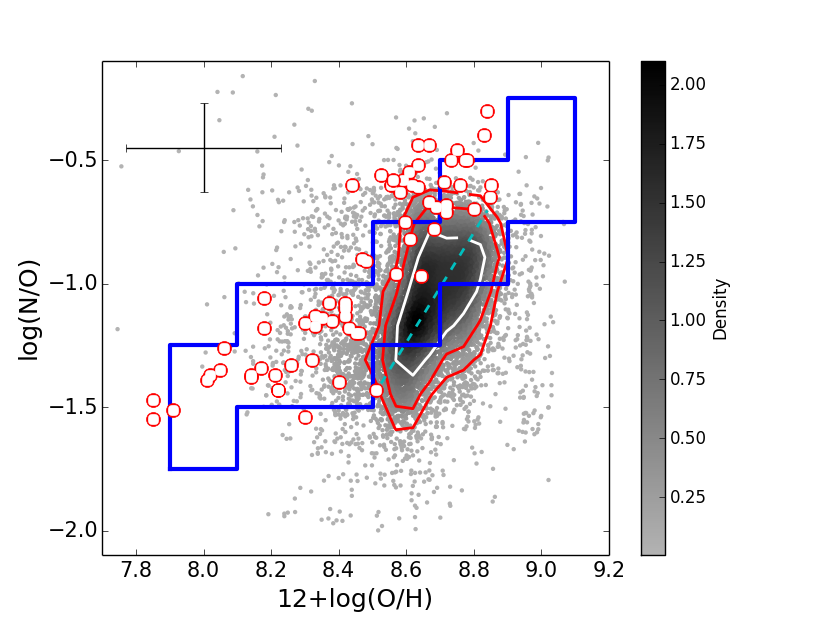}
   
   \caption{Relation between O/H and N/O as derived
from {\sc HCm} for all the star-forming \hii\ regions analysed in the CALIFA sample. 
Density of points over the mean value are coded by the colours.
The solid lines represent the $1\sigma$ (white), 2$\sigma$, and 3$\sigma$ (red) contours, while the
dashed line shows the linear fitting obtained in the lower panel of Fig.\ref{abs_comp}.
Finally, the cyan solid line encompasses the space of models used by {\sc HCm}
to calculate O/H without a previous determination of N/O that is taken from the position of objects
with a direct chemical abundance in \cite{hcm}.
The upper cross represents the typical errors associated with the derivation of abundances using {\sc HCm}.
The white points correspond to \hii\ regions in other O/H regimes whose abundances were 
calculated from the direct method taken from \cite{croxall15} and \cite{croxall16}}.

              \label{oh-no}%
    \end{figure}

   \begin{figure*}
   \centering
   \includegraphics[width=13cm]{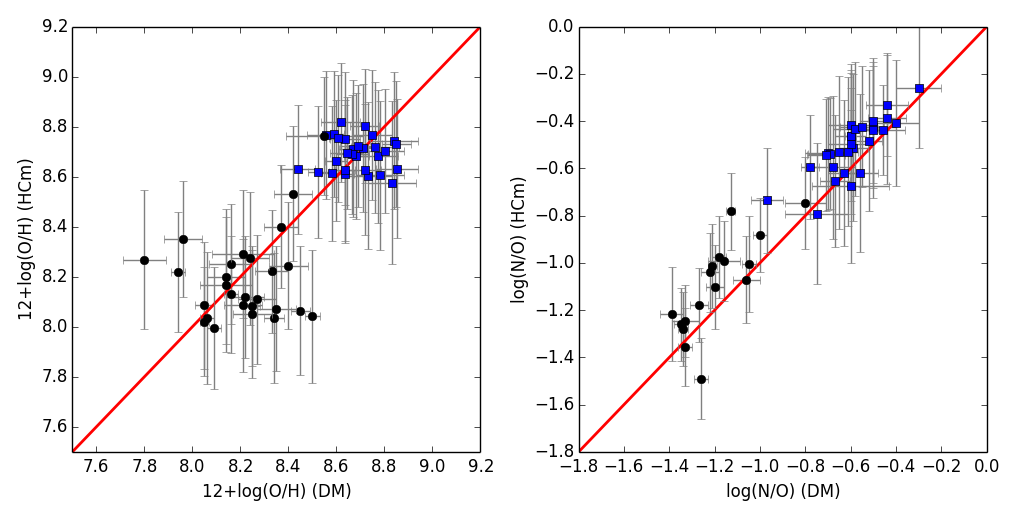}
   
   \caption{Comparison between abundances derived from the direct method and from {\sc HCm} in the
sample of \hii\ regions of M51 \citep{croxall15} as blue squares and M101 \citep{croxall16} 
as black circles for O/H (left) and N/O (right). The red solid lines represent the 1:1 relation in
both panels.}

              \label{comp_chaos}%
    \end{figure*}

\section{O/H-N/O for individual \hii\ regions}

The relation between O/H and N/O gives clues about
the chemical evolution of the sites where the ionised 
gas is observed. As O is a primary element it is mainly 
ejected to the ISM by massive stars. On the contrary, 
N can have either a primary or a secondary origin. This last
regime dominates the production of N in metal-rich \hii\ regions, 
as it is produced via the CNO cycle in less massive stars.
It is thus expected that N/O correlates with O/H for the \hii\ regions
of the spiral discs.

Figure \ref{oh-no} shows the relation between O/H and N/O
calculated using {\sc HCm} as described above
for the 8\,196 star-forming regions selected among the 350
analysed CALIFA galaxies. As can be seen there is a trend towards a growing N/O
for growing O/H as corresponds to a regime of secondary production of N in metal-rich
objects. 

We can compare the average position in the O/H-N/O plot of the CALIFA star-forming
regions analysed in this work with the line enclosing the regions and galaxies 
analysed in \cite{hcm} with electron temperatures, and whose O/H and N/O abundances are also derived 
by \cite{hcm} to constrain the grid of models used when
a previous N/O determination is not possible.
We see that the slope is higher for the CALIFA regions and that the
sample presents lower values of N/O for lower O/H values.
As seen in right panel of Fig. \ref{OH_comp_o3n2}, we can find some \hii\ regions with
oversolar O/H and N/O $<$ -1.6, but these can be considered as spurious given their low
number ($<$ 0.5\%) and they do not affect the results regarding slopes and characteristic values
described in the sections below.

We also added to Fig. \ref{oh-no} some \hii\ regions with a measurement of their abundances following
the direct method in other $Z$ ranges to check the consistency of the abundances derived in our sample.
The data at high $Z$ from NGC~5194 \citep{croxall15} mainly overlap our sample while
low-$Z$ data from M101 (\cite{croxall16} and references therein) show a certain overlap 
with our sample, but they are in agreement with the region of the plot corresponding
to other \hii\ regions with available electron temperatures. In Fig. \ref{comp_chaos} we show,
for these high- and low-$Z$ \hii\ regions,
the comparison between the abundances derived from the direct method and the abundances derived from
{\sc HCm} without using [\oiii] 4363 \AA, which is consistent with the derivation of the abundances in the
CALIFA sample. As can be seen, no systematic offset is found either for O/H or for N/O with
standard deviation of the residuals that is much lower than the associated uncertainties.

At the same time it can be seen that the relation between O/H and N/O 
shows a large dispersion. A given dispersion is also found with theoretical 
models as due to the different efficiencies of SFR in the different regions 
or galaxies as \cite{molla06} found. In that work the dispersion for the
 last time step calculated for their models is shown in their Fig.8 around 
the fitting given for those results. However, the dispersion found with our data 
seems larger than that  expected as due to different star formation efficiencies
It is, however, necessary to take into account that the time evolution also plays 
a role. As it is seen in Fig.9 from \cite{molla06}, different regions or galaxies 
evolve differently in the plane O/H-N/O, showing steeper tracks when the region/galaxy 
has higher efficiencies than those with low values, which show flat evolutionary tracks. 
Therefore, the data found below the  region of the plane shown by the 1 sigma contours probably have 
medium to high efficiencies to form stars, but are less evolved than those  in the "standard" locus; 
while the data above are probably regions with low efficiencies.

Besides, the gas interchange between a galaxy and the surrounding IGM, or between
different parts of a galaxy, has a very different influence on O/H and the ratio 
between the abundances of a primary and secondary element \citep{edmunds90}, so
the dispersion in the O/H-N/O diagram tends to be larger when gas flows exist in 
the galaxies.  

On the other hand, this plot illustrates the importance of the determination of N/O
for the chemical analysis in spiral galaxies. 
Firstly, as N/O does not follow the expected trend in its relation with O/H and
presents a large dispersion, all O/H determinations based
on [\nii] lines can lead to wrong results \citep{pmc09} and, secondly,
the range of variation in this sample is almost three times
larger in logarithmic scale for N/O ($\approx$ [-1.8,-0.5]) 
than for O/H ($\approx$ [8.4-8.9]).
In addition, in absence of the [\oiii] auroral line, the relative error
using {\sc HCm} is better for N/O (0.16 dex) than for O/H (0.22 dex).
Thus, the variations of N/O across galactic discs, can yield more accurate results
for the determination of spatial chemical variations across galactic discs.

   \begin{figure}
   \centering
   \includegraphics[width=7.6cm]{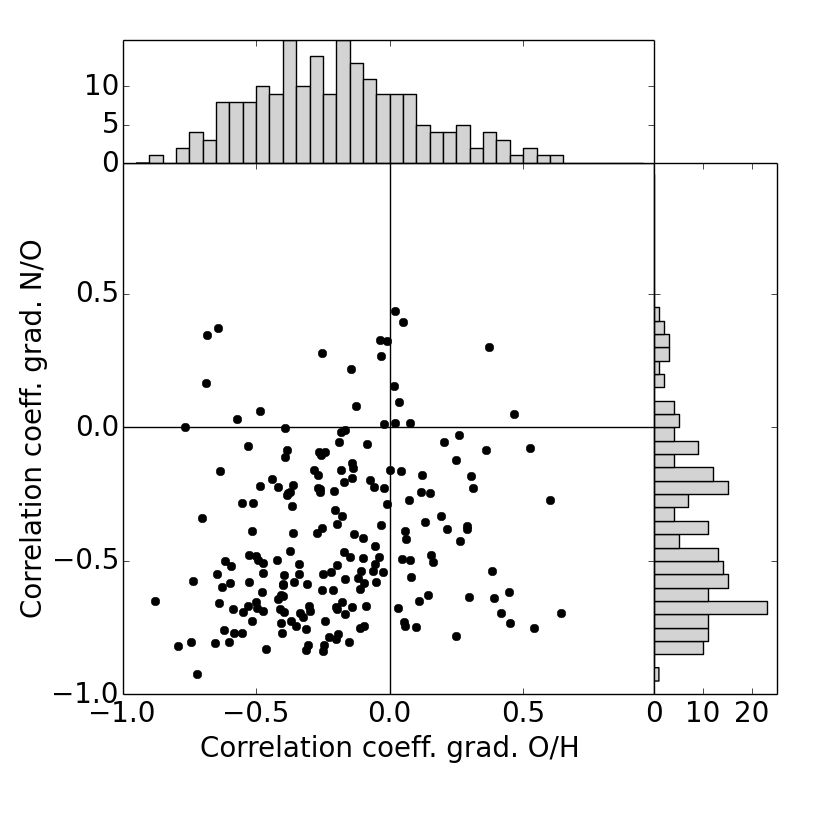}
   \includegraphics[width=7.6cm]{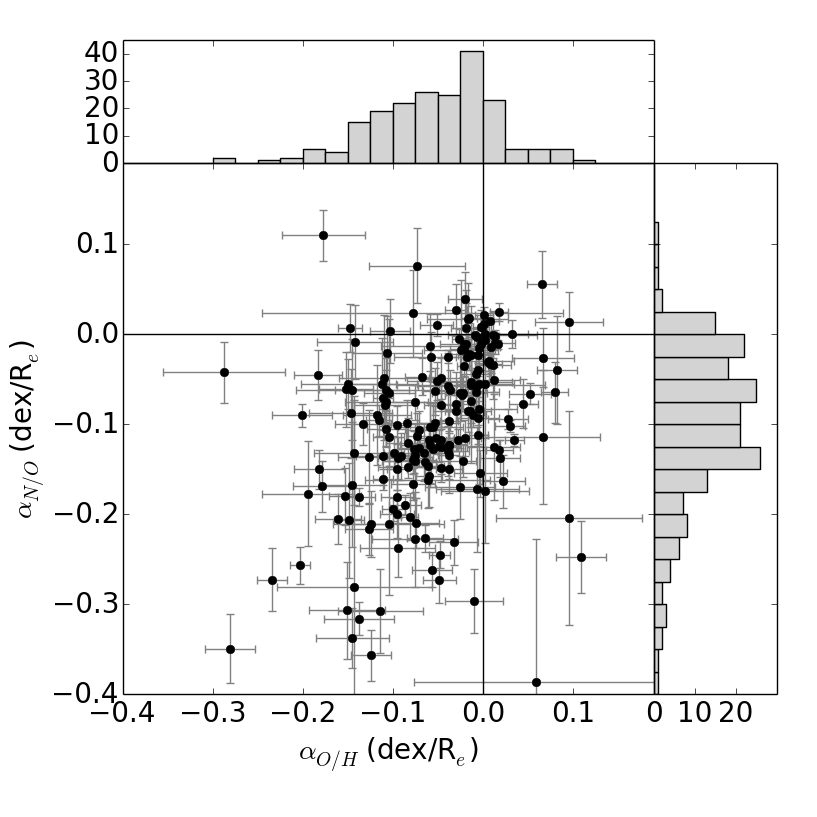}
   \includegraphics[width=7.6cm]{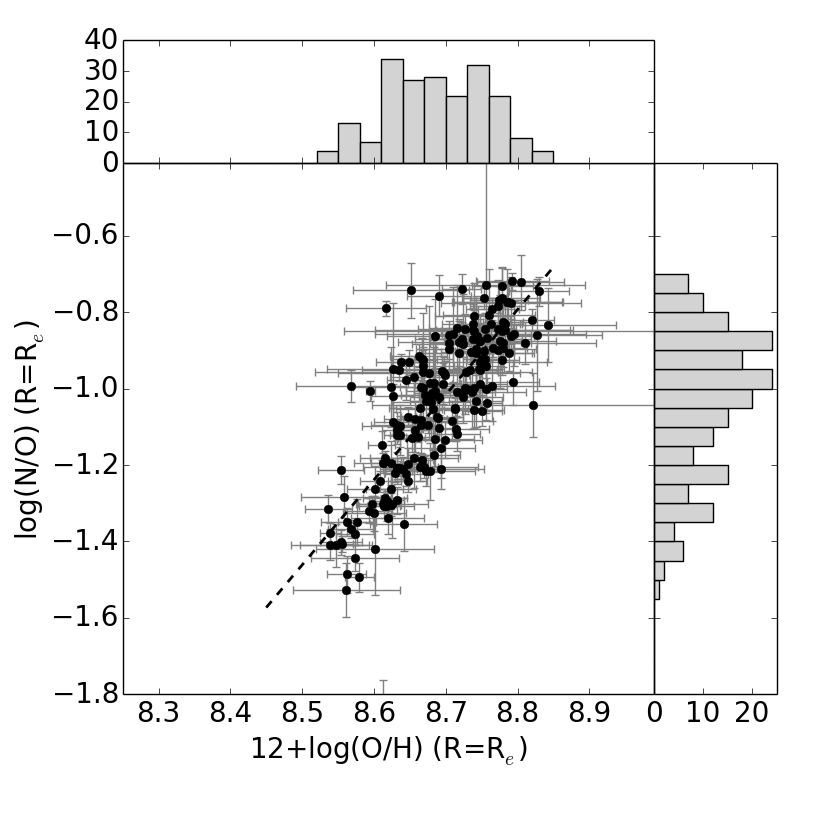}
   
   \caption{Histograms and relations of the properties derived from the
linear fittings through the radial scale both for O/H and N/O in those
galaxy discs of the CALIFA sample with enough star-forming \hii\ regions. From 
top to bottom: correlation coefficient, slope of the gradient, and predicted value 
at the effective radius of the galaxy.}
              \label{abs_comp}%
    \end{figure}

\section{Chemical abundance gradients in discs}

\subsection{Calculation of linear fittings}

By considering the resulting {\sc HCm} O/H and N/O abundance
ratios and the de-projected radial distances of the corresponding selected star-forming regions,
in our sample of 350 CALIFA galaxies, robust error-weighted
linear fittings were performed. 

The analysis was made at all de-projected galactocentric distances
normalised to the effective radius. The description about the measurement of
the effective radius in each object is given in \cite{S14_grad}.
Only those galaxies with ten or more selected \hii\ regions were considered for
the subsequent statistical analysis of the sample.
We checked  the impact of also performing linear fittings over
the median values of both O/H and N/O in different radial bins to avoid an
oversampling effect in certain radial positions in our results, but we did not find noticeable
variations in our results (e.g. the average O/H slopes changes in less than 0.01 dex/R$_e$).
In the last section, we present an analysis of the incidence in the results of 
the very well-known deviations from a strictly linear radial variation of the
chemical abundances (i.e. decrease in central positions or flattening at the 
outermost \hii\ regions). 
We also checked to what extent the calculation of abundances without
[\oii] emission lines in $\approx$30\% of the \hii\ regions
can affect our results, but these are not localised in specific radial
positions and/or galaxies, so the average slopes change in less than
0.01 dex/R$_e$.

\begin{table*}
\begin{minipage}{180mm}
\caption{List of the analysed CALIFA galaxies with their effective radii, inclinations, and the number of
\hii\ regions used to calculate their O/H and N/O slopes and the values at the effective radius of
the gradients.}

\begin{center}
\begin{tabular}{lccccccc}
\hline
\hline
Galaxy\footnote{The complete table will only be available in electronic
form} & R$_e$ & $i$ & \hii\ regions & $\alpha_{O/H}$ & 12+log(O/H) & $\alpha_{N/O}$ & log(N/O) \\
 &  (kpc)  & (\degree)  &   & (dex/R$_e$)  &  (at R$_e$)  &  (dex/R$_e$)  & (at R$_e$)  \\

\hline
ESO539-G014     &       -0.17   $\pm$   2.3     &       90      &       10      &       -0.006  $\pm$   0.046   &       8.602   $\pm$   0.083   &       -0.172  $\pm$   0.070   &       -1.420  $\pm$   0.120   \\
IC0159  &       0.19    $\pm$   0.2     &       40      &       41      &       -0.053  $\pm$   0.013   &       8.632   $\pm$   0.020   &       -0.064  $\pm$   0.026   &       -1.207  $\pm$   0.042   \\
IC0480  &       -0.02   $\pm$   0.5     &       87      &       20      &       0.003   $\pm$   0.000   &       8.632   $\pm$   0.009   &       0.000   $\pm$   0.002   &       -1.123  $\pm$   0.022   \\
IC0776  &       -0.54   $\pm$   0.3     &       57      &       39      &       0.011   $\pm$   0.049   &       8.574   $\pm$   0.062   &       -0.001  $\pm$   0.026   &       -1.444  $\pm$   0.034   \\
IC0995  &       -0.61   $\pm$   0.2     &       84      &       24      &       -0.003  $\pm$   0.030   &       8.619   $\pm$   0.051   &       -0.155  $\pm$   0.026   &       -1.337  $\pm$   0.042   \\
IC1151  &       -0.27   $\pm$   0.2     &       63      &       68      &       -0.028  $\pm$   0.013   &       8.636   $\pm$   0.019   &       -0.117  $\pm$   0.012   &       -1.208  $\pm$   0.020   \\
IC1256  &       0.24    $\pm$   0.2     &       55      &       47      &       -0.105  $\pm$   0.017   &       8.734   $\pm$   0.037   &       -0.115  $\pm$   0.027   &       -0.950  $\pm$   0.061   \\
IC1528  &       0.2     $\pm$   0.2     &       72      &       70      &       -0.099  $\pm$   0.011   &       8.713   $\pm$   0.018   &       -0.132  $\pm$   0.018   &       -1.053  $\pm$   0.036   \\
IC2095  &       -0.7    $\pm$   1.0     &       90      &       14      &       -0.017  $\pm$   0.045   &       8.562   $\pm$   0.074   &       0.017   $\pm$   0.039   &       -1.526  $\pm$   0.070   \\
IC2101  &       0.52    $\pm$   0.4     &       90      &       25      &       -0.030  $\pm$   0.012   &       8.666   $\pm$   0.025   &       0.027   $\pm$   0.029   &       -1.095  $\pm$   0.072   \\
IC2487  &       0.24    $\pm$   0.4     &       84      &       24      &       -0.234  $\pm$   0.017   &       8.755   $\pm$   0.029   &       -0.273  $\pm$   0.035   &       -1.001  $\pm$   0.063   \\
IC5309  &       0.13    $\pm$   0.2     &       63      &       25      &       -0.055  $\pm$   0.021   &       8.779   $\pm$   0.040   &       -0.127  $\pm$   0.012   &       -0.882  $\pm$   0.026   \\
MCG-02-02-040   &       0.08    $\pm$   0.4     &       84      &       22      &       -0.036  $\pm$   0.012   &       8.691   $\pm$   0.029   &       -0.062  $\pm$   0.023   &       -1.023  $\pm$   0.054   \\
MCG-02-03-015   &       0.32    $\pm$   0.7     &       75      &       12      &       0.095   $\pm$   0.081   &       8.676   $\pm$   0.127   &       -0.204  $\pm$   0.119   &       -0.959  $\pm$   0.176   \\
MCG-02-51-004   &       0.54    $\pm$   1.4     &       72      &       39      &       -0.108  $\pm$   0.021   &       8.699   $\pm$   0.034   &       -0.105  $\pm$   0.020   &       -0.965  $\pm$   0.035   \\
NGC0001 &       0.8     $\pm$   0.1     &       37      &       39      &       -0.026  $\pm$   0.031   &       8.760   $\pm$   0.064   &       -0.066  $\pm$   0.061   &       -0.808  $\pm$   0.121   \\
NGC0023 &       1.03    $\pm$   0.4     &       68      &       27      &       0.082   $\pm$   0.022   &       8.715   $\pm$   0.039   &       -0.040  $\pm$   0.060   &       -0.842  $\pm$   0.112   \\
NGC0036 &       0.64    $\pm$   0.4     &       51      &       25      &       -0.105  $\pm$   0.062   &       8.747   $\pm$   0.094   &       -0.211  $\pm$   0.079   &       -0.931  $\pm$   0.125   \\
NGC0165 &       0.43    $\pm$   0.2     &       35      &       28      &       -0.110  $\pm$   0.050   &       8.776   $\pm$   0.078   &       -0.048  $\pm$   0.069   &       -0.877  $\pm$   0.121   \\
NGC0171 &       0.02    $\pm$   0.2     &       52      &       37      &       -0.039  $\pm$   0.023   &       8.691   $\pm$   0.044   &       -0.130  $\pm$   0.028   &       -0.758  $\pm$   0.054   \\

\hline

\end{tabular}
\end{center}
\label{lista}
\end{minipage}
\end{table*}

In addition, to rule out environmental effects between the processes that can shape the
abundance gradient in galaxies, we did not consider those galaxies with some
level of interaction as determined from the visual inspection of their
optical images. More details on the interaction stage of each CALIFA galaxy 
are given in \cite{barrera}.

This leaves the total number of analysed galaxies
at 201. 
The list of non-interacting galaxies with at least 10 selected \hii\ regions,
with their effective radii, inclinations, resulting slopes in dex/$R_e$, and the values 
at the effective radius both for O/H and
N/O with their corresponding errors can be seen in Table \ref{lista}.

\subsection{Properties of the resulting fittings}

In this subsection we analyse the statistical properties of the distributions
of the resulting chemical abundance variations across the galactocentric
de-projected distances normalised to the effective radius in the 201 
CALIFA non-interacting galaxies with ten or more selected star-forming \hii\ regions.

In the upper panel of Fig. \ref{abs_comp} we show the resulting distributions
and the relation between the correlation coefficients of the linear fittings 
to O/H and N/O. It can be seen that most of the galaxies
present a negative coefficient for O/H and N/O, which is consistent with the classical idea
that spiral galaxies have larger abundances in their inner parts and lower values
at larger galactocentric distances. Accepting that the production of N 
for the metal content of these regions is mostly secondary, it is expected
that we also find a negative correlation coefficient for N/O in most of the spiral galaxies. 
However we also see that
a non-negligible fraction of the galaxies present a positive value of this
coefficient and that a large dispersion exists between O/H and N/O, in agreement with the large dispersion
found for individual \hii\ regions of the sample seen in Fig.\ref{oh-no}.

The middle panel of Fig.\ref{abs_comp} shows the 
distributions and relation for the O/H and N/O slopes of the 
resulting linear fittings normalised to the corresponding effective 
radii. Most of the galaxies present a negative gradient both for
O/H and N/O. However there is a large dispersion in the 
corresponding distributions with standard deviations of the same 
order as the average slopes, which are $\alpha_{O/H}$ = 
-0.053 $\pm$ 0.068 and $\alpha_{N/O}$ = -0.104 $\pm$ 0.096.

To evaluate whether this large variation is a consequence
of statistical fluctuations around a normal distribution, we performed
a Lilliefors test. In this test the null hypothesis is that the sample comes
from a Gaussian distribution and this cannot be rejected if the significance p value
is larger than 0.1. The resulting p values are 0.034 for O/H and 0.001 for N/O so,
apparently, the queue of flat and positive slopes seen in both distributions
are not the consequence of a statistical fluctuation. 

   \begin{figure*}[th]
   \centering
   \includegraphics[width=9cm]{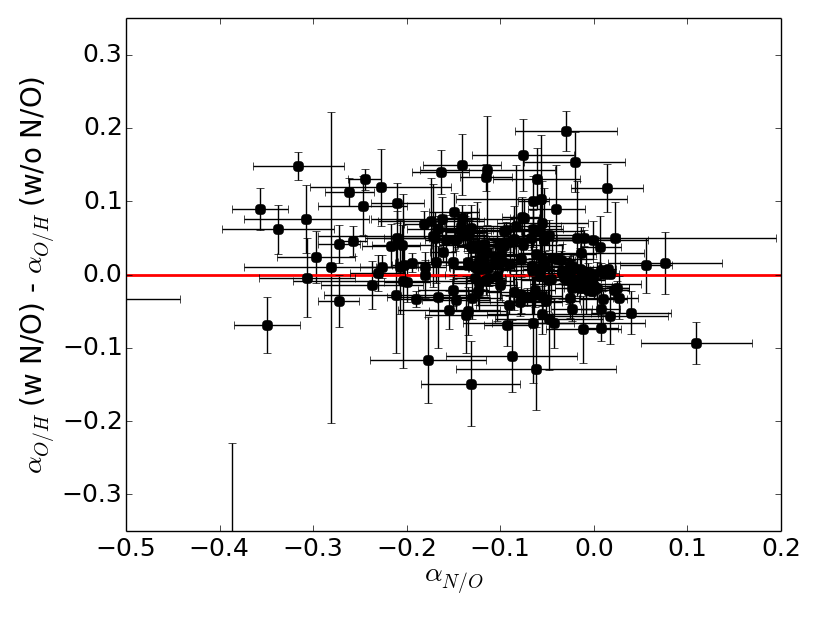}
   \includegraphics[width=9cm]{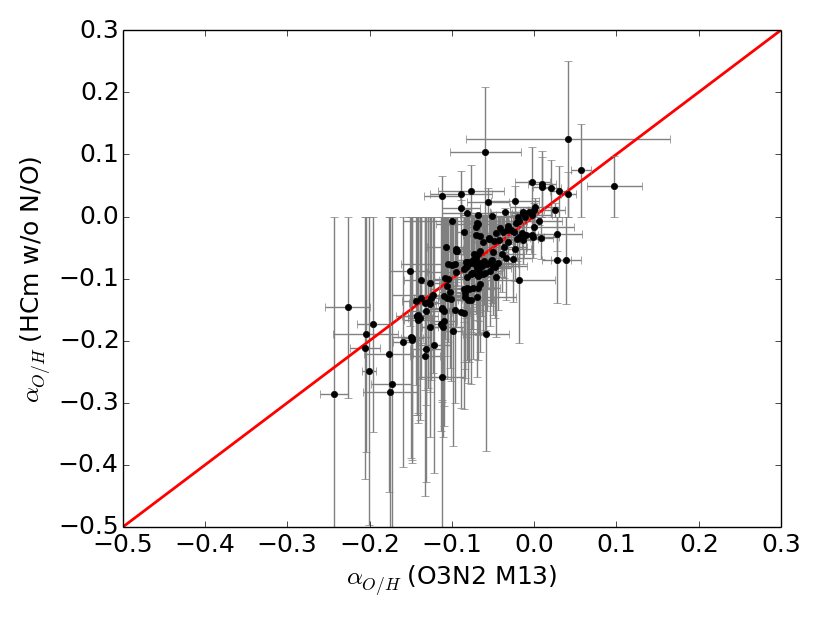}

   \caption{At left, difference as a function of the N/O slope of the O/H slopes calculated 
with a previous N/O calculation and the same slopes calculated assuming a given O/H-N/O relation 
for all the analysed CALIFA galaxies.
The red solid line represents the zero value.
At right, comparison between the O/H slopes calculated using this last 
approach with the same slopes calculated using the O3N2 parameter as
calibrated by \cite{m13}. The red solid line represents the 1:1 relation.
All slopes are expressed in dex/R$_e$}
    \label{Da_oh-alpha_no}
    \end{figure*}

\begin{table*}
\begin{minipage}{180mm}
\caption{Average slopes and values at the effective radius for
the linear fittings of O/H and N/O in the analysed CALIFA galaxies 
grouped by categories depending on the sign of the slopes.}

\begin{center}
\begin{tabular}{lcccccc}
\hline
\hline
     &  Number & \%  &  $\alpha_{O/H}$ & 12+log(O/H) & $\alpha_{N/O}$ & log(N/O) \\
     &     &       &   (dex/R$_e$)  &  (at R$_e$)    & (dex/R$_e$)   &  (at R$_e$)   \\
\hline
All                                               & 201 & 100 & -0.053 $\pm$ 0.068 & 8.688 $\pm$ 0.070  & -0.104 $\pm$ 0.096 & -1.041 $\pm$ 0.198 \\
$\alpha_{O/H}$ $<$ 0                    & 162 & 80.6 & -0.073 $\pm$ 0.057 & 8.697 $\pm$ 0.068 & -0.110 $\pm$ 0.086 & -1.026  $\pm$ 0.181 \\
$\alpha_{O/H}$ $<$ 0 (1$\sigma$)        & 130 & 64.7 & -0.088  $\pm$ 0.054 & 8.702 $\pm$ 0.066 & -0.121 $\pm$ 0.087 & -1.012 $\pm$ 0.174 \\
$\alpha_{O/H}$ $>$ 0                    & 39   &  19.4 & +0.032 $\pm$ 0.032 & 8.652 $\pm$ 0.069 & -0.079 $\pm$ 0.125 & -1.104 $\pm$ 0.250 \\
$\alpha_{O/H}$ $>$ 0 (1$\sigma$)        & 20   & 10.0 & +0.048 $\pm$ 0.033  & 8.652 $\pm$  0.067 & -0.062 $\pm$ 0.077 & -1.041 $\pm$ 0.216 \\
$\alpha_{N/O}$ $<$ 0                     & 186 & 92.5 &  -0.056 $\pm$ 0.068 & 8.690 $\pm$ 0.071 & -0.118 $\pm$ 0.090 & -1.036 $\pm$ 0.198 \\ 
$\alpha_{N/O}$ $<0$  (1$\sigma$)         & 156 & 77.6 &  -0.062 $\pm$ 0.068 & 8.695 $\pm$ 0.069 & -0.133 $\pm$ 0.088 & -1.030 $\pm$ 0.194 \\ 
$\alpha_{N/O}$ $>$ 0                     & 20   &  10.0 & -0.027 $\pm$ 0.065 & 8.678  $\pm$ 0066 & +0.025 $\pm$ 0.027 & -1.075 $\pm$ 0.205 \\
$\alpha_{N/O}$  $>$ 0 (1$\sigma$)         & 9    & 4.5 & -0.021 $\pm$ 0.061 & 8.674 $\pm$ 0.062    & +0.041 $\pm$ 0.033 & -1.107 $\pm$ 0.190 \\ 
\hline

\end{tabular}
\end{center}
\label{tab_slopes}
\end{minipage}
\end{table*}

   \begin{figure*}
   \centering
   \includegraphics[width=4.5cm]{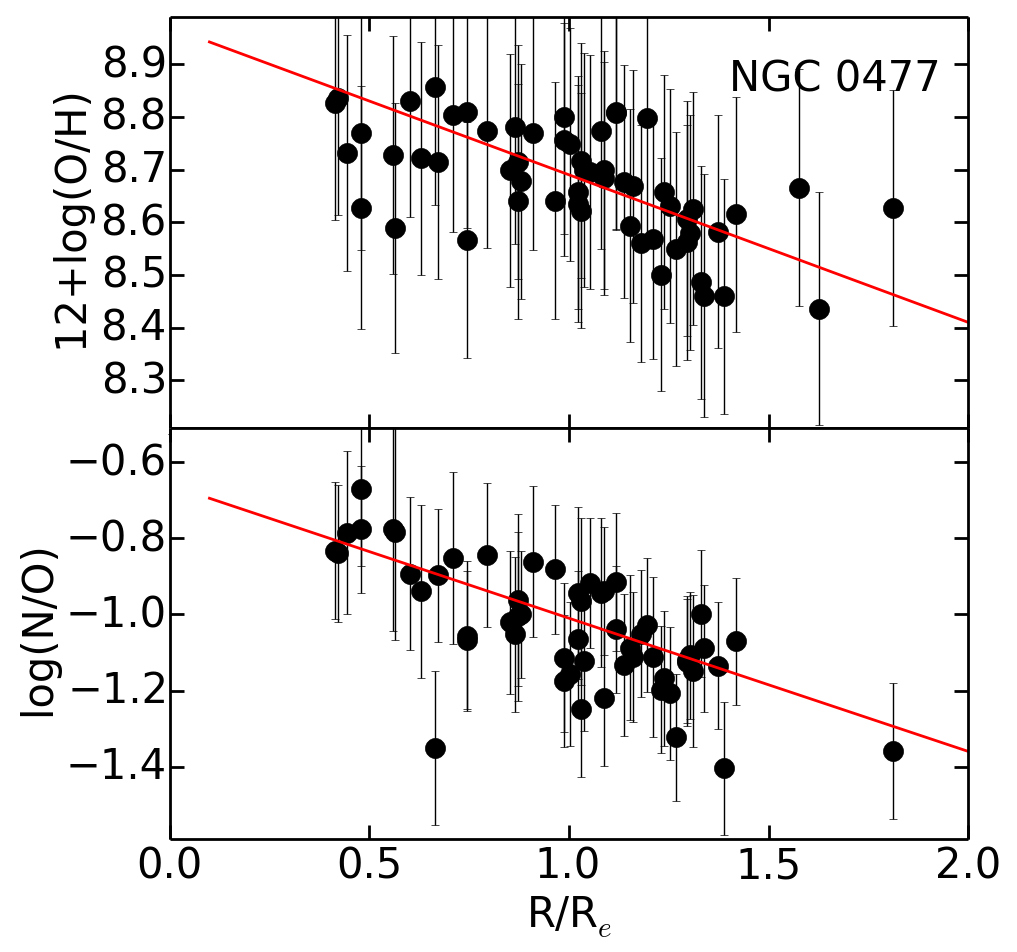}
   \includegraphics[width=4.5cm]{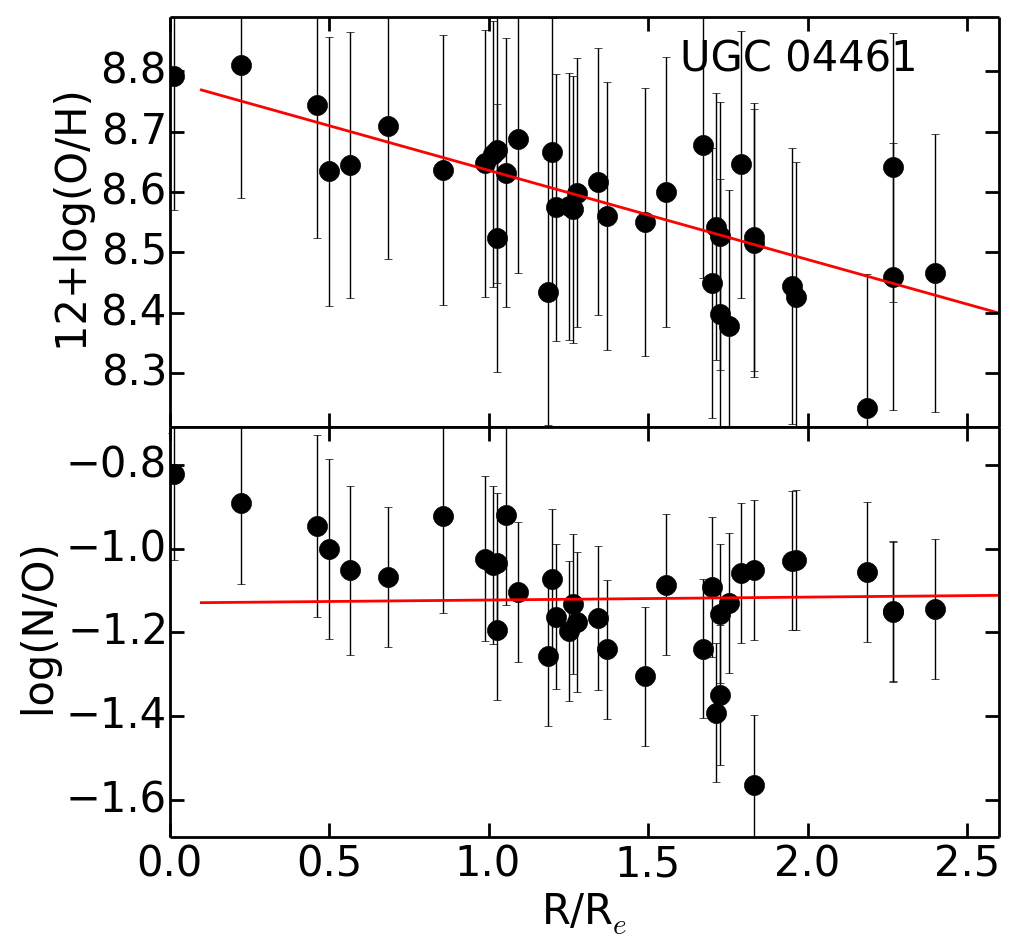}
   \includegraphics[width=4.5cm]{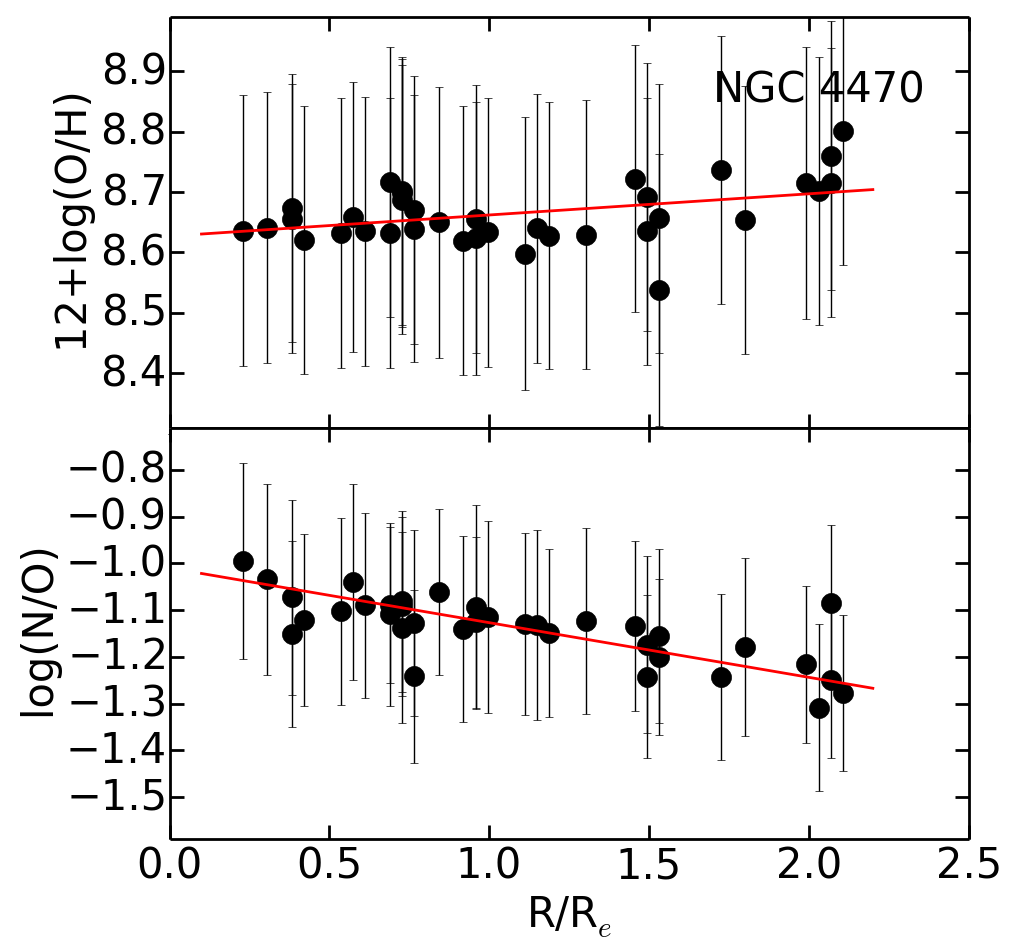}
   \includegraphics[width=4.5cm]{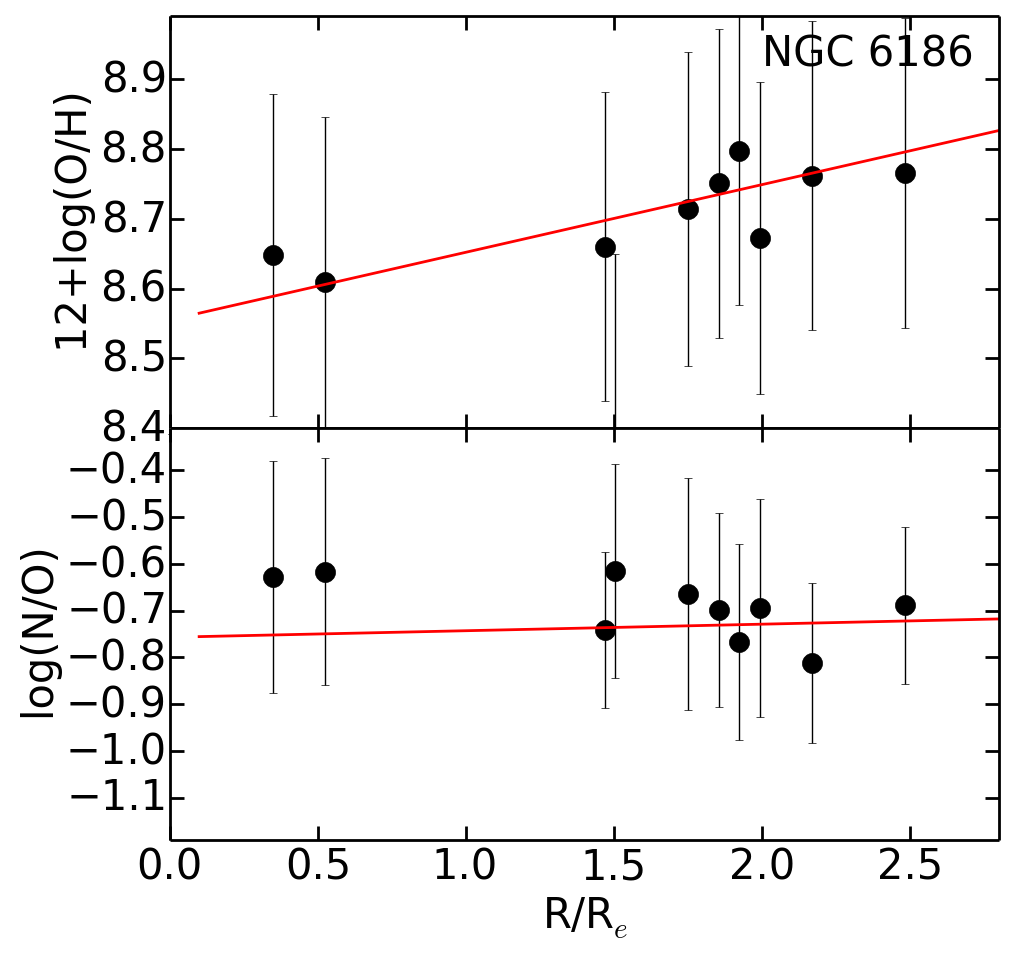}

   \caption{Four examples of different combinations of slopes in O/H and N/O 
in the sample of the CALIFA galaxies. From left to right: NGC~0477 presents negative gradients
both for O/H and N/O, UGC~04461 shows a negative
gradient of O/H, but the N/O gradient is flat. NGC~4470 has
positive O/H gradient and a negative N/O gradient and, finally, NGC~6186 shows a positive gradient both for O/H and N/O.}
              \label{grad_ex}%
    \end{figure*}

Although these results confirm the trend towards a negative radial gradient both
for O/H and N/O (e.g. \citealt{S14_grad, pcv04}), there is a non-negligible fraction
of objects that present a flat or even a positive gradient that is
already observed in the sample of CALIFA galaxies by \cite{S14_grad}.
On the contrary, no objects with inverted O/H or N/O gradients
are found by \cite{pcv04}, but the sample analysed 
there is much smaller.
The fraction of these positive
gradients are $\approx$ 19\% for O/H and 10\% for N/O. However
these fractions are reduced to 10\% and 4\%, respectively, when we only consider
those galaxies with a positive slope within the error. 
The number of galaxies and the average properties of the different 
subsamples as a function of the O/H and N/O slopes
are shown in Table \ref{tab_slopes}. 

Interestingly, the average slope for N/O is
more pronounced than for O/H. At same time, the number of objects with an inverted 
N/O gradient is lower than for O/H. These differences could be caused by two factors. Firstly, N/O is more precise
than O/H because the range of variation of N/O for the \hii\ regions in this sample is
much larger with a lower associated mean error (see Fig.\ref{oh-no}, i.e. the average 
slope of O/H represents a 5\% of the O/H range
for the CALIFA  \hii\ regions and a 2.5\% for N/O). Secondly, N/O tends to be unaffected
by processes related to hydrodynamical effects caused by the interchange of
(un-)processed material between different parts of the galaxies or even to the
surrounding IGM. This underlines once again the importance of the determination of N/O to calculate
chemical abundance gradients.

This N/O dependence can partially explain the lower average slope in this work 
as compared with \cite{S14_grad} and \cite{pcv04}. By using the O/H abundances 
derived by {\sc HCm}, but without a previous N/O determination (i.e. using a grid 
of models assuming a tight O/H-N/O relation as shown by the cyan line in Fig.\ref{oh-no}),
the resulting average slope for the same sample of galaxies is $\alpha_{O/H}$ = 
-0.069 dex/R$_e$ .
The difference between the O/H slope, which is derived in a galaxy using {\nii] lines when we consider a
previous N/O calculation, depends on the resulting N/O slope,
as can be seen in left panel of Fig.\ref{Da_oh-alpha_no}. 
This difference is negative for galaxies with a flat or inverted 
N/O gradient and positive for negative N/O slopes 
(i.e. when we calculate N/O, the resulting O/H slopes are flatter 
in those galaxies with a steeper N/O negative gradient).
On average, the mean O/H slope is 0.02 dex flatter when we consider a previous 
N/O estimation in the space of the grid of models for this sample. The difference is very small
if we compare the O/H slopes that are derived using this constrained grid of models with the
O/H slopes derived using a strong-line method that does not consider any N/O dependence. In right panel of Fig.\ref{Da_oh-alpha_no} we
show a comparison with the slopes obtained from the O3N2 parameter as calibrated
by \cite{m13}, and the residuals have an average lower than 0.01 dex/R$_e$.

On the other hand, the differences between the behaviours
of O/H and N/O radial variations are also evidenced by 
the not very high correlation between the slopes (Spearman’s
coefficient, $\rho_s =$ 0.39). Indeed, as it can be seen in 
Table \ref{tab_slopes}, the average N/O slope for those 
galaxies with a positive O/H gradient
is negative, although it is flatter than for the whole sample.
Similarly, the average O/H slope for those galaxies with 
a positive N/O gradient is negative although, again,  with 
a flatter value as compared with all galaxies.
The lack of a perfect correlation 
between O/H and N/O gradients is illustrated in 
Fig.\ref{grad_ex}, where four different combinations 
of gradient trends for O/H and N/O are shown.

Finally, the lower panel of Fig.\ref{abs_comp} shows the distributions
and scatter plot for the O/H and N/O values at the effective radius as calculated 
from the robust error-weighted linear fittings.
The average values are listed for all galaxies and as a function of the 
O/H and N/O slopes in Table \ref{tab_slopes}.
Contrary to the relation between
abundance ratios shown in Fig.\ref{oh-no} for individual \hii\ regions, 
the dispersion is very low and there is a high correlation coefficient 
($\rho_s$ = 0.80, while for individual \hii\ regions is 0.38) between the 
characteristic O/H and N/O values. 
A linear fitting to these points yields the following relation:
\[\log(N/O) = -20.39 + 2.23\cdot[12+\log(O/H)],\]
with a standard deviation of the residuals of 0.12 dex.
This linear fitting is also plotted in the lower panel of Fig.\ref{abs_comp}.

This very good correlation
between O/H and N/O values at the effective radius, in contrast to
the same relation for individual \hii\ regions or to the slopes of the same gradients, could
be the consequence that the typical expected values of $Z$ are not so sensitive
to possible internal variations across the discs, but mostly depend on other integrated properties
of the galaxies. This idea was already shown by \cite{pcv04}, where
a characteristic abundance value for spiral galaxies correlates much better with
some of the properties of the discs. According to these authors, this value corresponds to the
$Z$ predicted by the linear fitting at 0.4 times the isophotal radius (R$_{25}$). 
In our sample of selected CALIFA galaxies, the galactocentric distance 0.4$\times$R$_{25}$
corresponds on average to $\approx$ 2 times the effective radius. But, in what follows,
we consider the values
at the effective radius as typical values of the metal content of the galaxies studied here, to study possible correlations between the 
characteristic O/H and N/O at the effective radius and to study the
relation between these abundances 
and other integrated properties of the galaxies.This correlation between integrated properties and the fitted radial structure of
the spiral galaxies studied in the CALIFA survey has been also observed in
the stellar parameters by \cite{gonzalezd15}.

\section{Correlation with integrated galactic properties}

\subsection{Relation with luminosity and stellar mass}

\begin{figure*}[!htb]
   \centering
   \includegraphics[width=9cm]{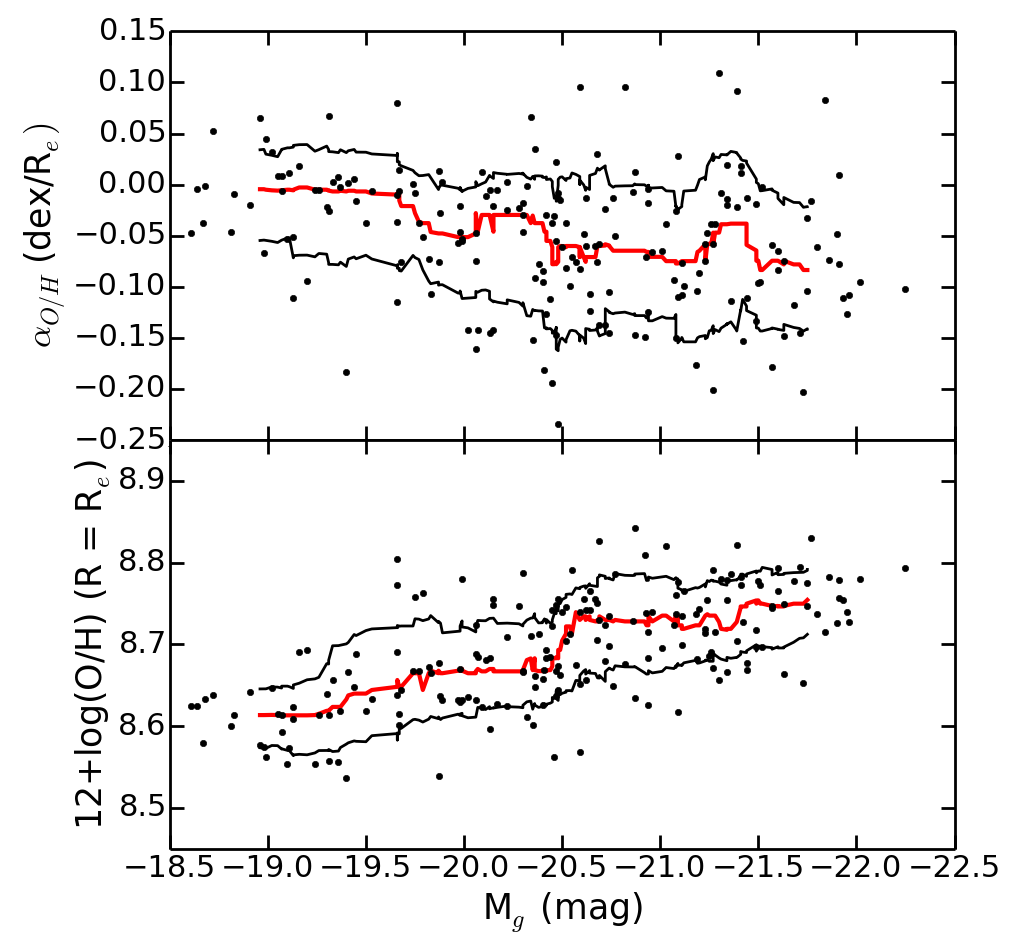}
   \includegraphics[width=9cm]{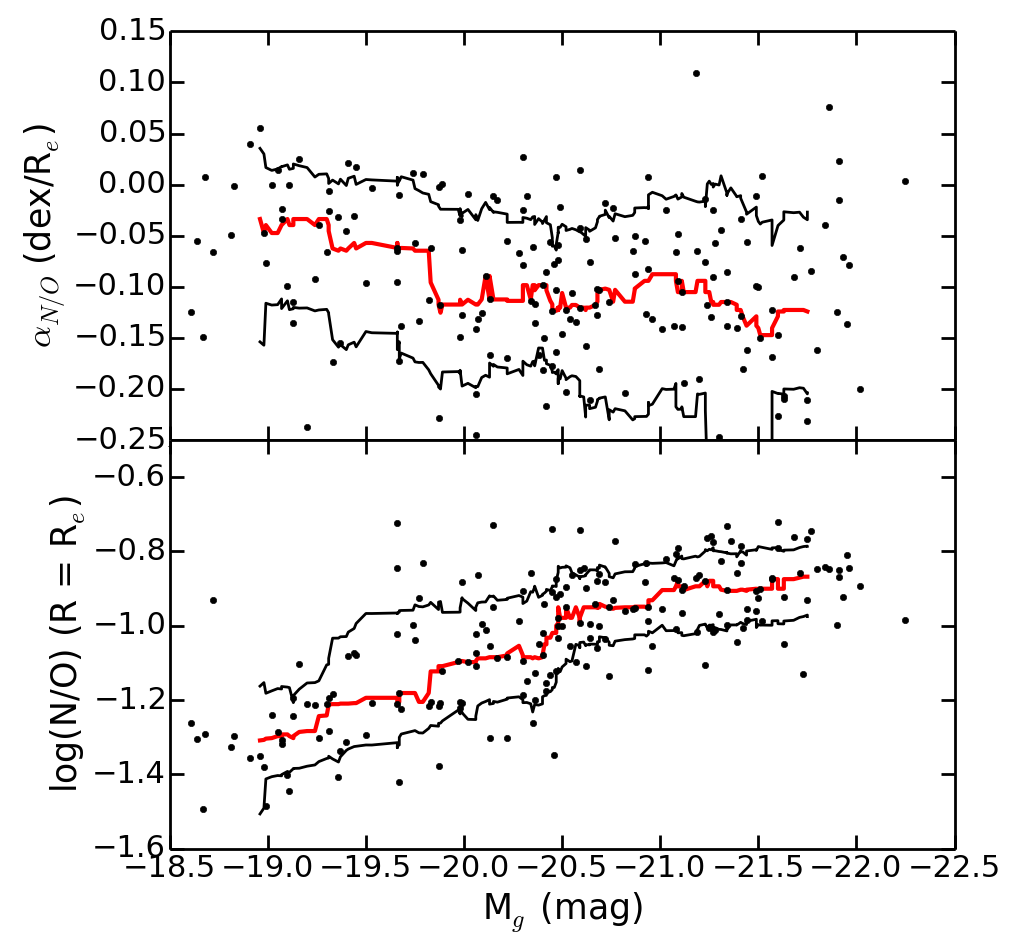}

   \caption{Relation between the absolute luminosity in the $g$ band and 
the derived slopes and characteristic values at the effective radius for O/H (left
panels) and N/O (right panels). The red solid line represents the running median for 
bins of 25 objects, and the black solid lines the $1\sigma$ above and
below the averages in the same bins.}
    \label{luminosity}
    \end{figure*}


   \begin{figure*}
   \centering
   \includegraphics[width=9cm]{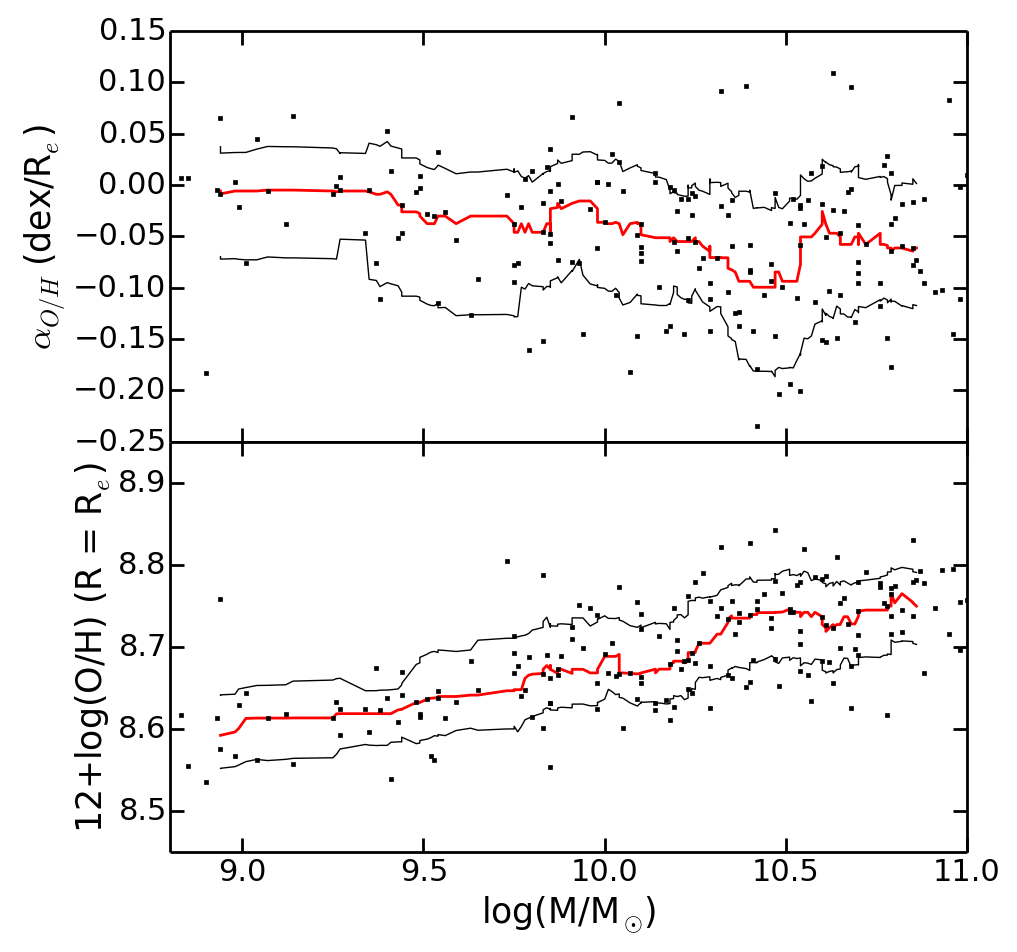}
   \includegraphics[width=9cm]{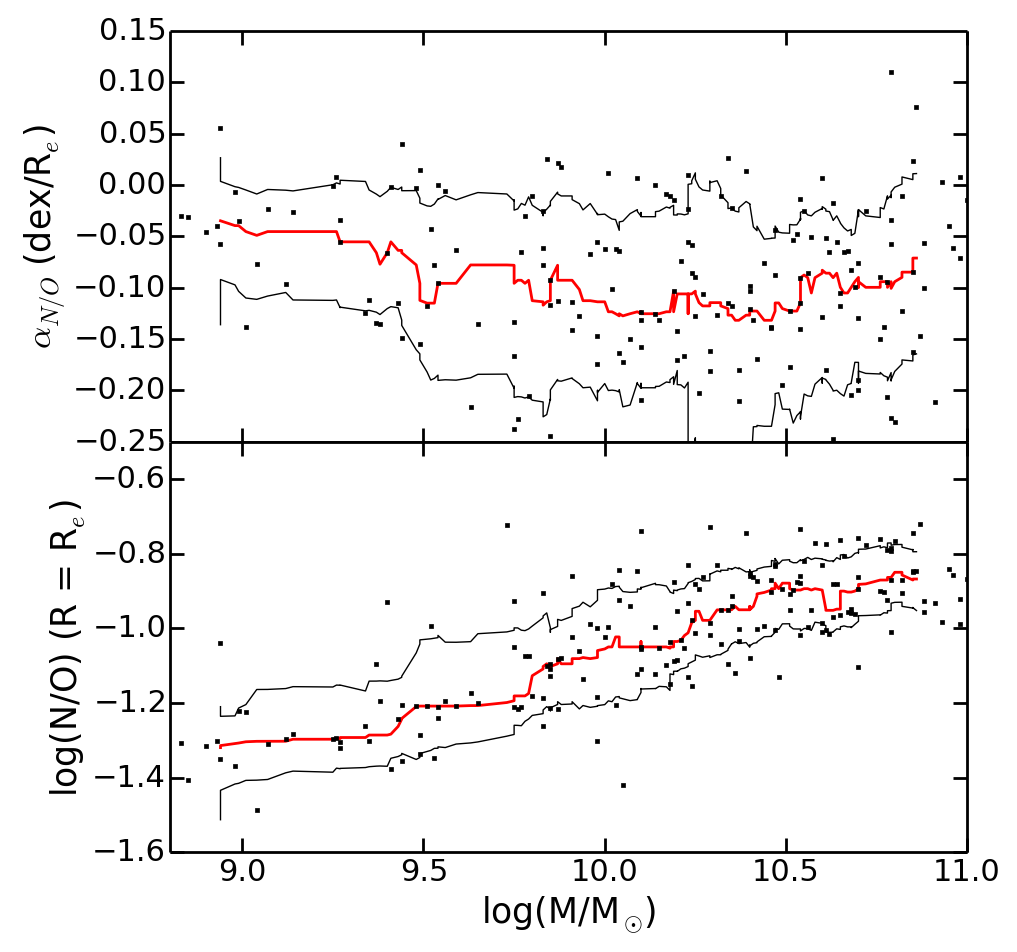}

   \caption{Relation between the total stellar mass and the derived slopes and 
characteristic values at the effective radius for O/H (left panels) and N/O
(right panels). The red and black solid lines have the same meaning as in Fig. \ref{luminosity}.}
    \label{mass}
    \end{figure*}

In this subsection we study to what extent is there a statistical dependence 
of the slopes or the characteristic abundance ratios on the total 
luminosity and the stellar mass of the 201 CALIFA galaxies with a derivations
of the O/H and N/O radial variations.

In Fig.\ref{luminosity} we show the derived slopes and the characteristic values both for O/H
and N/O along with the running medians for bins of 25 elements
as a function of the integrated absolute magnitude
in the SDSS filter $g$ as derived by \cite{walcher14} using growth curves. 
As can be seen, there is a slight trend towards flatter gradients  for both O/H and N/O
for lower luminosities. This is more evident
in the case of N/O, although, in both cases, the errors prevent us from being totally
confident with this result. For the less luminous galaxies of this sample
(M$_g$ $>$ -19.5) the mean slope is -0.014 dex/R$_e$ for O/H
and -0.055 dex/R$_e$ for N/O, while for the more luminous (M$_g$ $<$ -20.25) 
the mean slopes are -0.067 for O/H and -0.118 dex/R$_e$ for N/O.

Very similar results are found when we look at the relation with the total stellar
mass, which is shown in Fig.\ref{mass} both for O/H and N/O. The stellar mass derivations for this
sample are described in \cite{S13}.  The average slopes for galaxies with
log(M/M$_{\odot}$) $<$ 9.5 are less prominent for both O/H (-0.019 dex/R$_e$) and N/O (-0.061 dex/R$_e$) than for the rest of the sample. However, no clear 
trend is seen in this case for the high-mass galaxies. 
When we restrict the sample of galaxies to those with an inclination that is lower than 70$^o$
(113 objects), we still see the same trend towards sensibly shallower slopes for low luminosity/mass
objects (i.e. for M$_g >$ -19.5, $\alpha_{O/H}$ = -0.028 dex/R$_e$ 
and $\alpha_{N/O}$ = -0.059 dex/R$_e$).
Other works \citep[e.g.][]{S14_grad,ho15} have
not found any clear correlation between the average slope of the
O/H gradients, once normalised by the effective radius,  with the 
total luminosity or the total stellar mass. Our results 
point to a trend towards flatter slopes for less luminous and
less massive galaxies, but it is necessary to improve the statistical
significance of the results to confirm this trend.

In the lower panels of Fig.\ref{luminosity}, in the case of the integrated absolute $g$ luminosity,
and in Fig.\ref{mass}, for the total stellar mass, we can see the respective relations with the
obtained values of O/H and N/O at the effective radii.
As can be seen, there are tight and very low dispersion relations, confirming for this sample the
known connections between the total luminosity and/or stellar mass of a galaxy with
 $Z$ content (e.g. \citealt{lamareille04, t04}) and with N/O
\citep{pmc09,pm13}. In the case of O/H this result confirms that obtained by \cite{pcv04} who find a tighter correlation with luminosity, as a proxy
of the stellar mass, and with rotation velocity, as a proxy of the dynamical mass, when
a characteristic oxygen abundance value at a specific radial position from a fitting is assumed. 
This characteristic value was situated at 0.4 times the isophotal radius, instead 
of using an integrated value of $Z$. 

In our case a quadratic relation can be fitted between the total stellar masses
and the O/H values at the effective radius
\begin{equation}
y_1 = 8.1994 + 0.0.0168\cdot x - 0.0031\cdot x^2
,\end{equation}
\noindent where $x$ is log(M/M$_{\odot}$) and $y_1$ is
O/H in terms of 12+log(O/H). This fitting only has a standard deviation of
the residuals for the 201 studied galaxies of 0.03 dex, which is much lower
than the uncertainty associated with the O/H derivation and the
O/H characteristic values in each galaxy.

A similar fitting for N/O at the effective radius yields
\begin{equation}
y_2 = -8.0611 + 1.1581\cdot x -0.0456\cot x^2
,\end{equation}
\noindent where $x$ is log(M/M$_{\odot}$) and $y_2$ is
N/O in terms of log(N/O). For this fitting the standard deviation of the residuals
is also of 0.03 dex, which is much lower than the error associated with the N/O
derivation and the characteristic N/O values.
The quadratic fittings found in the relations between the total stellar 
mass and both the O/H and N/O characteristic values confirm the trend already 
observed (e.g. \citealt{t04,pil07}) of a flattening of the MZR at high
masses and luminosities.

\subsection{Relation with present-day star formation and colour}
\label{SF-colour}

   \begin{figure*}
   \centering
   \includegraphics[width=9cm]{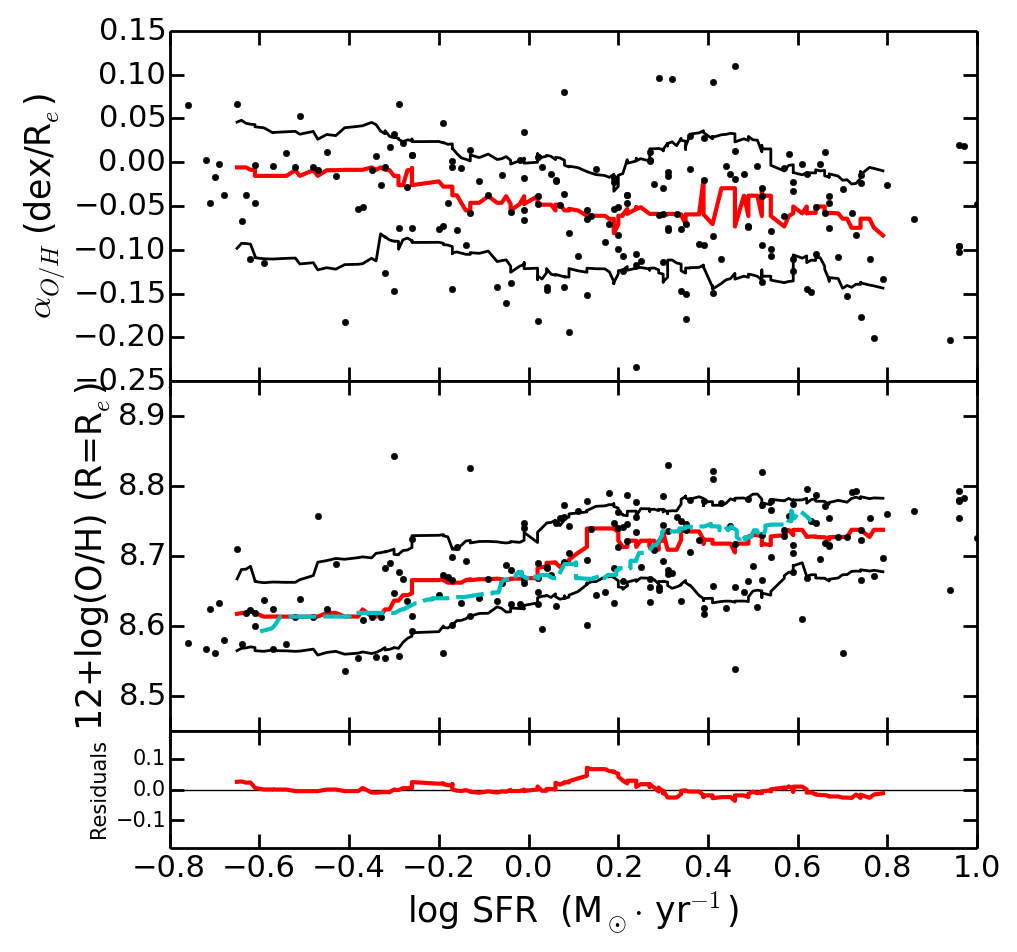}
   \includegraphics[width=9cm]{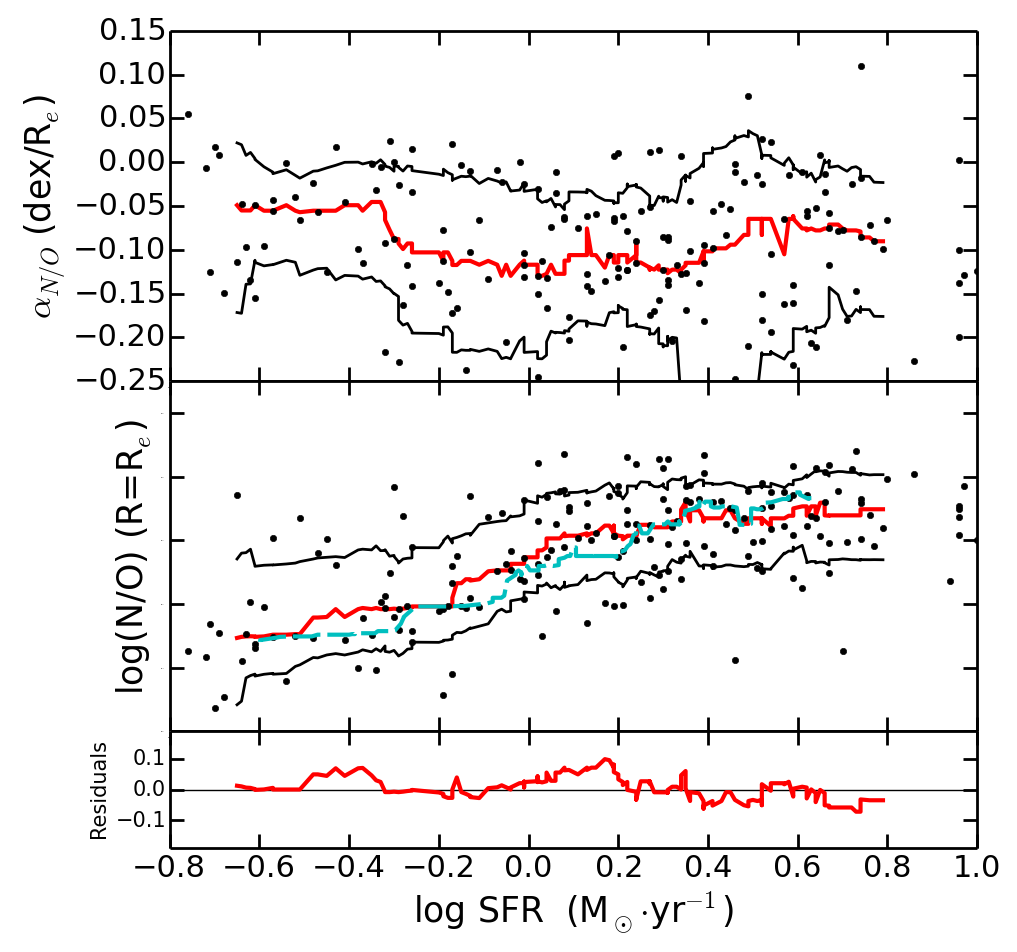}

   \caption{Relation between the integrated SFR and
the derived slopes and characteristic values at the effective radius 
for O/H (left panels) and N/O (right panels). 
The red and black solid lines in the upper and middle panels have the same meaning as in Fig. \ref{luminosity}.
In the middle panels, the dashed cyan line represents the running median 
assuming the observed relation between the SFR and stellar mass. In the lower panel,
 the residuals are indicated in red and the running medians are indicated in cyan.}

    \label{sfr}
    \end{figure*}

 \begin{figure*}[!htb]
   \centering
   \includegraphics[width=9cm]{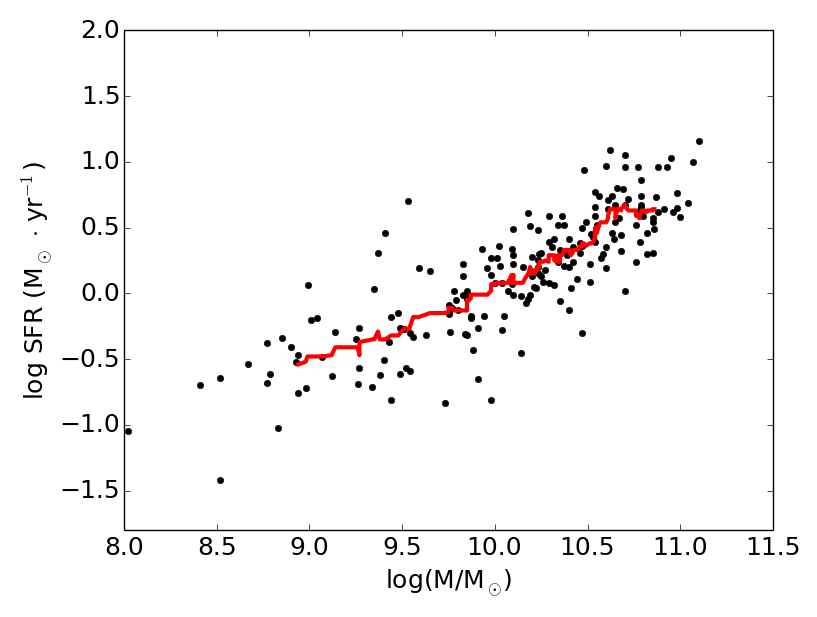}
   \includegraphics[width=9cm]{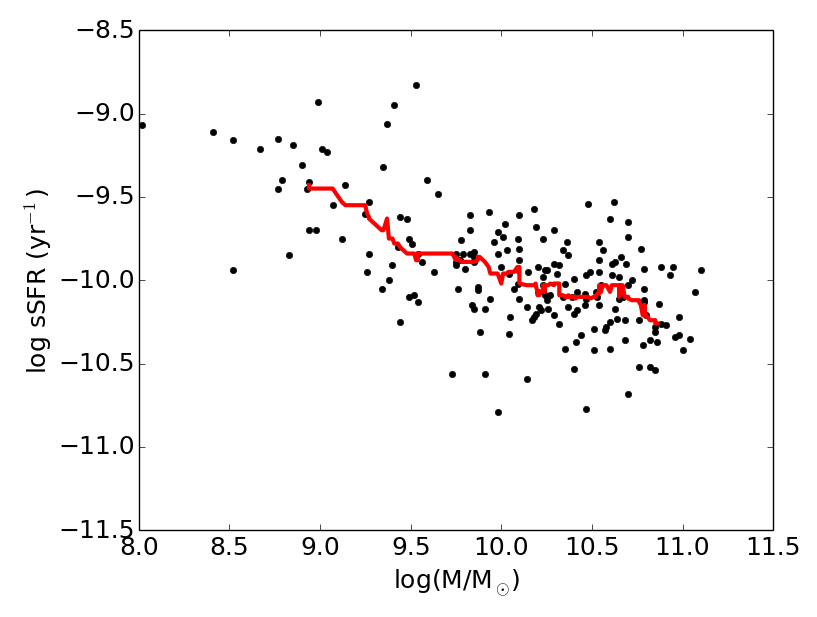}

   \caption{Relation between the total stellar mass and the present-day 
integrated SFR (at left) and the sSFR (at right) for the 201 selected spiral non-interacting CALIFA galaxies. 
The red solid lines represent the running median for bins of 25 objects.
}
 \label{MS}   
 \end{figure*}

   \begin{figure*}
   \centering
    \includegraphics[width=9cm]{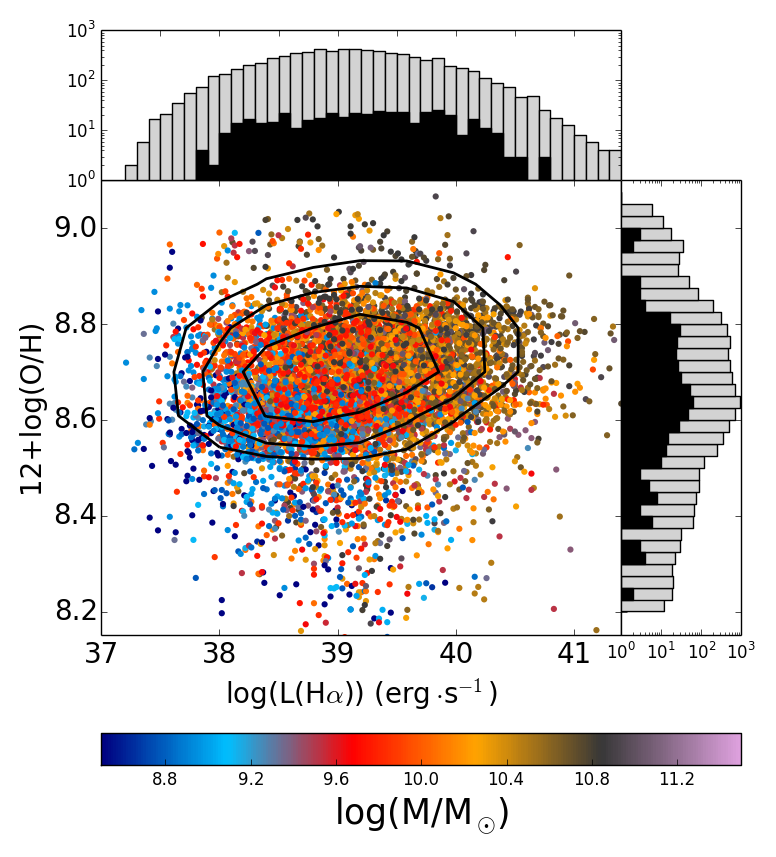}
    \includegraphics[width=9cm]{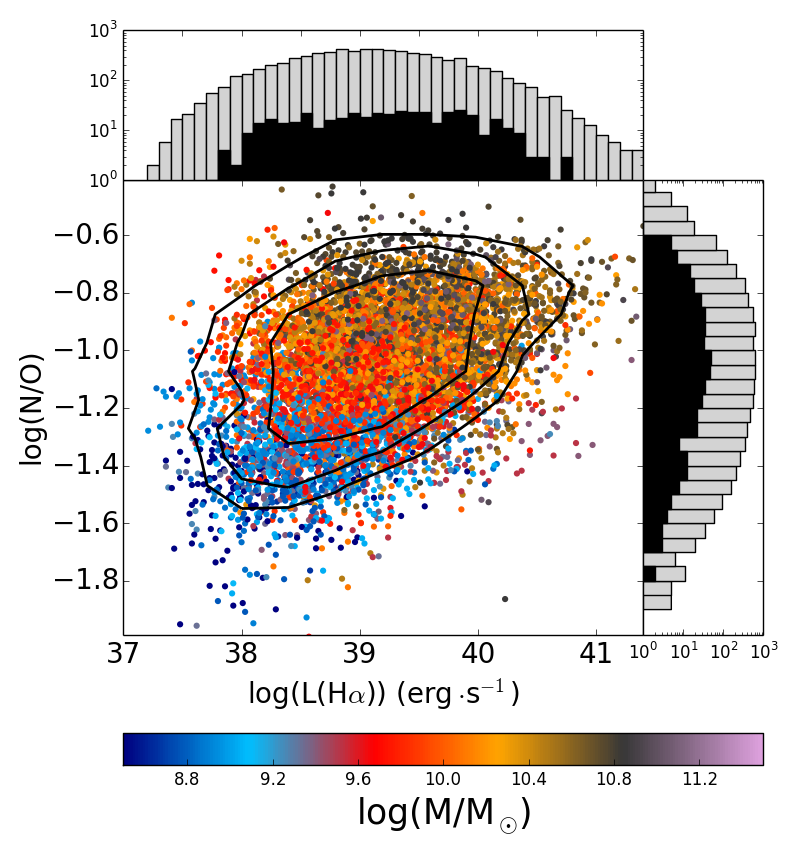}

   \caption{Relation between the H$\alpha$ luminosity
and O/H, at left, and N/O, at right, for the selected \hii\
regions of the galaxies for which a linear gradient was calculated. 
The colour encodes the total stellar mass of the galaxy where each \hii\ region is 
situated.
The histograms show in logarithmic scales in all axes 
the distribution for all objects represented in the plot (white bars)
and for those \hii\ regions in a galaxy with 
an inverted gradient (black bars). Colours indicate the density of points and 
solid lines represent the 1$\sigma$, 2$\sigma$, and 3$\sigma$ contours.}
\label{lha}
\end{figure*}

   \begin{figure*}
   \centering
   \includegraphics[width=9cm]{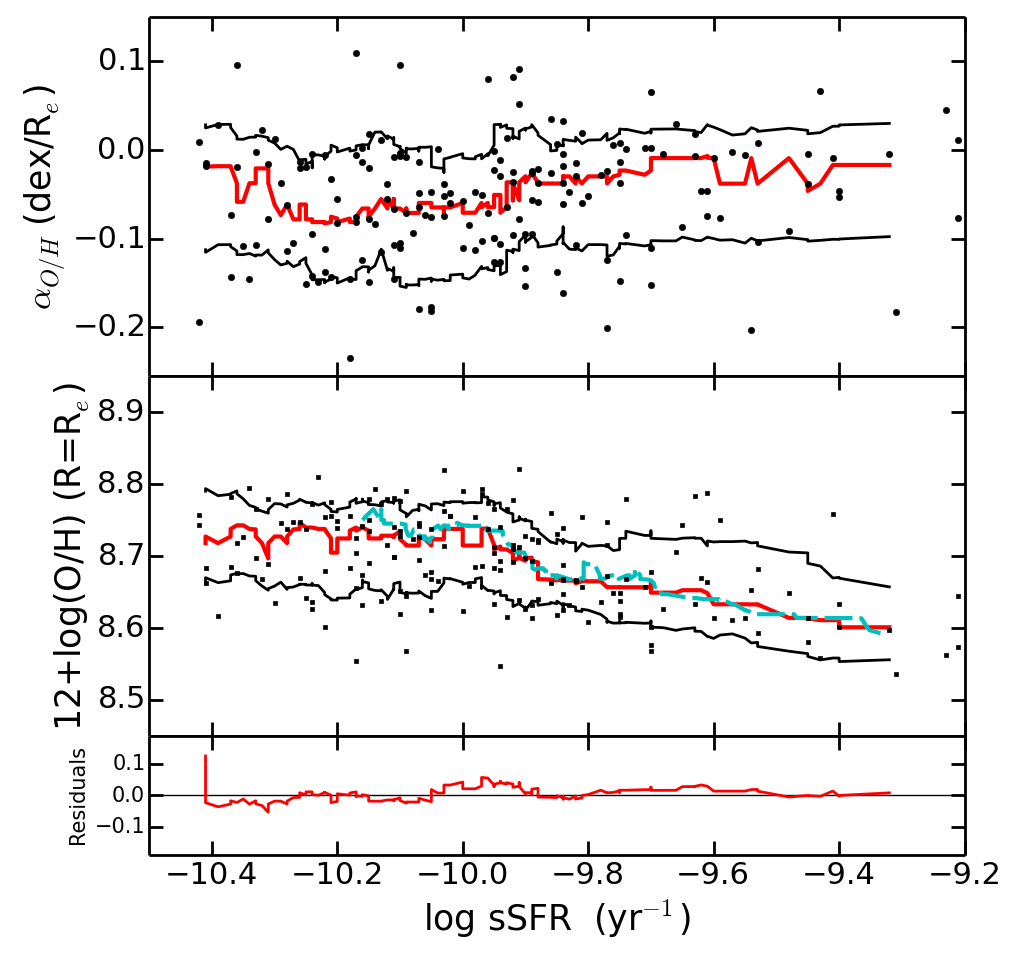}
   \includegraphics[width=9cm]{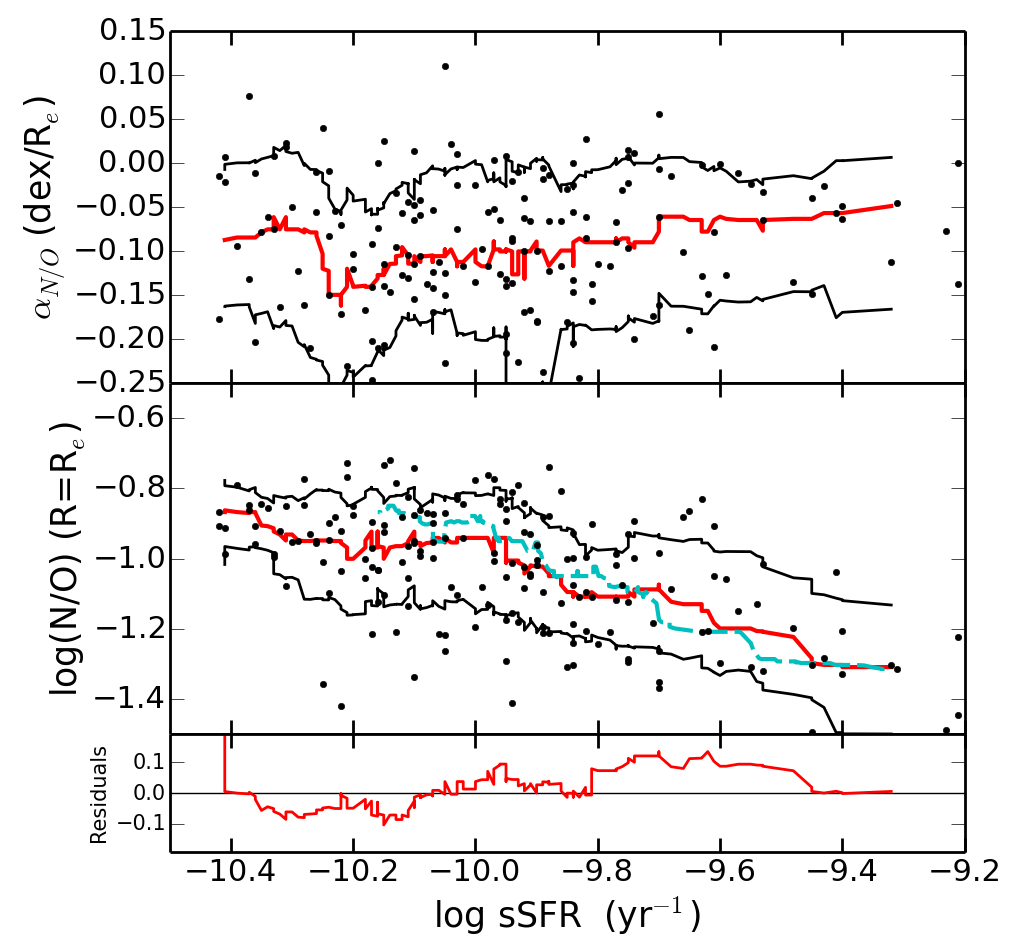}

   \caption{Relation between the integrated sSFR and
the derived slopes and characteristic values at the effective radius 
for O/H (left panels) and N/O (right panels). 
The red and black solid lines in the upper and middle panels have the same meaning as in Fig. \ref{luminosity}.
In the middle panels, the dashed cyan line represents the running median 
assuming the observed relation between the SFR and stellar mass. In the lower panel, 
 the residuals are indicated in red and running medians are indicated in cyan.}
    \label{ssfr}
    \end{figure*}

   \begin{figure*}
   \centering
   \includegraphics[width=8cm]{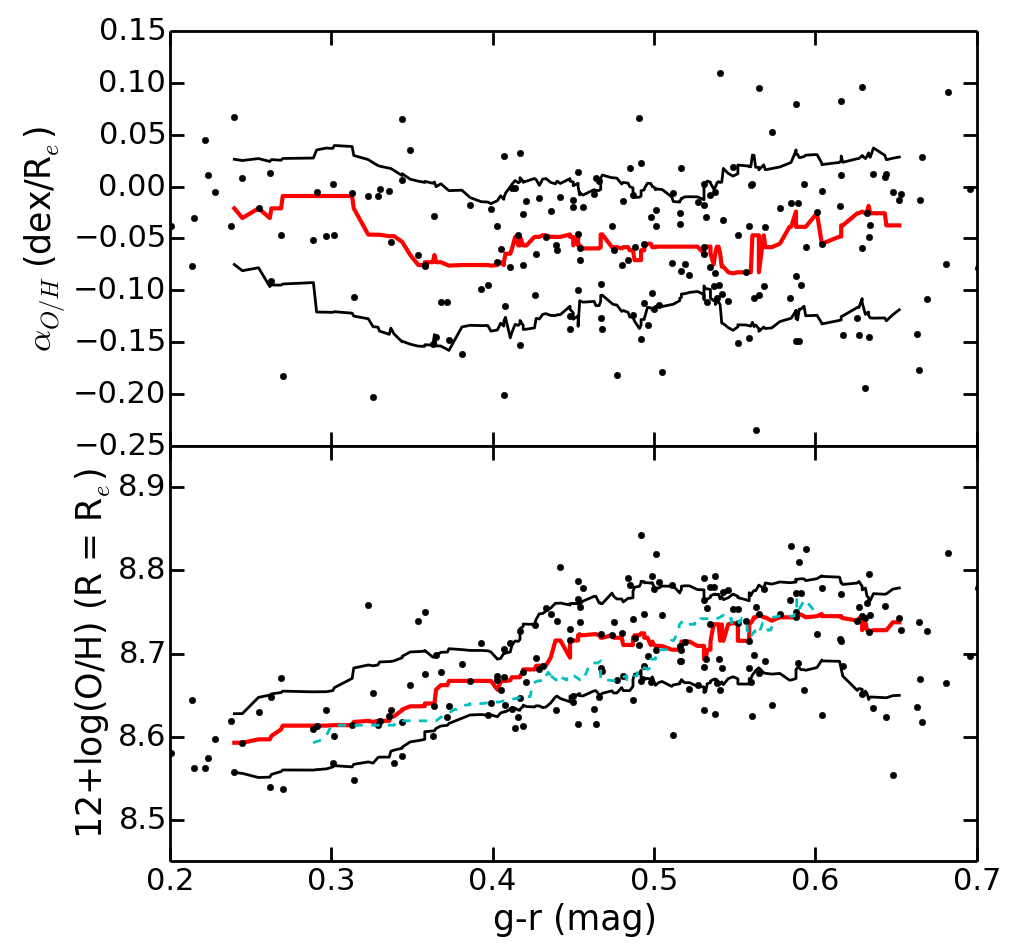}
   \includegraphics[width=8cm]{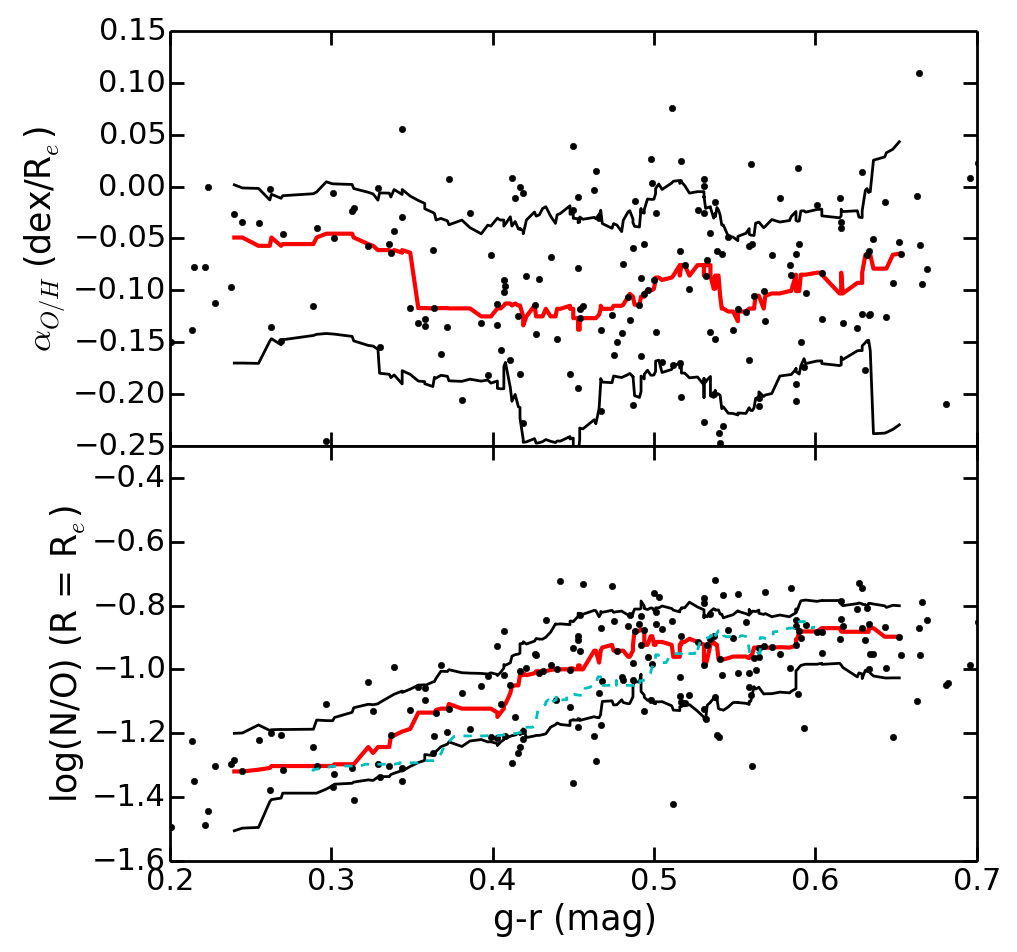}

   \caption{Relation between the colour index $g-r$ and the derived slopes and 
the characteristic values at the effective radius for O/H (left panels) and N/O
(right panels). 
The red and black solid lines have the same meaning as in Fig. \ref{luminosity}.
The dashed cyan lines represent the running median, assuming the observed 
relation between $g-r$ and stellar mass for the same galaxies.}
    \label{g-r}
    \end{figure*}

Regarding the relation between the metal content of a galaxy and
its total SFR, it has been suggested, for certain samples of galaxies, 
that, using the integrated emission of star-forming galaxies, there is a
dependence between $Z$ and SFR.
This dependence points towards lower $Z$ values for higher SFRs in a specific mass bin,
thereby reducing the dispersion of the MZR in the so-called fundamental 
metallicity relation (FMR; e.g. \citealt{mannucci10,lara10}).
According to \cite{pm13} from the analysis of SDSS data, the SFR
dependence is not observed with N/O, so this could be indicative that the
SFR-$Z$ for each mass bin is due to the presence of inflows of metal-poor gas and/or outflows 
of enriched gas, that do not affect  N/O, but can considerably alter both the SFR and $Z$.

It is thus interesting to examine whether the slopes or the characteristic values of the chemical abundances
calculated for the sample of CALIFA galaxies depend on their integrated SFR.
The integrated SFR were calculated using the extinction-corrected integrated
\ha\ fluxes following the same procedure as described in \cite{S13}.

In the upper panels of Fig.\ref{sfr} we see the derived O/H and N/O slopes, with their 
corresponding running medians, as a function of log SFR. 
As in the case of stellar mass, there is a very slight, not statistically significant,
difference between the O/H slopes of low and high SFR (i.e. $\alpha_{O/H}$
= -0.043 dex/R$_e$ for SFRs lower than the mean value, 10$^{0.2}$ M$_{\odot}$/yr,
and -0.062 for higher SFRs). A similar trend is obtained for N/O,  but
the difference between the two bins is lower than 0.01 dex/R$_e$.

Much more clearly, in the middle panels of the same figures, we see that there is a
trend both for O/H and N/O to find higher characteristic values at the
effective radius for galaxies with a higher integrated SFR.
To check whether this relation is contrary to the FMR, we examine
the relation between SFR and the stellar mass. In the left panel of Fig.\ref{MS} we show
this relation for the sample of CALIFA galaxies selected for the analysis
of radial gradients.
As in the case of the galaxies of the main sequence of star formation, as shown in
\cite{brinchmann04}, the SFR correlates with the
stellar mass of the galaxies, as expected from the fact that larger galaxies have
larger gas repositories.
Taking the mass-SFR relation that is obtained from the corresponding running median for
this sample, we used the observed MZR and MNOR plotted in
Fig. \ref{mass} to compare the resulting SFR-O/H or SFR-N/O
with the same observed relations.
As can be seen in the lower panels of Fig.\ref{sfr}, the residuals between the SFR-O/H 
(at left) and SFR-N/O (at right) and the same relation considering the observed M-SFR
relation are very small.
The standard deviations of these residuals are 0.03 dex for both plots with no mean deviation 
at any SFR. This M-SFR relation also explains the slight slope difference observed
for O/H, as low-mass galaxies have, on average, slightly flatter slopes.

We can therefore conclude that
no additional dependence between O/H or N/O and SFR is 
observed for this sample of galaxies contrary to the
results pointing to the existence of the FMR. This agrees with the results
in \cite{S13} for individual CALIFA \hii\ regions, for
which no additional relation between O/H and SFR is found, supporting a
scenario of gas recycling faster than gas accretion from inflows
or mass loss by outflows.

Although a certain relation between O/H (or N/O) and the SFR is observed
for individual \hii\ regions,
this is again a consequence of the M-SFR relation.
In Fig.\ref{lha}, a certain correlation between the O/H of the individual \hii\
regions and their H$\alpha$ luminosity ($\rho_s$ = 0.17) can be seen, which is even clearer
for N/O ($\rho_s$ = 0.42). These correlations can be explained as due to the stellar mass
of the galaxies hosting the \hii\ regions. The average L(H$\alpha$) is lower
for the less massive galaxies than for the more massive galaxies.
The mean L(H$\alpha$) = 10$^{38.89}$ erg/s for M$_*$
$<$ 10$^{10.22}$ M$_{\odot}$ and 10$^{39.43}$ for higher masses,
as M=10$^{10.22}$ M$_{\odot}$ is the median stellar mass
of the 201 galaxies with an estimate of the abundance gradient.

Very similar results are obtained when we analyse the dependence
between the slopes of the gradients or the characteristic abundance
values on the specific star formation rate (sSFR = SFR/M), which can
be considered an indicator of the ratio between present-day and past SFR.
There is also evidence in the literature that  specific mass bin galaxies
with higher sSFR have on average lower $Z$ (e.g. \citealt{ellison08}).
The resulting median slopes and characteristic values can be seen in
Fig.\ref{ssfr}.

As in the case of the mass or the SFR, a slight trend towards flatter O/H and N/O gradients for galaxies with a lower sSFR is found
(i.e. $\alpha_{O/H}$ = -0.029 dex/R$_e$ ($\alpha_{N/O}$ = -0.085) for the bin
of sSFR that is lower than the median (10$^{-9.97}$ yr$^{-1}$ and
$\alpha_{O/H}$ = -0.059 dex/R$_e$ 
($\alpha_{N/O}$ = -0.105) for the upper sSFR bin. 

Regarding characteristic values,
a clear trend towards lower values both for O/H and N/O 
for higher values of sSFR is seen in the
middle panels of the same figure.
However, as in the case of SFR, we see that this relation is owing
to the previously studied MZR in combination with the relation between
mass and sSFR observed in right panel of Fig.\ref{MS}. This sample
of galaxies, again, shows  a clear relation in this plot
as the sSFR is lower for galaxies of higher stellar mass.
By comparing, in the lower panels of Fig.\ref{sfr}, the relations
O/H - sSFR or N/O - sSFR with the running medians of the MZR
using the M - sSFR relation, we find that the standard deviation
of the residuals is lower than 0.02 dex in both cases.

The same behaviour between slopes or characteristic values is observed
in Fig. \ref{g-r}, where we analyse these quantities as a function of the integrated photometric colours
$g-r$ \citep{walcher14}, which can be interpreted as a proxy of the relation between ongoing and past
star formation, as sSFR. In the respective upper panels, no evident relation
is found between the O/H and N/O slopes and the integrated colours. On the
contrary, in the lower panels bluer galaxies tend to
have lower abundances and redder objects tend to have higher abundances.
Again, as in the case of SFR and sSFR, this result must be interpreted in light of the MZR and
the relations of the colour with stellar mass for this specific 
non-interacting CALIFA sample with many \hii\ regions.
When we assume the observed relation between mass and $g-r$ colour index in the
figures, we cannot see significant differences with the observed pattern,
which demonstrates that no additional dependence exists between colour and $Z$ or N/O.

The observed empirical relation between colours and masses in this sample 
goes in the same direction as pointed out for the CALIFA sample by \cite{gonzalezd15}, 
who show that on average less massive galaxies are bluer and more massive galaxies are redder.
Some works have suggested that a correlation between N/O and the integrated colour can be
indicative of the star formation history owing to the delay in the ejection of N to the ISM (e.g.
dwarf irregular galaxies, \citealt{vh06, lopezs10}). However, in our sample, as most of
the observed N has a secondary origin, the correlation with colour is totally understood in terms
of the MNOR and the M-sSFR relation.

\subsection{Relation with morphological type}

In this subsection we analyse the possible link between the calculated
slopes and the characteristic O/H and N/O values with the morphological
type for the 201 galaxies of our CALIFA sample of non-interacting galaxies.
The morphological classification was performed by eye, based on 
independent analysis by members of the CALIFA collaboration. The results are compatible with other photometric indexes and are 
described in detail in \cite{walcher14}.

The results for the averages of slopes and characteristic abundance values at the effective
radius for the different morphological types are listed in Table \ref{bar_slopes}
and are also represented in Fig.\ref{morph}.
As can be seen in the respective upper panels, there is a slight 
trend towards shallower gradients both for O/H and N/O for late-type
galaxies ($\alpha_{O/H}$ = -0.018 dex/R$_e$ and $\alpha_{N/O}$ = -0.065 dex/R$_e$ for Sd galaxies).
At the same time, for O/H there is also a trend towards shallower gradients for
early-type galaxies ($\alpha_{O/H}$ = -0.022 dex/R$_e$ in Sa galaxies).
This result agrees with the trend found in \cite{S14_grad}, who 
found a slightly shallower gradient on average for very early spiral 
galaxies and also with the results found by \cite{vce92} and \cite{zaritsky94} 
who find steeper gradients in the intermediate-type galaxies.
A similar result is found for the slope of the gradient of the stellar
properties by \cite{gonzalezd15} in the CALIFA sample.

Regarding the abundance values at the effective radius, considered
as characteristic of the metal content of each galaxy, there is an evident decrease
in the O/H value, and this is even more evident for N/O when we move from
early- to late-type galaxies (i.e. O/H is reduced in 0.14 dex and N/O in 0.35 dex
from the early to late types). This same pattern was already pointed
out by \cite{rh94}.

However, this result must be inspected in
light of the relation between the typical metal content of a galaxy and its stellar
mass, which is explored in subsections above. In this sample, there is a bias in the sense that 
late-type galaxies tend to have lower masses on average than early-type galaxies.
In the lower panels of Fig.\ref{morph}, we also show the median values
for O/H and N/O at the average stellar mass of each morphological type group. 
As can be seen, no apparent difference appears between the expected values at 
the average stellar mass and the found value at a specific Hubble type
(i.e. the mean difference is lower than 0.02 dex both for O/H and N/O).

Then we can conclude that the observed relation between slopes or
characteristic values with the morphological types are due to the combined effect 
of the MZR or  MNOR and the mass-morphological type relation of this sample. This 
relation has been also observed for the CALIFA sample of galaxies by \cite{gonzalezd15}
by means of the analysis of the spatial distribution of the stellar properties of
the galaxies.

   \begin{figure*}
   \centering
   \includegraphics[width=9cm]{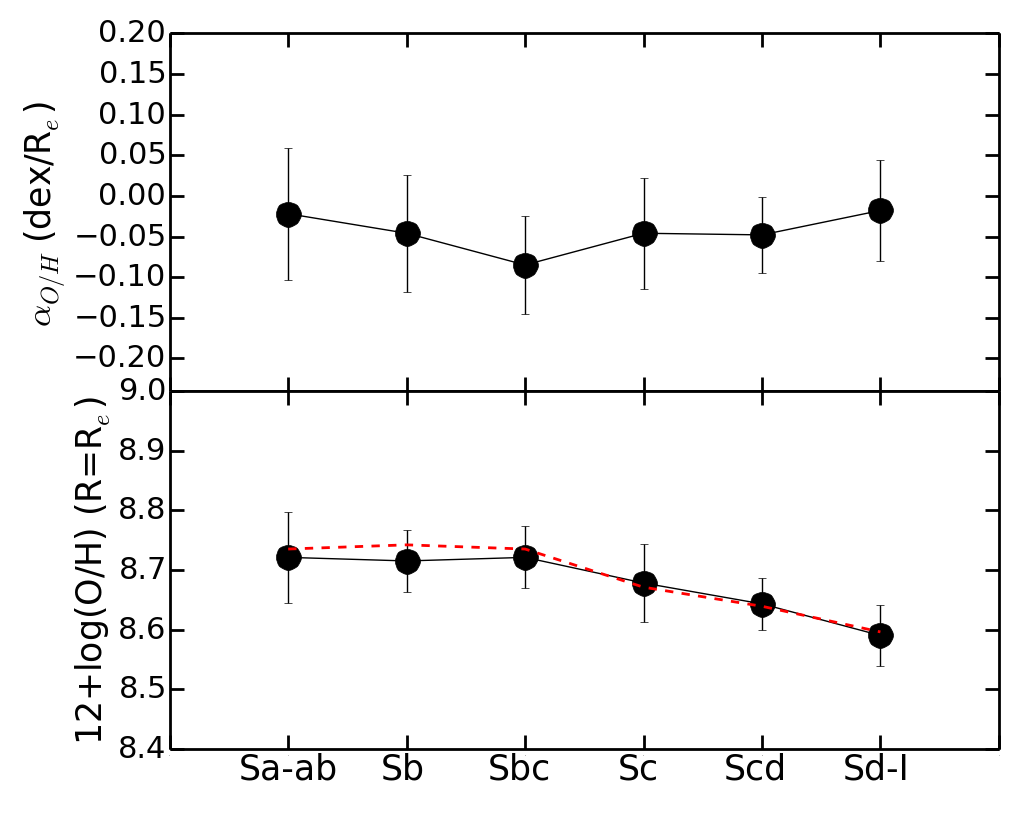}
  \includegraphics[width=9cm]{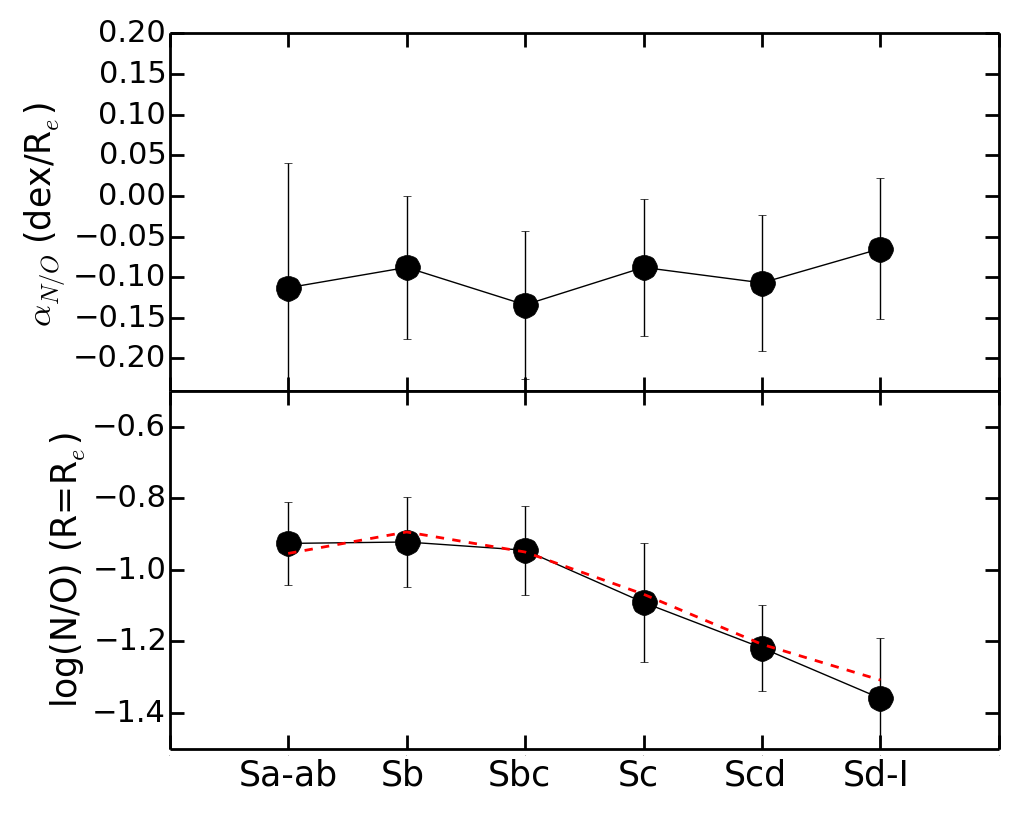}

   \caption{Average slopes and value at the effective radius of the linear fittings 
in the non-interacting analysed galaxies as a function of the morphological type for O/H (left 
panels) and N/O (right panels).
The black solid lines connects the average points for each morphological type, while the red dashed lines 
correspond to the average typical abundance ratios found at the average stellar mass
of each type.}
\label{morph}%
    \end{figure*}

\subsection{Relation with the presence of a bar}

It has been proposed that the presence of a bar can be an element that is important
to explain the flattening of gradients of chemical abundances in spiral galaxies.
\cite{vce92} and \cite{zaritsky94} find this trend in their samples of 
spiral galaxies and they claim that this could be the consequence of the presence 
of gas flows driven by the bars that make the metal distribution homogenise 
throughout the discs.

The inspection of our sample for the presence of bars was performed
by eye by members of the CALIFA collaboration, and it is described elsewhere 
in detail in \cite{walcher14}. Three different groups were defined, following 
the classical scheme: (A) galaxies with no bar; (AB) galaxies that may have a 
bar, but it is not clearly visible; and (B) clearly barred galaxies. 
The visual classification was cross-checked with an automatic search for bars
and this produced very similar results.

The results for the averages of slopes and characteristic abundance values at the effective
radius for the three analysed categories are listed in Table \ref{bar_slopes}.
As can be seen, no significant differences exist between the average slopes
either for O/H or for N/O as a function of the presence of a bar in this sample of
galaxies, although there is a slight trend towards flatter O/H gradients in barred
galaxies.  The difference in the N/O slopes of barred and unbarred galaxies
is even smaller.
This slight difference within the errors is still obtained when galaxies
with an inclinations $>$ 70$^o$ are ruled out, as the detection of a bar is more difficult
in these objects.

The possible abundance gradient flattening as a function of the presence
of a bar is not observed by \cite{S14_grad, marino15}, who use some of the same data but a different methodology for the calculation of O/H.
A similar result is obtained by \cite{sanchezb14} for a
subsample of face-on spiral galaxies, where no difference is found 
for the gradients of stellar age and $Z$ as a function of the presence of a bar.

Regarding the characteristic chemical abundance ratios at the effective radius, 
there are not apparent large differences
between the O/H values found for barred and unbarred galaxies, but a slight trend
towards higher N/O values for barred galaxies is found. This difference is not
statistically significant, but it cannot be because of any bias in the stellar masses of
the three subsamples defined as a function of the presence of a bar.  The mean stellar mass for the three subgroups are
log(M/M$_{\odot}$) = 10.01 for A galaxies, 10.08 for AB, and 10.23 for B, which
would represent an enhancement of 0.01 dex for O/H and 0.04 dex for N/O,
according to the results described in the above subsections. On the other hand,
these results agree with the differences of chemical abundances in the
centres of barred galaxies found by \cite{florido}, which point
to an enhancement of N/O in barred galaxies, while they do not find any
difference when the studied ratio is O/H. This difference in N/O and not in O/H
in the centres of barred galaxies could be due to a different star formation
efficiency in the inner parts of galaxies as a consequence of the influence of
bar-driven flows of gas.


\begin{table*}
\begin{minipage}{180mm}
\caption{Average slopes and values at the effective radius for
the linear fittings of O/H and N/O in the analysed CALIFA galaxies 
for categories attending the observed morphology and the presence of a bar.}

\begin{center}
\begin{tabular}{ccccccc}
\hline
\hline
     &  Number & \%  &  $\alpha_{O/H}$ & 12+log(O/H) & $\alpha_{N/O}$ & log(N/O) \\
Type     &     &       &   (dex/R$_e$)  & (at R$_e$)   & (dex/R$_e$)   & (at R$_e$)       \\
\hline
\multicolumn{7}{c}{\it Morphology} \\
\hline
Sa-ab              & 17 & 8.5   & -0.022 $\pm$ 0.081 & 8.721 $\pm$ 0.076 & -0.113 $\pm$ 0.154 & -0.926 $\pm$ 0.154 \\
Sb                 & 44 & 21.9  & -0.046 $\pm$ 0.072 & 8.715 $\pm$ 0.052 & -0.088 $\pm$ 0.088 & -0.922 $\pm$ 0.125 \\
Sbc                & 58 & 28.9  & -0.085 $\pm$ 0.060 & 8.721 $\pm$ 0.052 & -0.134 $\pm$ 0.091 & -0.945 $\pm$ 0.124 \\
Sc                 & 37 & 18.4  & -0.046 $\pm$ 0.068 & 8.678 $\pm$ 0.066 & -0.088 $\pm$ 0.084 & -1.091 $\pm$ 0.166 \\
Scd                & 26 & 12.9  & -0.048 $\pm$ 0.047 & 8.643 $\pm$ 0.044 & -0.107 $\pm$ 0.084 & -1.278 $\pm$ 0.120 \\
Sd-I               & 19 & 9.4   & -0.018 $\pm$ 0.062 & 8.590 $\pm$ 0.051 & -0.065 $\pm$ 0.087 & -1.359 $\pm$ 0.169 \\

\hline
\multicolumn{7}{c}{\it Presence of a bar} \\
\hline
A                 & 89 & 44.3  & -0.068 $\pm$ 0.067 & 8.680 $\pm$ 0.068 & -0.102 $\pm$ 0.091 & -1.071 $\pm$ 0.213 \\
AB & 53 & 26.4   & -0.061 $\pm$ 0.075 & 8.678 $\pm$ 0.067 & -0.109 $\pm$ 0.099 & -1.061 $\pm$ 0.176 \\
B   & 59 &  29.4 & -0.041 $\pm$ 0.061 & 8.710 $\pm$ 0.073 & -0.100 $\pm$ 0.101 & -0.979 $\pm$ 0.182 \\
\hline

\end{tabular}
\end{center}
\label{bar_slopes}
\end{minipage}
\end{table*}

\section{An insight into flat and inverted gradients}

A non-negligible fraction of the 201 non-interacting galaxies
in the CALIFA survey studied in this paper show flat or positive gradients
of O/H and/or N/O. 
Given the uncertainties for the  emission-line fluxes that we used, as well as for those
 from the methodology used to derive the abundances, many of these slopes 
have large associated errors, but the fraction of galaxies with an inverted slope varies
from 10\% up to 19\% in the case of O/H and from 4.5\% up to
10\% in the case of N/O depending on the errors of the slopes. We do not find a complete coincidence between
the galaxies that present inverted gradients of O/H and N/O; i.e. only
7 out of the 20 galaxies with an inverted N/O gradient also
have an inverted O/H gradient.

As shown in previous sections, there is a slight, although
not statistically significant, trend towards flatter gradients
in less luminous and massive galaxies and in the earliest
morphological types. 
However, there is not any clear difference
between the average integrated properties of the galaxies with an
inverted gradient as compared with the whole sample.
The average absolute $g$ luminosity and stellar mass for
the 201 galaxies are -20.41 mag and 10$^{10.09}$ M$_{\odot}$, respectively, 
while these numbers are only marginally different for
galaxies with an inverted O/H (-20.13 mag and 10$^{9.97}$ M$_{\odot}$)
or N/O gradient  (-20.36 mag and 10$^{10.07}$ M$_{\odot}$).

For the rest of the integrated properties, the observed
relations with abundance slopes appear to be caused by the 
relation with the stellar mass.
Therefore, in these 
other properties, there is not much difference between the whole sample and the subset of galaxies with an
inverted O/H or N/O gradient. The average colours, morphological
type, and classes as a function of the presence of a bar are almost identical
for galaxies with inverted and negative gradients.

It has been proposed that the flows of gas throughout and across 
the galaxy discs can sensibly alter the observed radial distribution 
of the metals  (e.g. \citealt{kewley10}).
In the case of our sample, interacting objects were ruled out, but we should not discard the effect of minor mergers or past interactions in a recent period,
after which the usual gradient shape has not yet been recovered  (e.g. \citealt{miralles14}).
Positive abundance gradients have been also observed in low-mass gas-rich
galaxies where a radial gas redistribution has possibly taken place
(e.g. NGC~2915, \citealt{werk10}).

In this sense, a flat or inverted gradient can be the evidence of an absence of correlation
of the radial abundance distribution  
to a greater extent than the existence of a homogeneous $Z$ throughout the disc. For
our sample the mean correlation coefficient of the calculated linear fittings 
for galaxies with a negative gradient 
is -0.29 for O/H and -0.52 for N/O, while for inverted gradients the mean correlation coefficient is +0.20 for O/H 
and only +0.04 for N/O. This implies that positive gradients observed for N/O are 
mostly a consequence of a random abundance distribution throughout the discs. 
On the contrary, this is not so clearly observed in the case of O/H, once
taken N/O into account for the derivation of O/H.

The simulations support the existence of positive gradients of $Z$ 
after an interaction both in the local Universe \citep{rupke10} and at larger 
$z$ \citep{tissera16} and the drop of $Z$ in 
the centres of galaxies after the fall of unprocessed gas (e.g. \citealt{lee04}).
The fall of metal-poor gas from the exterior to explain the inverted gradients 
could be supported by the fact that the characteristic average O/H at the effective 
radius in these galaxies is lower. In Table \ref{tab_slopes} we see that in these galaxies O/H 
is 0.04 dex lower on average. 
Besides, the N/O characteristic value is almost identical to inverted O/H
gradient, which would partially support the argument of the infall of
metal-poor gas, as N/O tio is not expected to change substantially 
when a gas interchange is produced with the IGM medium \citep{edmunds90}. 
On the contrary, galaxies with inverted N/O gradient present sensibly lower N/O
characteristic values, which could be indicative of other causes for the inversion of
the gradient, which is more related to the variation of the star formation efficient.

However, the influence of these gas flows should also be observed in the derived SFR and
neither for the integrated SFR nor for sSFR is any clear difference seen between galaxies with an 
inverted gradient and the whole sample, as shown in subsection \ref{SF-colour}.
At same time, we do not observed that the mean H$\alpha$ luminosity
of the \hii\ regions in galaxies with an inverted gradient is higher
(L(H$\alpha$ = 10$^{39.03}$ erg/s for regions in galaxies
with positive O/H slope and 10$^{39.13}$ erg/s for all \hii\ regions)
as seen in Fig. \ref{lha}. The mean O/H and N/O are also the same in
both sets of \hii\ regions.

An inner gas transportation and redistribution from the galaxies centres towards 
the outer positions could be also supported by the observation of the flattening 
of the $Z$ gradient beyond 2$\cdot$R$_e$ \citep[e.g.][]{bresolin09,
rosales11,marino12,S14_grad}. Additionally, it has been observed that a decrease of $Z$
in the inner regions of spiral galaxies in the  radial range 0.3-0.5 R$_e$ 
\citep[e.g.][]{rosales11,S12b,S14_grad} is possibly related to an accumulation of gas in 
rings, which can make he resulting slopes of the $Z$ gradients flatter.
For this work, no restriction was made in the fitted radial range 
Nevertheless, these two factors can have a certain influence 
on the fraction of observed flat or inverted gradients in our sample. By restricting
the analysed \hii\ regions to the range 0.5 - 2.1 · R$_e$, the number of galaxies
with at least 10 \hii\ regions is reduced to 147. For these the average slopes are 
-0.065 dex/R$_e$ for O/H and -0.115 dex/R$_e$ for N/O, which are slightly more pronounced 
than for the whole galaxy. The respective fractions of objects with an inverted gradient 
oscillates depending on the error from 9\% up to 17\% in the case of O/H and from 2\% up to 
12\% for N/O.

Another factor that can have an influence on the calculations and thus can bias 
the results towards flatter abundance gradients is the inclination of the galaxies. 
\cite{S14_grad} only use objects with an inclination that is lower than 70$^o$  to 
sample \hii\ regions at all galactocentric distances and to avoid a bias towards the 
regions at outer distances. We calculated the inclinations using the ratio
of the semi-axis ($b/a$) measured over the photometric growth curves given
by \cite{walcher14} and we used the same restriction as \cite{S14_grad}, which
reduces the sample of analysed galaxies to 113. The average slopes for these 
are -0.063 dex/R$_e$ for O/H and -0.125 dex/R$_e$ for N/O. The respective fractions 
of objects with an inverted gradient in this subsample ranges from 6\% up to 15\% for 
O/H and from 1\% up to 4\% for N/O. By combining this subsample of objects with an 
inclination that is lower than 70$^o$ with the fitting in the radial range 0.5-2.1 R$_e$  
(98 galaxies), the average slope is of -0.062 dex/R$_e$ for O/H and -0.113 dex/R$_e$  
for N/O. The fraction of galaxies with an inverted gradients ranges from 11\% up to 19\% 
for O/H and from 1\% up to 8\% for N/O. For this same subsample, by making calculations 
assuming the O/H values derived by {\sc HCm} without a previous determination
of N/O and assuming a typical O/H-N/O relation, we obtained an average O/H slope of 
-0.095 dex/R$_e$ and a very high p value in the Lilliefors normality test (0.94); this
approximates the results obtained by \cite{S14_grad} very accurately.

\section{Summary and conclusions}

In this work we studied the O/H and N/O abundance ratios  
of the ionised gas phase in individual 
\hii\ regions and the radial distributions of these abundances across the discs in a sample 
of 351 galaxies using the spatially resolved IFS data from the CALIFA 
survey \cite{S12}.

The abundances were derived in the selected star-forming \hii\ regions
using the semi-empirical code {\sc HCm} \citep{hcm}, 
based on photoionisation models. This code permits the derivation of 
O/H, N/O and log $U,$ covering a wide range of input values and is totally 
consistent with the direct method when the resulting abundances are compared
in samples of objects with measured electron temperature, as in \cite{hcm}.
This method takes the dependence of [\nii] emission lines on nitrogen 
abundances into account before its use to derive $Z,$ so it is
adequate to independently study O/H and N/O without considering any previous assumption
of their relation.

The analysis of the O/H-N/O relation for the selected
star-forming regions reveals a large dispersion
with a higher slope in relation to the sample of objects used to calibrate
the O/H-N/O in the models when N/O cannot be calculated, but this analysis is 
in agreement with a general scenario of higher N/O for higher O/H in the regime 
of production of secondary N.
Nevertheless, it cannot be ruled out that the discrepancies found
between the CALIFA \hii\ regions and the \hii\ regions with the direct 
method can be due to emission-line measurements or 
uncertainties in the methodology  used.
The range of variation of
N/O in this sample [-1.8,-0,5] is much larger than for O/H [8.4,8.9]
with a similar associated error, which\ supports the idea that N/O can be a much more
accurate indicator of the chemical status of an object than O/H.

The analysis of the radial distributions both for O/H and N/O 
normalised to the effective radius in those non-interacting galaxies 
with at least 10 star-forming \hii\ regions (201 galaxies) shows that 
most of the galaxies present a negative gradient, although these galaxies have a large 
dispersion with a mean slope $\alpha$ = -0.053 $\pm$ 0.068 dex/R$_e$ for O/H
and $\alpha$ = -0.104 $\pm$ 0.096 dex/R$_e$ for N/O. 
In fact, a non-negligible fraction of galaxies present flat or even inverted 
gradients (at least a 10\% for O/H and 4\% for N/O). A correlation between O/H 
and N/O slopes exists ($\rho_s$ = 0.39) but with large dispersion and even different 
combinations of O/H and N/O gradients can co-exist. This fact reveals the importance 
of a previous determination of N/O when [\nii] lines are used to derive O/H.

Contrary to individual \hii\ regions for which the relation between O/H and N/O shows
a very large dispersion and a relatively low correlation ($\rho_s$ = 0.37), the characteristic O/H 
and N/O values (i.e. the values of the linear fittings at the effective radius) are well 
correlated ($\rho_s$ = 0.80) and have a very low dispersion (i.e. the standard deviation
of the residuals is of 0.12 dex); this supports the idea that the global chemical conditions 
of a galaxy is mostly related with their integrated properties regardless of its inner spatial 
variations. This has been also observed for the CALIFA sample of galaxies for the stellar
population properties \citep{gonzalezd15}.

When analysing the distribution of slopes as a function of the galactic 
integrated properties, we see that galaxies with low stellar masses and absolute
$g$ luminosities tend to have flatter gradients, once normalised by the
effective radius  for O/H and N/O, but this 
result is not statistically significant. The same occurs with the morphological 
type, as early and late spiral galaxies have flatter gradients than intermediate-type objects. In addition, no significant variation of the median slopes is found 
as a function of the presence of a bar. These results agree with the analysis made 
for other samples of these CALIFA galaxies using other strong-line methods for \hii\ 
regions \citep{S14_grad, marino15}, in the analysis spaxel-to-spaxel \citep{smenguiano16} 
or even from the analysis of the stellar abundances \citep{sanchezb14}.

Regarding the characteristic O/H and N/O values at the effective radius, 
 the MZR and  MNOR are recovered with very low dispersion (i.e. 
the standard deviation of the residuals to a quadrant fitting is around 0.03 dex 
for both abundance ratios), which confirms that these global relations are better 
obtained when the characteristic value of a galaxy, independent of the inner 
spatial variations, is used. We also observe clear relations between these 
characteristic abundances for each galaxy and other integrated properties, 
but all of these can be interpreted in light of the MZR or the MNOR. 
In this way, although clear relations are observed between both O/H or N/O
with the integrated present-day SFR and sSFR, these relations result from the M-SFR 
relation followed by all these objects. The observed M-SFR-$Z$-N/O
relation makes it possible to observe a certain correlation between L(H$\alpha$) and O/H
(or N/O) for the individual \hii\ regions of the galaxies in this sample.
This absence of an additional dependence between O/H and SFR, as well as with N/O
in the regime of secondary production of nitrogen, agrees with the results found by \cite{S13}
for the sample of \hii\ regions in CALIFA galaxies.
We also observe clear correlations between abundances with integrated colours
and morphology but, again, these are understood in terms of a
sequence of earlier types and redder colours for more massive galaxies in this
sample as also seen by \citep{gonzalezd15}.

Finally, we explored the nature of those objects with flat and inverted gradients of $Z$
and N/O, but we did not find any special feature in these objects that can explain the 
inversion of the chemical gradient in the whole radial range. No significant variations 
in the average slopes or the fraction of inverted gradients are found when additional 
restrictions are taken that relate to the fitted radial range or the inclination of the 
galaxies, although these further reduce the proportion of these gradients. A deeper inspection of these galaxies considering the past history of minor interactions,
gas accretion, and radial flows should be considered to understand that causes
of these flat radial abundance distributions.

\begin{acknowledgements}
This study makes uses of the data provided by the Calar Alto Legacy 
Integral Field Area (CALIFA) survey (\url{http://califa.caha.es/}). 
CALIFA is the first legacy survey being performed at Calar Alto. 
The CALIFA collaboration would like to thank the IAA-CSIC and MPIA-MPG 
as major partners at the observatory, and CAHA itself, for the unique 
access to telescope time and support in manpower and infrastructures. 
The CALIFA collaboration also thanks the CAHA staff for their dedication 
to this project. 
EPM, JMV, CK, SP, and JIP acknowledge support from the Spanish MICINN through grants
AYA2010-21887-C04-01 and AYA2013-47742-C4-1-P and the Junta de Andaluc\'\i a for
grant EXC/2011 FQM-7058. 
RGB, RGD, and EP acknowledge support from grants AYA2014-57490-P and JA-FQM-2828.
Support for LG is provided by the Ministry of Economy, Development, and Tourism's Millennium Science Initiative through grant IC120009, awarded to The Millennium Institute of Astrophysics, MAS. LG acknowledges support by CONICYT through FONDECYT grant 3140566.
This research made use of python (\url{http://www.python.org}), of Matplotlib
\citep{hunter2007}, a suite of open-source python modules that provides a framework
for creating scientific plots. We also thank Dr. Christophe Morisset for many discussions and constructive
comments that have helped to improve the original manuscript.
\end{acknowledgements}

\bibliographystyle{aa}
\bibliography{califa_grads}

\begin{thebibliography}{94}
\expandafter\ifx\csname natexlab\endcsname\relax\def\natexlab#1{#1}\fi

\bibitem[{{Alloin} {et~al.}(1979){Alloin}, {Collin-Souffrin}, {Joly}, \&
  {Vigroux}}]{alloin}
{Alloin}, D., {Collin-Souffrin}, S., {Joly}, M., \& {Vigroux}, L. 1979, \aap,
  78, 200

\bibitem[{{Amor{\'{\i}}n} {et~al.}(2010){Amor{\'{\i}}n}, {P{\'e}rez-Montero},
  \& {V{\'{\i}}lchez}}]{amorin10}
{Amor{\'{\i}}n}, R.~O., {P{\'e}rez-Montero}, E., \& {V{\'{\i}}lchez}, J.~M.
  2010, \apjl, 715, L128

\bibitem[{{Baldwin} {et~al.}(1981){Baldwin}, {Phillips}, \& {Terlevich}}]{BPT}
{Baldwin}, J.~A., {Phillips}, M.~M., \& {Terlevich}, R. 1981, \pasp, 93, 5

\bibitem[{{Barrera-Ballesteros} {et~al.}(2015){Barrera-Ballesteros},
  {S{\'a}nchez}, {Garc{\'{\i}}a-Lorenzo}, {Falc{\'o}n-Barroso}, {Mast},
  {Garc{\'{\i}}a-Benito}, {Husemann}, {van de Ven}, {Iglesias-P{\'a}ramo},
  {Rosales-Ortega}, {P{\'e}rez-Torres}, {M{\'a}rquez}, {Kehrig}, {Marino},
  {Vilchez}, {Galbany}, {L{\'o}pez-S{\'a}nchez}, {Walcher}, \& {Califa
  Collaboration}}]{barrera}
{Barrera-Ballesteros}, J.~K., {S{\'a}nchez}, S.~F., {Garc{\'{\i}}a-Lorenzo},
  B., {et~al.} 2015, \aap, 579, A45

\bibitem[{{Bresolin} {et~al.}(2012){Bresolin}, {Kennicutt}, \&
  {Ryan-Weber}}]{bresolin12}
{Bresolin}, F., {Kennicutt}, R.~C., \& {Ryan-Weber}, E. 2012, \apj, 750, 122

\bibitem[{{Bresolin} {et~al.}(2009){Bresolin}, {Ryan-Weber}, {Kennicutt}, \&
  {Goddard}}]{bresolin09}
{Bresolin}, F., {Ryan-Weber}, E., {Kennicutt}, R.~C., \& {Goddard}, Q. 2009,
  \apj, 695, 580

\bibitem[{{Brinchmann} {et~al.}(2004){Brinchmann}, {Charlot}, {White},
  {Tremonti}, {Kauffmann}, {Heckman}, \& {Brinkmann}}]{brinchmann04}
{Brinchmann}, J., {Charlot}, S., {White}, S.~D.~M., {et~al.} 2004, \mnras, 351,
  1151

\bibitem[{{Bryant} {et~al.}(2015){Bryant}, {Owers}, {Robotham}, {Croom},
  {Driver}, {Drinkwater}, {Lorente}, {Cortese}, {Scott}, {Colless}, {Schaefer},
  {Taylor}, {Konstantopoulos}, {Allen}, {Baldry}, {Barnes}, {Bauer},
  {Bland-Hawthorn}, {Bloom}, {Brooks}, {Brough}, {Cecil}, {Couch}, {Croton},
  {Davies}, {Ellis}, {Fogarty}, {Foster}, {Glazebrook}, {Goodwin}, {Green},
  {Gunawardhana}, {Hampton}, {Ho}, {Hopkins}, {Kewley}, {Lawrence},
  {Leon-Saval}, {Leslie}, {McElroy}, {Lewis}, {Liske}, {L{\'o}pez-S{\'a}nchez},
  {Mahajan}, {Medling}, {Metcalfe}, {Meyer}, {Mould}, {Obreschkow}, {O'Toole},
  {Pracy}, {Richards}, {Shanks}, {Sharp}, {Sweet}, {Thomas}, {Tonini}, \&
  {Walcher}}]{sami}
{Bryant}, J.~J., {Owers}, M.~S., {Robotham}, A.~S.~G., {et~al.} 2015, \mnras,
  447, 2857

\bibitem[{{Bundy} {et~al.}(2015){Bundy}, {Bershady}, {Law}, {Yan}, {Drory},
  {MacDonald}, {Wake}, {Cherinka}, {S{\'a}nchez-Gallego}, {Weijmans}, {Thomas},
  {Tremonti}, {Masters}, {Coccato}, {Diamond-Stanic}, {Arag{\'o}n-Salamanca},
  {Avila-Reese}, {Badenes}, {Falc{\'o}n-Barroso}, {Belfiore}, {Bizyaev},
  {Blanc}, {Bland-Hawthorn}, {Blanton}, {Brownstein}, {Byler}, {Cappellari},
  {Conroy}, {Dutton}, {Emsellem}, {Etherington}, {Frinchaboy}, {Fu}, {Gunn},
  {Harding}, {Johnston}, {Kauffmann}, {Kinemuchi}, {Klaene}, {Knapen},
  {Leauthaud}, {Li}, {Lin}, {Maiolino}, {Malanushenko}, {Malanushenko}, {Mao},
  {Maraston}, {McDermid}, {Merrifield}, {Nichol}, {Oravetz}, {Pan}, {Parejko},
  {Sanchez}, {Schlegel}, {Simmons}, {Steele}, {Steinmetz}, {Thanjavur},
  {Thompson}, {Tinker}, {van den Bosch}, {Westfall}, {Wilkinson}, {Wright},
  {Xiao}, \& {Zhang}}]{manga}
{Bundy}, K., {Bershady}, M.~A., {Law}, D.~R., {et~al.} 2015, \apj, 798, 7

\bibitem[{{Cardelli} {et~al.}(1989){Cardelli}, {Clayton}, \&
  {Mathis}}]{Cardelli}
{Cardelli}, J.~A., {Clayton}, G.~C., \& {Mathis}, J.~S. 1989, \apj, 345, 245

\bibitem[{{Carton} {et~al.}(2015){Carton}, {Brinchmann}, {Wang}, {Bigiel},
  {Cormier}, {van der Hulst}, {J{\'o}zsa}, {Serra}, \& {Verheijen}}]{carton15}
{Carton}, D., {Brinchmann}, J., {Wang}, J., {et~al.} 2015, \mnras, 451, 210

\bibitem[{{Cid Fernandes} {et~al.}(2013){Cid Fernandes}, {P{\'e}rez},
  {Garc{\'{\i}}a Benito}, {Gonz{\'a}lez Delgado}, {de Amorim}, {S{\'a}nchez},
  {Husemann}, {Falc{\'o}n Barroso}, {S{\'a}nchez-Bl{\'a}zquez}, {Walcher}, \&
  {Mast}}]{CF13}
{Cid Fernandes}, R., {P{\'e}rez}, E., {Garc{\'{\i}}a Benito}, R., {et~al.}
  2013, \aap, 557, A86

\bibitem[{{Cid Fernandes} {et~al.}(2011){Cid Fernandes}, {Stasi{\'n}ska},
  {Mateus}, \& {Vale Asari}}]{CF11}
{Cid Fernandes}, R., {Stasi{\'n}ska}, G., {Mateus}, A., \& {Vale Asari}, N.
  2011, \mnras, 413, 1687

\bibitem[{{Contini} {et~al.}(2012){Contini}, {Garilli}, {Le F{\`e}vre},
  {Kissler-Patig}, {Amram}, {Epinat}, {Moultaka}, {Paioro}, {Queyrel}, {Tasca},
  {Tresse}, {Vergani}, {L{\'o}pez-Sanjuan}, \& {Perez-Montero}}]{massiv}
{Contini}, T., {Garilli}, B., {Le F{\`e}vre}, O., {et~al.} 2012, \aap, 539, A91

\bibitem[{{Cresci} {et~al.}(2010){Cresci}, {Mannucci}, {Maiolino}, {Marconi},
  {Gnerucci}, \& {Magrini}}]{cresci10}
{Cresci}, G., {Mannucci}, F., {Maiolino}, R., {et~al.} 2010, \nat, 467, 811

\bibitem[{{Croxall} {et~al.}(2016){Croxall}, {Pogge}, {Berg}, {Skillman}, \&
  {Moustakas}}]{croxall16}
{Croxall}, K., {Pogge}, R.~W., {Berg}, D.~A., {Skillman}, E.~D., \&
  {Moustakas}, J. 2016, ArXiv e-prints

\bibitem[{{Croxall} {et~al.}(2015){Croxall}, {Pogge}, {Berg}, {Skillman}, \&
  {Moustakas}}]{croxall15}
{Croxall}, K.~V., {Pogge}, R.~W., {Berg}, D.~A., {Skillman}, E.~D., \&
  {Moustakas}, J. 2015, \apj, 808, 42

\bibitem[{{Diaz}(1989)}]{diaz89}
{Diaz}, A.~I. 1989, in Evolutionary Phenomena in Galaxies, ed. J.~E. {Beckman}
  \& B.~E.~J. {Pagel}, 377--397

\bibitem[{{Dopita} \& {Evans}(1986)}]{de86}
{Dopita}, M.~A. \& {Evans}, I.~N. 1986, \apj, 307, 431

\bibitem[{{Edmunds}(1990)}]{edmunds90}
{Edmunds}, M.~G. 1990, \mnras, 246, 678

\bibitem[{{Ellison} {et~al.}(2008){Ellison}, {Patton}, {Simard}, \&
  {McConnachie}}]{ellison08}
{Ellison}, S.~L., {Patton}, D.~R., {Simard}, L., \& {McConnachie}, A.~W. 2008,
  \apjl, 672, L107

\bibitem[{{Evans} \& {Dopita}(1985)}]{ed85}
{Evans}, I.~N. \& {Dopita}, M.~A. 1985, \apjs, 58, 125

\bibitem[{{Falc{\'o}n-Barroso} {et~al.}(2011){Falc{\'o}n-Barroso},
  {S{\'a}nchez-Bl{\'a}zquez}, {Vazdekis}, {Ricciardelli}, {Cardiel}, {Cenarro},
  {Gorgas}, \& {Peletier}}]{Falcon}
{Falc{\'o}n-Barroso}, J., {S{\'a}nchez-Bl{\'a}zquez}, P., {Vazdekis}, A.,
  {et~al.} 2011, \aap, 532, A95

\bibitem[{{Ferland} {et~al.}(2013){Ferland}, {Porter}, {van Hoof}, {Williams},
  {Abel}, {Lykins}, {Shaw}, {Henney}, \& {Stancil}}]{cloudy}
{Ferland}, G.~J., {Porter}, R.~L., {van Hoof}, P.~A.~M., {et~al.} 2013, \rmxaa,
  49, 137

\bibitem[{{Florido} {et~al.}(2015){Florido}, {Zurita}, {P{\'e}rez},
  {P{\'e}rez-Montero}, {Coelho}, \& {Gadotti}}]{florido}
{Florido}, E., {Zurita}, A., {P{\'e}rez}, I., {et~al.} 2015, \aap, 584, A88

\bibitem[{{Garc{\'{\i}}a-Benito} {et~al.}(2015){Garc{\'{\i}}a-Benito},
  {Zibetti}, {S{\'a}nchez}, {Husemann}, {de Amorim}, {Castillo-Morales}, {Cid
  Fernandes}, {Ellis}, {Falc{\'o}n-Barroso}, {Galbany}, {Gil de Paz},
  {Gonz{\'a}lez Delgado}, {Lacerda}, {L{\'o}pez-Fernandez}, {de
  Lorenzo-C{\'a}ceres}, {Lyubenova}, {Marino}, {Mast}, {Mendoza}, {P{\'e}rez},
  {Vale Asari}, {Aguerri}, {Ascasibar}, {Bekerait*error*{\.e}},
  {Bland-Hawthorn}, {Barrera-Ballesteros}, {Bomans}, {Cano-D{\'{\i}}az},
  {Catal{\'a}n-Torrecilla}, {Cortijo}, {Delgado-Inglada}, {Demleitner},
  {Dettmar}, {D{\'{\i}}az}, {Florido}, {Gallazzi}, {Garc{\'{\i}}a-Lorenzo},
  {Gomes}, {Holmes}, {Iglesias-P{\'a}ramo}, {Jahnke}, {Kalinova}, {Kehrig},
  {Kennicutt}, {L{\'o}pez-S{\'a}nchez}, {M{\'a}rquez}, {Masegosa}, {Meidt},
  {Mendez-Abreu}, {Moll{\'a}}, {Monreal-Ibero}, {Morisset}, {del Olmo},
  {Papaderos}, {P{\'e}rez}, {Quirrenbach}, {Rosales-Ortega}, {Roth},
  {Ruiz-Lara}, {S{\'a}nchez-Bl{\'a}zquez}, {S{\'a}nchez-Menguiano}, {Singh},
  {Spekkens}, {Stanishev}, {Torres-Papaqui}, {van de Ven}, {Vilchez},
  {Walcher}, {Wild}, {Wisotzki}, {Ziegler}, {Alves}, {Barrado}, {Quintana}, \&
  {Aceituno}}]{GB15}
{Garc{\'{\i}}a-Benito}, R., {Zibetti}, S., {S{\'a}nchez}, S.~F., {et~al.} 2015,
  \aap, 576, A135

\bibitem[{{Garnett} \& {Shields}(1987)}]{gs87}
{Garnett}, D.~R. \& {Shields}, G.~A. 1987, \apj, 317, 82

\bibitem[{{Garnett} {et~al.}(1997){Garnett}, {Shields}, {Skillman}, {Sagan}, \&
  {Dufour}}]{garnett97}
{Garnett}, D.~R., {Shields}, G.~A., {Skillman}, E.~D., {Sagan}, S.~P., \&
  {Dufour}, R.~J. 1997, \apj, 489, 63

\bibitem[{{Gonz{\'a}lez Delgado} {et~al.}(2015){Gonz{\'a}lez Delgado},
  {Garc{\'{\i}}a-Benito}, {P{\'e}rez}, {Cid Fernandes}, {de Amorim},
  {Cortijo-Ferrero}, {Lacerda}, {L{\'o}pez Fern{\'a}ndez}, {Vale-Asari},
  {S{\'a}nchez}, {Moll{\'a}}, {Ruiz-Lara}, {S{\'a}nchez-Bl{\'a}zquez},
  {Walcher}, {Alves}, {Aguerri}, {Bekerait{\'e}}, {Bland-Hawthorn}, {Galbany},
  {Gallazzi}, {Husemann}, {Iglesias-P{\'a}ramo}, {Kalinova},
  {L{\'o}pez-S{\'a}nchez}, {Marino}, {M{\'a}rquez}, {Masegosa}, {Mast},
  {M{\'e}ndez-Abreu}, {Mendoza}, {del Olmo}, {P{\'e}rez}, {Quirrenbach}, \&
  {Zibetti}}]{gonzalezd15}
{Gonz{\'a}lez Delgado}, R.~M., {Garc{\'{\i}}a-Benito}, R., {P{\'e}rez}, E.,
  {et~al.} 2015, \aap, 581, A103

\bibitem[{{Ho} {et~al.}(2015){Ho}, {Kudritzki}, {Kewley}, {Zahid}, {Dopita},
  {Bresolin}, \& {Rupke}}]{ho15}
{Ho}, I.-T., {Kudritzki}, R.-P., {Kewley}, L.~J., {et~al.} 2015, \mnras, 448,
  2030

\bibitem[{Hunter(2007)}]{hunter2007}
Hunter, J.~D. 2007, Computing In Science \& Engineering, 9, 90

\bibitem[{{Kauffmann} {et~al.}(2003){Kauffmann}, {Heckman}, {Tremonti},
  {Brinchmann}, {Charlot}, {White}, {Ridgway}, {Brinkmann}, {Fukugita}, {Hall},
  {Ivezi{\'c}}, {Richards}, \& {Schneider}}]{kauffman03}
{Kauffmann}, G., {Heckman}, T.~M., {Tremonti}, C., {et~al.} 2003, \mnras, 346,
  1055

\bibitem[{{Kehrig} {et~al.}(2012){Kehrig}, {Monreal-Ibero}, {Papaderos},
  {V{\'{\i}}lchez}, {Gomes}, {Masegosa}, {S{\'a}nchez}, {Lehnert}, {Cid
  Fernandes}, {Bland-Hawthorn}, {Bomans}, {Marquez}, {Mast}, {Aguerri},
  {L{\'o}pez-S{\'a}nchez}, {Marino}, {Pasquali}, {Perez}, {Roth},
  {S{\'a}nchez-Bl{\'a}zquez}, \& {Ziegler}}]{Kehrig12}
{Kehrig}, C., {Monreal-Ibero}, A., {Papaderos}, P., {et~al.} 2012, \aap, 540,
  A11

\bibitem[{{Kelz} {et~al.}(2006){Kelz}, {Verheijen}, {Roth}, {Bauer}, {Becker},
  {Paschke}, {Popow}, {S{\'a}nchez}, \& {Laux}}]{kelz}
{Kelz}, A., {Verheijen}, M.~A.~W., {Roth}, M.~M., {et~al.} 2006, \pasp, 118,
  129

\bibitem[{{Kewley} {et~al.}(2001){Kewley}, {Dopita}, {Sutherland}, {Heisler},
  \& {Trevena}}]{kewley01}
{Kewley}, L.~J., {Dopita}, M.~A., {Sutherland}, R.~S., {Heisler}, C.~A., \&
  {Trevena}, J. 2001, \apj, 556, 121

\bibitem[{{Kewley} {et~al.}(2006){Kewley}, {Groves}, {Kauffmann}, \&
  {Heckman}}]{kewley06}
{Kewley}, L.~J., {Groves}, B., {Kauffmann}, G., \& {Heckman}, T. 2006, \mnras,
  372, 961

\bibitem[{{Kewley} {et~al.}(2010){Kewley}, {Rupke}, {Zahid}, {Geller}, \&
  {Barton}}]{kewley10}
{Kewley}, L.~J., {Rupke}, D., {Zahid}, H.~J., {Geller}, M.~J., \& {Barton},
  E.~J. 2010, \apjl, 721, L48

\bibitem[{{K{\"o}ppen} \& {Hensler}(2005)}]{kh05}
{K{\"o}ppen}, J. \& {Hensler}, G. 2005, \aap, 434, 531

\bibitem[{{Lamareille} {et~al.}(2004){Lamareille}, {Mouhcine}, {Contini},
  {Lewis}, \& {Maddox}}]{lamareille04}
{Lamareille}, F., {Mouhcine}, M., {Contini}, T., {Lewis}, I., \& {Maddox}, S.
  2004, \mnras, 350, 396

\bibitem[{{Lara-L{\'o}pez} {et~al.}(2010){Lara-L{\'o}pez}, {Cepa},
  {Bongiovanni}, {P{\'e}rez Garc{\'{\i}}a}, {Ederoclite}, {Casta{\~n}eda},
  {Fern{\'a}ndez Lorenzo}, {Povi{\'c}}, \& {S{\'a}nchez-Portal}}]{lara10}
{Lara-L{\'o}pez}, M.~A., {Cepa}, J., {Bongiovanni}, A., {et~al.} 2010, \aap,
  521, L53

\bibitem[{{Lee} {et~al.}(2004){Lee}, {Salzer}, \& {Melbourne}}]{lee04}
{Lee}, J.~C., {Salzer}, J.~J., \& {Melbourne}, J. 2004, \apj, 616, 752

\bibitem[{{Lequeux} {et~al.}(1979){Lequeux}, {Peimbert}, {Rayo}, {Serrano}, \&
  {Torres-Peimbert}}]{lequeux79}
{Lequeux}, J., {Peimbert}, M., {Rayo}, J.~F., {Serrano}, A., \&
  {Torres-Peimbert}, S. 1979, \aap, 80, 155

\bibitem[{{Lilly} {et~al.}(2007){Lilly}, {Le F{\`e}vre}, {Renzini}, {Zamorani},
  {Scodeggio}, {Contini}, {Carollo}, {Hasinger}, {Kneib}, {Iovino}, {Le Brun},
  {Maier}, {Mainieri}, {Mignoli}, {Silverman}, {Tasca}, {Bolzonella},
  {Bongiorno}, {Bottini}, {Capak}, {Caputi}, {Cimatti}, {Cucciati}, {Daddi},
  {Feldmann}, {Franzetti}, {Garilli}, {Guzzo}, {Ilbert}, {Kampczyk}, {Kovac},
  {Lamareille}, {Leauthaud}, {Borgne}, {McCracken}, {Marinoni}, {Pello},
  {Ricciardelli}, {Scarlata}, {Vergani}, {Sanders}, {Schinnerer}, {Scoville},
  {Taniguchi}, {Arnouts}, {Aussel}, {Bardelli}, {Brusa}, {Cappi}, {Ciliegi},
  {Finoguenov}, {Foucaud}, {Franceschini}, {Halliday}, {Impey}, {Knobel},
  {Koekemoer}, {Kurk}, {Maccagni}, {Maddox}, {Marano}, {Marconi}, {Meneux},
  {Mobasher}, {Moreau}, {Peacock}, {Porciani}, {Pozzetti}, {Scaramella},
  {Schiminovich}, {Shopbell}, {Smail}, {Thompson}, {Tresse}, {Vettolani},
  {Zanichelli}, \& {Zucca}}]{zcosmos}
{Lilly}, S.~J., {Le F{\`e}vre}, O., {Renzini}, A., {et~al.} 2007, \apjs, 172,
  70

\bibitem[{{L{\'o}pez-S{\'a}nchez} \& {Esteban}(2010)}]{lopezs10}
{L{\'o}pez-S{\'a}nchez}, {\'A}.~R. \& {Esteban}, C. 2010, \aap, 517, A85

\bibitem[{{L{\'o}pez-S{\'a}nchez} {et~al.}(2015){L{\'o}pez-S{\'a}nchez},
  {Westmeier}, {Esteban}, \& {Koribalski}}]{lopezs15}
{L{\'o}pez-S{\'a}nchez}, {\'A}.~R., {Westmeier}, T., {Esteban}, C., \&
  {Koribalski}, B.~S. 2015, \mnras, 450, 3381

\bibitem[{{Mannucci} {et~al.}(2010){Mannucci}, {Cresci}, {Maiolino}, {Marconi},
  \& {Gnerucci}}]{mannucci10}
{Mannucci}, F., {Cresci}, G., {Maiolino}, R., {Marconi}, A., \& {Gnerucci}, A.
  2010, \mnras, 408, 2115

\bibitem[{{Marino} {et~al.}(2012){Marino}, {Gil de Paz}, {Castillo-Morales},
  {Mu{\~n}oz-Mateos}, {S{\'a}nchez}, {P{\'e}rez-Gonz{\'a}lez}, {Gallego},
  {Zamorano}, {Alonso-Herrero}, \& {Boissier}}]{marino12}
{Marino}, R.~A., {Gil de Paz}, A., {Castillo-Morales}, A., {et~al.} 2012, \apj,
  754, 61

\bibitem[{{Marino} {et~al.}(2016){Marino}, {Gil de Paz}, {S{\'a}nchez},
  {S{\'a}nchez-Bl{\'a}zquez}, {Cardiel}, {Castillo-Morales}, {Pascual},
  {V{\'{\i}}lchez}, {Kehrig}, {Moll{\'a}}, {Mendez-Abreu},
  {Catal{\'a}n-Torrecilla}, {Florido}, {Perez}, {Ruiz-Lara}, {Ellis},
  {L{\'o}pez-S{\'a}nchez}, {Gonz{\'a}lez Delgado}, {de Lorenzo-C{\'a}ceres},
  {Garc{\'{\i}}a-Benito}, {Galbany}, {Zibetti}, {Cortijo}, {Kalinova}, {Mast},
  {Iglesias-P{\'a}ramo}, {Papaderos}, {Walcher}, \&
  {Bland-Hawthorn}}]{marino15}
{Marino}, R.~A., {Gil de Paz}, A., {S{\'a}nchez}, S.~F., {et~al.} 2016, \aap,
  585, A47

\bibitem[{{Marino} {et~al.}(2013){Marino}, {Rosales-Ortega}, {S{\'a}nchez},
  {Gil de Paz}, {V{\'{\i}}lchez}, {Miralles-Caballero}, {Kehrig},
  {P{\'e}rez-Montero}, {Stanishev}, {Iglesias-P{\'a}ramo}, {D{\'{\i}}az},
  {Castillo-Morales}, {Kennicutt}, {L{\'o}pez-S{\'a}nchez}, {Galbany},
  {Garc{\'{\i}}a-Benito}, {Mast}, {Mendez-Abreu}, {Monreal-Ibero}, {Husemann},
  {Walcher}, {Garc{\'{\i}}a-Lorenzo}, {Masegosa}, {Del Olmo Orozco},
  {Mour{\~a}o}, {Ziegler}, {Moll{\'a}}, {Papaderos},
  {S{\'a}nchez-Bl{\'a}zquez}, {Gonz{\'a}lez Delgado}, {Falc{\'o}n-Barroso},
  {Roth}, {van de Ven}, \& {Califa Team}}]{m13}
{Marino}, R.~A., {Rosales-Ortega}, F.~F., {S{\'a}nchez}, S.~F., {et~al.} 2013,
  \aap, 559, A114

\bibitem[{{Mattsson}(2009)}]{mattsson09}
{Mattsson}, L. 2009, in IAU Symposium, Vol. 254, IAU Symposium, ed.
  J.~{Andersen}, {Nordstr{\"o}ara}, B.~{m}, \& J.~{Bland-Hawthorn}, 43P

\bibitem[{{Miralles-Caballero} {et~al.}(2014){Miralles-Caballero},
  {D{\'{\i}}az}, {Rosales-Ortega}, {P{\'e}rez-Montero}, \&
  {S{\'a}nchez}}]{miralles14}
{Miralles-Caballero}, D., {D{\'{\i}}az}, A.~I., {Rosales-Ortega}, F.~F.,
  {P{\'e}rez-Montero}, E., \& {S{\'a}nchez}, S.~F. 2014, \mnras, 440, 2265

\bibitem[{{Moll{\'a}} {et~al.}(2009){Moll{\'a}}, {Garc{\'{\i}}a-Vargas}, \&
  {Bressan}}]{popstar}
{Moll{\'a}}, M., {Garc{\'{\i}}a-Vargas}, M.~L., \& {Bressan}, A. 2009, \mnras,
  398, 451

\bibitem[{{Moll{\'a}} {et~al.}(2006){Moll{\'a}}, {V{\'{\i}}lchez},
  {Gavil{\'a}n}, \& {D{\'{\i}}az}}]{molla06}
{Moll{\'a}}, M., {V{\'{\i}}lchez}, J.~M., {Gavil{\'a}n}, M., \& {D{\'{\i}}az},
  A.~I. 2006, \mnras, 372, 1069

\bibitem[{{Moran} {et~al.}(2012){Moran}, {Heckman}, {Kauffmann}, {Dav{\'e}},
  {Catinella}, {Brinchmann}, {Wang}, {Schiminovich}, {Saintonge},
  {Gracia-Carpio}, {Tacconi}, {Giovanelli}, {Haynes}, {Fabello}, {Hummels},
  {Lemonias}, \& {Wu}}]{moran12}
{Moran}, S.~M., {Heckman}, T.~M., {Kauffmann}, G., {et~al.} 2012, \apj, 745, 66

\bibitem[{{Moustakas} \& {Kennicutt}(2006)}]{mk06}
{Moustakas}, J. \& {Kennicutt}, Jr., R.~C. 2006, \apjs, 164, 81

\bibitem[{{P{\'e}rez-Montero}(2014)}]{hcm}
{P{\'e}rez-Montero}, E. 2014, \mnras, 441, 2663

\bibitem[{{P{\'e}rez-Montero} \& {Contini}(2009)}]{pmc09}
{P{\'e}rez-Montero}, E. \& {Contini}, T. 2009, \mnras, 398, 949

\bibitem[{{P{\'e}rez-Montero} {et~al.}(2013){P{\'e}rez-Montero}, {Contini},
  {Lamareille}, {Maier}, {Carollo}, {Kneib}, {Le F{\`e}vre}, {Lilly},
  {Mainieri}, {Renzini}, {Scodeggio}, {Zamorani}, {Bardelli}, {Bolzonella},
  {Bongiorno}, {Caputi}, {Cucciati}, {de la Torre}, {de Ravel}, {Franzetti},
  {Garilli}, {Iovino}, {Kampczyk}, {Knobel}, {Kova{\v c}}, {Le Borgne}, {Le
  Brun}, {Mignoli}, {Pell{\`o}}, {Peng}, {Presotto}, {Ricciardelli},
  {Silverman}, {Tanaka}, {Tasca}, {Tresse}, {Vergani}, \& {Zucca}}]{pm13}
{P{\'e}rez-Montero}, E., {Contini}, T., {Lamareille}, F., {et~al.} 2013, \aap,
  549, A25

\bibitem[{{P{\'e}rez-Montero} \& {D{\'{\i}}az}(2005)}]{pmd05}
{P{\'e}rez-Montero}, E. \& {D{\'{\i}}az}, A.~I. 2005, \mnras, 361, 1063

\bibitem[{{P{\'e}rez-Montero} \& {V{\'{\i}}lchez}(2009)}]{pmv09}
{P{\'e}rez-Montero}, E. \& {V{\'{\i}}lchez}, J.~M. 2009, \mnras, 400, 1721

\bibitem[{{P{\'e}rez-Montero} {et~al.}(2011){P{\'e}rez-Montero},
  {V{\'{\i}}lchez}, {Cedr{\'e}s}, {H{\"a}gele}, {Moll{\'a}}, {Kehrig},
  {D{\'{\i}}az}, {Garc{\'{\i}}a-Benito}, \& {Mart{\'{\i}}n-Gord{\'o}n}}]{pm11}
{P{\'e}rez-Montero}, E., {V{\'{\i}}lchez}, J.~M., {Cedr{\'e}s}, B., {et~al.}
  2011, \aap, 532, A141

\bibitem[{{Pichon} {et~al.}(2011){Pichon}, {Pogosyan}, {Kimm}, {Slyz},
  {Devriendt}, \& {Dubois}}]{pichon11}
{Pichon}, C., {Pogosyan}, D., {Kimm}, T., {et~al.} 2011, \mnras, 418, 2493

\bibitem[{{Pilyugin} \& {Grebel}(2016)}]{pg16}
{Pilyugin}, L.~S. \& {Grebel}, E.~K. 2016, \mnras, 457, 3678

\bibitem[{{Pilyugin} {et~al.}(2007){Pilyugin}, {Thuan}, \&
  {V{\'{\i}}lchez}}]{pil07}
{Pilyugin}, L.~S., {Thuan}, T.~X., \& {V{\'{\i}}lchez}, J.~M. 2007, \mnras,
  376, 353

\bibitem[{{Pilyugin} {et~al.}(2004){Pilyugin}, {V{\'{\i}}lchez}, \&
  {Contini}}]{pcv04}
{Pilyugin}, L.~S., {V{\'{\i}}lchez}, J.~M., \& {Contini}, T. 2004, \aap, 425,
  849

\bibitem[{{Queyrel} {et~al.}(2012){Queyrel}, {Contini}, {Kissler-Patig},
  {Epinat}, {Amram}, {Garilli}, {Le F{\`e}vre}, {Moultaka}, {Paioro}, {Tasca},
  {Tresse}, {Vergani}, {L{\'o}pez-Sanjuan}, \& {Perez-Montero}}]{queyrel12}
{Queyrel}, J., {Contini}, T., {Kissler-Patig}, M., {et~al.} 2012, \aap, 539,
  A93

\bibitem[{{Rich} {et~al.}(2012){Rich}, {Torrey}, {Kewley}, {Dopita}, \&
  {Rupke}}]{rich12}
{Rich}, J.~A., {Torrey}, P., {Kewley}, L.~J., {Dopita}, M.~A., \& {Rupke},
  D.~S.~N. 2012, \apj, 753, 5

\bibitem[{{Roberts} \& {Haynes}(1994)}]{rh94}
{Roberts}, M.~S. \& {Haynes}, M.~P. 1994, \araa, 32, 115

\bibitem[{{Rosales-Ortega} {et~al.}(2011){Rosales-Ortega}, {D{\'{\i}}az},
  {Kennicutt}, \& {S{\'a}nchez}}]{rosales11}
{Rosales-Ortega}, F.~F., {D{\'{\i}}az}, A.~I., {Kennicutt}, R.~C., \&
  {S{\'a}nchez}, S.~F. 2011, \mnras, 415, 2439

\bibitem[{{Rosales-Ortega} {et~al.}(2012){Rosales-Ortega}, {S{\'a}nchez},
  {Iglesias-P{\'a}ramo}, {D{\'{\i}}az}, {V{\'{\i}}lchez}, {Bland-Hawthorn},
  {Husemann}, \& {Mast}}]{Fabian12}
{Rosales-Ortega}, F.~F., {S{\'a}nchez}, S.~F., {Iglesias-P{\'a}ramo}, J.,
  {et~al.} 2012, \apjl, 756, L31

\bibitem[{{Roth} {et~al.}(2005){Roth}, {Kelz}, {Fechner}, {Hahn}, {Bauer},
  {Becker}, {B{\"o}hm}, {Christensen}, {Dionies}, {Paschke}, {Popow}, {Wolter},
  {Schmoll}, {Laux}, \& {Altmann}}]{Roth}
{Roth}, M.~M., {Kelz}, A., {Fechner}, T., {et~al.} 2005, \pasp, 117, 620

\bibitem[{{Rupke} {et~al.}(2010){Rupke}, {Kewley}, \& {Chien}}]{rupke10}
{Rupke}, D.~S.~N., {Kewley}, L.~J., \& {Chien}, L.-H. 2010, \apj, 723, 1255

\bibitem[{{S{\'a}nchez} {et~al.}(2006){S{\'a}nchez}, {Garc{\'{\i}}a-Lorenzo},
  {Jahnke}, {Mediavilla}, {Gonz{\'a}lez-Serrano}, {Christensen}, \&
  {Wisotzki}}]{S06}
{S{\'a}nchez}, S.~F., {Garc{\'{\i}}a-Lorenzo}, B., {Jahnke}, K., {et~al.} 2006,
  \nar, 49, 501

\bibitem[{{S{\'a}nchez} {et~al.}(2012{\natexlab{a}}){S{\'a}nchez}, {Kennicutt},
  {Gil de Paz}, {van de Ven}, {V{\'{\i}}lchez}, {Wisotzki}, {Walcher}, {Mast},
  {Aguerri}, {Albiol-P{\'e}rez}, {Alonso-Herrero}, {Alves}, {Bakos},
  {Bart{\'a}kov{\'a}}, {Bland-Hawthorn}, {Boselli}, {Bomans},
  {Castillo-Morales}, {Cortijo-Ferrero}, {de Lorenzo-C{\'a}ceres}, {Del Olmo},
  {Dettmar}, {D{\'{\i}}az}, {Ellis}, {Falc{\'o}n-Barroso}, {Flores},
  {Gallazzi}, {Garc{\'{\i}}a-Lorenzo}, {Gonz{\'a}lez Delgado}, {Gruel},
  {Haines}, {Hao}, {Husemann}, {Igl{\'e}sias-P{\'a}ramo}, {Jahnke}, {Johnson},
  {Jungwiert}, {Kalinova}, {Kehrig}, {Kupko}, {L{\'o}pez-S{\'a}nchez},
  {Lyubenova}, {Marino}, {M{\'a}rmol-Queralt{\'o}}, {M{\'a}rquez}, {Masegosa},
  {Meidt}, {Mendez-Abreu}, {Monreal-Ibero}, {Montijo}, {Mour{\~a}o},
  {Palacios-Navarro}, {Papaderos}, {Pasquali}, {Peletier}, {P{\'e}rez},
  {P{\'e}rez}, {Quirrenbach}, {Rela{\~n}o}, {Rosales-Ortega}, {Roth},
  {Ruiz-Lara}, {S{\'a}nchez-Bl{\'a}zquez}, {Sengupta}, {Singh}, {Stanishev},
  {Trager}, {Vazdekis}, {Viironen}, {Wild}, {Zibetti}, \& {Ziegler}}]{S12}
{S{\'a}nchez}, S.~F., {Kennicutt}, R.~C., {Gil de Paz}, A., {et~al.}
  2012{\natexlab{a}}, \aap, 538, A8

\bibitem[{{S{\'a}nchez} {et~al.}(2014){S{\'a}nchez}, {Rosales-Ortega},
  {Iglesias-P{\'a}ramo}, {Moll{\'a}}, {Barrera-Ballesteros}, {Marino},
  {P{\'e}rez}, {S{\'a}nchez-Blazquez}, {Gonz{\'a}lez Delgado}, {Cid Fernandes},
  {de Lorenzo-C{\'a}ceres}, {Mendez-Abreu}, {Galbany}, {Falcon-Barroso},
  {Miralles-Caballero}, {Husemann}, {Garc{\'{\i}}a-Benito}, {Mast}, {Walcher},
  {Gil de Paz}, {Garc{\'{\i}}a-Lorenzo}, {Jungwiert}, {V{\'{\i}}lchez},
  {J{\'{\i}}lkov{\'a}}, {Lyubenova}, {Cortijo-Ferrero}, {D{\'{\i}}az},
  {Wisotzki}, {M{\'a}rquez}, {Bland-Hawthorn}, {Ellis}, {van de Ven}, {Jahnke},
  {Papaderos}, {Gomes}, {Mendoza}, \& {L{\'o}pez-S{\'a}nchez}}]{S14_grad}
{S{\'a}nchez}, S.~F., {Rosales-Ortega}, F.~F., {Iglesias-P{\'a}ramo}, J.,
  {et~al.} 2014, \aap, 563, A49

\bibitem[{{S{\'a}nchez} {et~al.}(2013){S{\'a}nchez}, {Rosales-Ortega},
  {Jungwiert}, {Iglesias-P{\'a}ramo}, {V{\'{\i}}lchez}, {Marino}, {Walcher},
  {Husemann}, {Mast}, {Monreal-Ibero}, {Cid Fernandes}, {P{\'e}rez},
  {Gonz{\'a}lez Delgado}, {Garc{\'{\i}}a-Benito}, {Galbany}, {van de Ven},
  {Jahnke}, {Flores}, {Bland-Hawthorn}, {L{\'o}pez-S{\'a}nchez}, {Stanishev},
  {Miralles-Caballero}, {D{\'{\i}}az}, {S{\'a}nchez-Blazquez}, {Moll{\'a}},
  {Gallazzi}, {Papaderos}, {Gomes}, {Gruel}, {P{\'e}rez}, {Ruiz-Lara},
  {Florido}, {de Lorenzo-C{\'a}ceres}, {Mendez-Abreu}, {Kehrig}, {Roth},
  {Ziegler}, {Alves}, {Wisotzki}, {Kupko}, {Quirrenbach}, {Bomans}, \& {Califa
  Collaboration}}]{S13}
{S{\'a}nchez}, S.~F., {Rosales-Ortega}, F.~F., {Jungwiert}, B., {et~al.} 2013,
  \aap, 554, A58

\bibitem[{{S{\'a}nchez} {et~al.}(2011){S{\'a}nchez}, {Rosales-Ortega},
  {Kennicutt}, {Johnson}, {Diaz}, {Pasquali}, \& {Hao}}]{S11}
{S{\'a}nchez}, S.~F., {Rosales-Ortega}, F.~F., {Kennicutt}, R.~C., {et~al.}
  2011, \mnras, 410, 313

\bibitem[{{S{\'a}nchez} {et~al.}(2012{\natexlab{b}}){S{\'a}nchez},
  {Rosales-Ortega}, {Marino}, {Iglesias-P{\'a}ramo}, {V{\'{\i}}lchez},
  {Kennicutt}, {D{\'{\i}}az}, {Mast}, {Monreal-Ibero}, {Garc{\'{\i}}a-Benito},
  {Bland-Hawthorn}, {P{\'e}rez}, {Gonz{\'a}lez Delgado}, {Husemann},
  {L{\'o}pez-S{\'a}nchez}, {Cid Fernandes}, {Kehrig}, {Walcher}, {Gil de Paz},
  \& {Ellis}}]{S12b}
{S{\'a}nchez}, S.~F., {Rosales-Ortega}, F.~F., {Marino}, R.~A., {et~al.}
  2012{\natexlab{b}}, \aap, 546, A2

\bibitem[{{S{\'a}nchez-Bl{\'a}zquez} {et~al.}(2014){S{\'a}nchez-Bl{\'a}zquez},
  {Rosales-Ortega}, {M{\'e}ndez-Abreu}, {P{\'e}rez}, {S{\'a}nchez}, {Zibetti},
  {Aguerri}, {Bland-Hawthorn}, {Catal{\'a}n-Torrecilla}, {Cid Fernandes}, {de
  Amorim}, {de Lorenzo-Caceres}, {Falc{\'o}n-Barroso}, {Galazzi},
  {Garc{\'{\i}}a Benito}, {Gil de Paz}, {Gonz{\'a}lez Delgado}, {Husemann},
  {Iglesias-P{\'a}ramo}, {Jungwiert}, {Marino}, {M{\'a}rquez}, {Mast},
  {Mendoza}, {Moll{\'a}}, {Papaderos}, {Ruiz-Lara}, {van de Ven}, {Walcher}, \&
  {Wisotzki}}]{sanchezb14}
{S{\'a}nchez-Bl{\'a}zquez}, P., {Rosales-Ortega}, F.~F., {M{\'e}ndez-Abreu},
  J., {et~al.} 2014, \aap, 570, A6

\bibitem[{{S{\'a}nchez-Menguiano} {et~al.}(2016){S{\'a}nchez-Menguiano},
  {S{\'a}nchez}, {P{\'e}rez}, {Garc{\'{\i}}a-Benito}, {Husemann}, {Mast},
  {Mendoza}, {Ruiz-Lara}, {Ascasibar}, {Bland-Hawthorn}, {Cavichia},
  {D{\'{\i}}az}, {Florido}, {Galbany}, {G{\'o}nzalez Delgado}, {Kehrig},
  {Marino}, {M{\'a}rquez}, {Masegosa}, {M{\'e}ndez-Abreu}, {Moll{\'a}}, {Del
  Olmo}, {P{\'e}rez}, {S{\'a}nchez-Bl{\'a}zquez}, {Stanishev}, {Walcher},
  {L{\'o}pez-S{\'a}nchez}, \& {Califa Collaboration}}]{smenguiano16}
{S{\'a}nchez-Menguiano}, L., {S{\'a}nchez}, S.~F., {P{\'e}rez}, I., {et~al.}
  2016, \aap, 587, A70

\bibitem[{{Skillman}(1989)}]{skillman89}
{Skillman}, E.~D. 1989, \apj, 347, 883

\bibitem[{{Storey} \& {Hummer}(1995)}]{storey95}
{Storey}, P.~J. \& {Hummer}, D.~G. 1995, \mnras, 272, 41

\bibitem[{{Thuan} {et~al.}(2010){Thuan}, {Pilyugin}, \& {Zinchenko}}]{thuan10}
{Thuan}, T.~X., {Pilyugin}, L.~S., \& {Zinchenko}, I.~A. 2010, \apj, 712, 1029

\bibitem[{{Tissera} {et~al.}(2016){Tissera}, {Pedrosa}, {Sillero}, \&
  {Vilchez}}]{tissera16}
{Tissera}, P.~B., {Pedrosa}, S.~E., {Sillero}, E., \& {Vilchez}, J.~M. 2016,
  \mnras, 456, 2982

\bibitem[{{Tremonti} {et~al.}(2004){Tremonti}, {Heckman}, {Kauffmann},
  {Brinchmann}, {Charlot}, {White}, {Seibert}, {Peng}, {Schlegel}, {Uomoto},
  {Fukugita}, \& {Brinkmann}}]{t04}
{Tremonti}, C.~A., {Heckman}, T.~M., {Kauffmann}, G., {et~al.} 2004, \apj, 613,
  898

\bibitem[{{Troncoso} {et~al.}(2014){Troncoso}, {Maiolino}, {Sommariva},
  {Cresci}, {Mannucci}, {Marconi}, {Meneghetti}, {Grazian}, {Cimatti},
  {Fontana}, {Nagao}, \& {Pentericci}}]{troncoso14}
{Troncoso}, P., {Maiolino}, R., {Sommariva}, V., {et~al.} 2014, \aap, 563, A58

\bibitem[{{van Zee} \& {Haynes}(2006)}]{vh06}
{van Zee}, L. \& {Haynes}, M.~P. 2006, \apj, 636, 214

\bibitem[{{Vazdekis} {et~al.}(2010){Vazdekis}, {S{\'a}nchez-Bl{\'a}zquez},
  {Falc{\'o}n-Barroso}, {Cenarro}, {Beasley}, {Cardiel}, {Gorgas}, \&
  {Peletier}}]{Vazdekis}
{Vazdekis}, A., {S{\'a}nchez-Bl{\'a}zquez}, P., {Falc{\'o}n-Barroso}, J.,
  {et~al.} 2010, \mnras, 404, 1639

\bibitem[{{Veilleux} \& {Osterbrock}(1987)}]{VO87}
{Veilleux}, S. \& {Osterbrock}, D.~E. 1987, \apjs, 63, 295

\bibitem[{{Vila-Costas} \& {Edmunds}(1992)}]{vce92}
{Vila-Costas}, M.~B. \& {Edmunds}, M.~G. 1992, \mnras, 259, 121

\bibitem[{{Walcher} {et~al.}(2014){Walcher}, {Wisotzki}, {Bekerait{\'e}},
  {Husemann}, {Iglesias-P{\'a}ramo}, {Backsmann}, {Barrera Ballesteros},
  {Catal{\'a}n-Torrecilla}, {Cortijo}, {del Olmo}, {Garcia Lorenzo},
  {Falc{\'o}n-Barroso}, {Jilkova}, {Kalinova}, {Mast}, {Marino},
  {M{\'e}ndez-Abreu}, {Pasquali}, {S{\'a}nchez}, {Trager}, {Zibetti},
  {Aguerri}, {Alves}, {Bland-Hawthorn}, {Boselli}, {Castillo Morales}, {Cid
  Fernandes}, {Flores}, {Galbany}, {Gallazzi}, {Garc{\'{\i}}a-Benito}, {Gil de
  Paz}, {Gonz{\'a}lez-Delgado}, {Jahnke}, {Jungwiert}, {Kehrig}, {Lyubenova},
  {M{\'a}rquez Perez}, {Masegosa}, {Monreal Ibero}, {P{\'e}rez}, {Quirrenbach},
  {Rosales-Ortega}, {Roth}, {Sanchez-Blazquez}, {Spekkens}, {Tundo}, {van de
  Ven}, {Verheijen}, {Vilchez}, \& {Ziegler}}]{walcher14}
{Walcher}, C.~J., {Wisotzki}, L., {Bekerait{\'e}}, S., {et~al.} 2014, \aap,
  569, A1

\bibitem[{{Werk} {et~al.}(2011){Werk}, {Putman}, {Meurer}, \&
  {Santiago-Figueroa}}]{werk11}
{Werk}, J.~K., {Putman}, M.~E., {Meurer}, G.~R., \& {Santiago-Figueroa}, N.
  2011, \apj, 735, 71

\bibitem[{{Werk} {et~al.}(2010){Werk}, {Putman}, {Meurer}, {Thilker}, {Allen},
  {Bland-Hawthorn}, {Kravtsov}, \& {Freeman}}]{werk10}
{Werk}, J.~K., {Putman}, M.~E., {Meurer}, G.~R., {et~al.} 2010, \apj, 715, 656

\bibitem[{{Zaritsky} {et~al.}(1994){Zaritsky}, {Kennicutt}, \&
  {Huchra}}]{zaritsky94}
{Zaritsky}, D., {Kennicutt}, Jr., R.~C., \& {Huchra}, J.~P. 1994, \apj, 420, 87

\end{thebibliography}

\end{document}